\documentclass[12pt,preprint]{aastex}

\shortauthors{Koo et al.}
\shorttitle{Luminous Bulges of $z \sim 1$ Galaxies}

\newcommand{\etal } {{\it et al.}\ }
\newcommand{\kms}{${\rm km~s^{-1}}$}

\begin{document}
\title{The DEEP Groth Strip Survey VIII: The 
Evolution of Luminous Field Bulges at Redshift {\rm z} $\sim 1$
\altaffilmark{1}
}

\author{David C. Koo\altaffilmark{2},
Luc Simard\altaffilmark{2,3}, 
Christopher N. A. Willmer\altaffilmark{2,4},
Karl Gebhardt\altaffilmark{2,5,6},
Rychard J. Bouwens\altaffilmark{2},
Guinevere Kauffmann\altaffilmark{7},
Timothy Crosby\altaffilmark{8},
S. M. Faber\altaffilmark{2},
Justin Harker\altaffilmark{2},
Vicki L. Sarajedini\altaffilmark{2,11},
Nicole P. Vogt\altaffilmark{2,9},
Benjamin J. Weiner\altaffilmark{2},
Andrew J. Phillips\altaffilmark{2}, 
Myungshin Im\altaffilmark{2,10}, \&
K. L. Wu\altaffilmark{2,12}
}

\altaffiltext{1} {Based on observations obtained at the W. M. Keck
Observatory, which is operated jointly by the University of California
and the California Institute of Technology and on observations made
with the NASA/ESA {\it Hubble Space Telescope (HST)} obtained from the
data Archive at the Space Telescope Science Institute (STScI), which
is operated by the Association of Universities for Research in
Astronomy, Inc., under NASA contract NAS5-26555. These observations
are associated with proposals GTO 5090 and GTO 5109.}

\altaffiltext{2} {UCO/Lick Observatory, Department of Astronomy and 
Astrophysics,
University of California, Santa Cruz, CA 95064}
\email{ koo@ucolick.org, cnaw@ucolick.org, bouwens@ucolick.org,
 faber@ucolick.org, jharker@ucolick.org, bjw@ucolick.org, phillips@ucolick.org}

\altaffiltext{3} {Present address: Herzberg Institute of Astrophysics, National Research Council of Canada, 5071 West Saanich Road, Victoria, BC, V9E 2E7 
Canada}
\email {Luc.Simard@nrc-cnrc.gc.ca}

\altaffiltext{4} {On leave from Observatorio Nacional, MCT, CNPq, Rio de Janeiro, Brazil}

\altaffiltext{5}{Hubble Fellow}

\altaffiltext{6} {Department of Astronomy,
University of Texas, Austin, TX 78712}
\email {gebhardt@hoku.as.utexas.edu}

\altaffiltext{7} {Max-Planck Institut f\"ur Astrophysik, D-85740
Garching, Germany} 
\email{gamk@MPA-Garching.MPE.DE }

\altaffiltext{8} {Lincoln Laboratory,
Massachusetts Institute of Technology,
244 Wood Street,
Lexington, MA 02420-9108}
\email {crosby@mit.edu }

\altaffiltext{9} {Department of Astronomy, New Mexico State University, Las Cruces, NM 88003-8001}
\email{nicole@nmsu.edu}

\altaffiltext{10} {Astronomy Program, School of Earth and Environmental Sciences, 
Seoul National University, Seoul, South Korea}
\email {mim@astro8.snu.ac.kr}

\altaffiltext{11} {Astronomy Department, University of Florida, 211 Bryant Space
Science Center, Gainesville, FL 32611}
\email {vicki@astro.ufl.edu}

\altaffiltext{12} {Dept. of Chemistry and Physics, University of
Tampa,
401 West Kennedy Blvd. Tampa, FL 33606
}
\email{klwu@ut.edu}

\begin{abstract}
We present a candidate sample of luminous bulges (including
ellipticals) found within the Groth Strip Survey (GSS), with
spectroscopic redshifts of $0.73< z < 1.04$ from the Keck
Telescope. This work is distinguished by its use of 2-D two-component
decomposition photometry from Hubble Space Telescope ({\it HST})
images to separate the bulge from any disk before applying the sample
selection and to measure disk-free colors.  We define a statistically
complete sample of 86 bulges with $r^{1/4}$ profiles and luminosities
brighter than $I_{AB} = 24$.  Although larger samples of distant
early-type galaxies exist, this is the largest and most homogeneous
sample of {\it bulges} at $z \sim 1$ with spectroscopy.  A brighter
subset of 52 objects with added structural constraints defines our
``quality sample'' that is used to explore bulge luminosities and
colors.

We find that 85\% of luminous ($M_B < -19$) {\it field} bulges at
redshift $z \sim 0.8$ are nearly as red ($U-B \sim 0.50$) as local
E/S0's.  Almost all (90\%) of these very red bulges reside in
galaxies with the morphologies of normal early-type or spiral
galaxies.  Moreover, the slope of the color-luminosity
relation is shallow ($-0.04 \pm 0.04$) and the intrinsic $U-B$ color
dispersion is small ($\sigma \lesssim 0.03$ mag), suggesting roughly
coeval formation.  All three results are similar to that seen among
early-type {\it cluster} galaxies at the same epoch.

Yet we also measured $\sim 1$ mag increase in surface brightness.
Since simple passive evolution of a single-burst stellar population
results in redder colors as the galaxy fades, the observed constancy
of very red colors at high redshift suggests more complex histories.
One alternative starts with a {\it metal-rich} (twice solar),
early-formation ($z \sim 1.5-2.0$) population that is later polluted
with small amounts ($\sim$5\% by total mass) of star formation over an
extended period of several Gyr.  This ``drizzling'' history is
supported by our finding spectroscopic evidence for continued star
formation ([O II] emission lines) among 80\% of luminous high redshift
galaxies that have very red colors in both their bulges and disks.
Although some very red ($U-B \geq 0.25$) disks are found, almost all
disks have the same or bluer colors than their accompanying bulges,
regardless of the bulge-disk ratio and bulge luminosity. This
result matches the results of semi-analytic hierarchical galaxy
formation models, in which massive bulges are assembled from major
mergers of large disks with accompanying disks forming later from gas
infall.

Finally, we measure the integrated very-red ($U-B \geq 0.25$) bulge
light at $z \sim 0.8$ to be $\sim 7\times 10^7L_{\odot}
Mpc^{-3}$. This amount is roughly one-third of the restframe $B$
luminosity for all GSS galaxies at that redshift. The uncertainties in
both local and our distant bulge luminosity densities remain too
large to settle the issue of whether a large fraction of bulges
were formed or assembled after a redshift $z \sim 1$.
  
Blue ($U-B < 0$) bulge candidates are present, but only as a minor
(8\%) population.  In general, such candidates have luminosities and
surface brightnesses {\it lower} than that of the very red bulges;
have large disk fractions by luminosity; and have emission linewidths
typically less than 100 \kms. These properties are all {\it
inconsistent} with those predicted for star-forming progenitors of the
luminous bulges of today, i.e., the blue photo-bulges are not
genuine blue ellipticals or bulges.  Moreover, over 60\% of the
bulge candidates that are not very red appear to reside in galaxies
with morphologies suggestive of interactions and mergers.  Thus our
deeper, more extensive, and less disk-contaminated observations
challenge prior claims by other groups that 30\% to 50\% of field
bulges or ellipticals are in a blue, star-forming phase at redshifts
$z < 1$.

We conclude, with the caveat that {\it luminous} ellipticals and
bulges at $z \sim 1$ have $r^{1/4}$ light profiles, that they, as do
luminous early-type cluster galaxies at the same redshift, are already
dominated by metal-rich, old stellar populations that have been fading
from a formation epoch earlier than redshift $z > 1.5$. Only small
amounts of residual star formation are needed to explain both the
absence of bluening of bulges to today and the presence of emission
lines seen in the Keck spectra of the very-red distant galaxies.

\end{abstract}


\section {Introduction}

\subsection {Background}

As reviewed by \citet*{wyse97},
the ages of bulges (defined here to be the equivalent of 
the term spheroids that include
ellipticals \footnote {
E, E/S0,  and S0 morphological
types together as a class are often called
spheroidal galaxies or more succinctly designated in this paper
as E-S0, to avoid
the ambiguity of the E/S0 designation and confusion with bulges.} 
and the bulges of S0's and spirals) remain
an important unsolved problem in stellar populations and galaxy
formation. Moreover, the formation of bulges is  now of enhanced
interest given the discovery of the tight relationship between the
masses and velocity dispersions of local bulges and the masses of
black holes in galactic nuclei \citep{magorrian98, ferrarese00, gebhardt00}.

To explore the ages, formation mechanisms, and evolution of bulges,
astronomers have taken two basic observational approaches --- 1) to
study the fossil records imprinted in the luminosities, colors,
kinematics, spatial distribution, and chemical abundances of stars
and in gas distributions in {\it local} bulges
and 2)
to study the more global properties (structure and stellar
populations) of galaxies distant enough in lookback time to
reveal  the evolution and even perhaps formation of bulges $in \  situ$.

As one example of the latter approach, $HST$ data of distant galaxies
were compared to plausible formation scenarios by \citet*{bouwens99}.  
They specified three basic models: 1) a secular evolution
model in which bulges first form 2 Gyr after disks; 2) a simultaneous
formation model in which bulge formation commences at the formation
time of the disks; and 3) an early bulge formation model in which bulges
and field E-S0's form before disks.
Models 1) and 2) both predict that a large fraction of distant
bulges are luminous and very blue, while model 3) predicts mainly
very red bulges. By examining the colors and bulge-to-total
ratios (B/T) for about 60 galaxies in the literature with redshifts
$0.3 < z \lesssim 1$, \citet{bouwens99}  found  that they were {\it 
unable to differentiate among the models}.  The larger sample and
higher-redshift  needed for discrimination is met by the new 
sample presented here. 
Our new data
unambiguously exclude models 1) and 2), with   only model 3) matching well enough to be viable.

Another area of controversy is whether E-S0's were 1) predominantly
formed in a rapid burst of star formation at high redshifts (e.g., $z
> 2$), or instead 2) formed their stars mainly at later epochs
(redshifts $z < 1$) via merging.  As comprehensively reviewed by
\citet{schade99}, the evidence is extensive but inconclusive.
Clusters show consistent results from different studies: a tight
color-magnitude relation for E-S0's and very red colors that persist
to quite high redshifts ($z \sim 1$).  These findings support the
early bulge formation scenario, at least for some cluster galaxies
(see \citealt{dokkum01} giving a more complicated model).  Studies of
field populations, on the other hand, show no such consistency.
Several studies favor scenarios with extensive and recent
evolution by claiming that 30\% to 50\% of E-S0's are blue at high
redshifts ($z < 1$) or that the volume density of elliptical and red
galaxies was 2 or 3 times lower in the past.  Other
studies find little evidence for such recent dramatic evolution.

\citet{schade99} tried to  address  this issue. Based on 11 ellipticals
with spectroscopic redshifts $z \sim 1$, they measured luminosity
evolution that matches that of passively evolving cluster galaxies and
found no evidence for a major decline in volume density since $z \sim
1$. On the other hand, they had two results that are inconsistent
with a strictly old stellar population: blue colors for their
ellipticals and strong [O II] emission lines.

In a more recent  work using HST optical and near-infrared ground photometry
of E-S0's  \citep {ellis01}, the authors find that 
the centers of non-peculiar spirals with prominent bulges are
redder than the colors of the surrounding disks.
This is one of the robust
predictions of hierarchical models, namely  
that disks form after bulge formation.
These central colors, presumably dominated by the bulge,
are, however, bluer than those of 
most pure ellipticals at the same redshifts (up to
$z \sim 1$). This result in the optical appears to contradict
the robust prediction of hierarchical galaxy formation models that
spiral bulges should on average be older (i.e., redder) than
pure ellipticals  \citep*{kauffmann96a, baugh96}.  
On the other hand,
while the central colors {\it in the near infrared} of spiral bulges
remain bluer than
most ellipticals at low redshifts ($ z < 0.6$), \cite{ellis01} find that they
become as red or redder than that of ellipticals at higher
redshifts. Ellis \etal  surmise that this difference in relative colors in the
optical and near infrared could be explained by star formation
in bulges that occurs through bursts rather than more continuous
activity. They also speculate that the match in redshifts of this change in
behavior to that found for the disappearance of barred spirals
\citep{abraham99} might support the secular formation of at least some bulges at
low redshifts.

Further evidence for continued formation of field E-S0's since
$z \sim 1$ comes from two other surveys.  \cite{stanford04} find that  roughly half of the 
early-type galaxies (may include some early spirals) 
found to just beyond $z \sim 1$  and identified by morphology using the HST 
near-infrared (NICMOS) images, are bluer than predicted by passive evolution of  an early burst.
Another work   finds  strong internal 
spatial variations in the 
colors of more than 30\% of the faint E-S0's in the HDF
\citep*{menanteau01}. They do not find such variations
in cluster galaxies and estimate ``that at $z \sim 1$, about half
the field spheroidals  must be undergoing recent episodes of
star-formation,'' a result qualitatively expected in some  hierarchical models
of elliptical formation.

\subsection {Present Work}

To readdress these issues on {\it field} bulge formation, 
the first phase of the  DEEP {\footnote {Deep Extragalactic Evolutionary
Probe: see URL http://deep.ucolick.org/}}
survey has focused on several pilot programs that 
rely on a redshift survey of over 1000 faint (median
$I_{AB} \sim 22.3$) field galaxies. These data have been taken with
the first generation of spectrographs on the W. M. Keck 10 m telescopes
and are complemented with $HST$ imaging and ground-based multicolor
photometry \citep{koo98}.  
The second phase of  DEEP (DEEP2) is  a much
more extensive survey of about 50,000 galaxies reaching similar 
limits of
$R_{AB} \sim 24$ and exploiting multicolor photometry to isolate
galaxies with redshifts $z \gtrsim 0.7$ \citep{faber03, davis03}. 

As part of phase one,  DEEP has recently completed the
acquisition and reduction of 604 redshifts in the Groth Strip Survey
(see \ref{samplechar}).
The present work is one of four papers addressing
the nature of early-type galaxies and bulges at high
redshifts $z \sim 1$.  In one companion paper, \citet[][:GSS9]{gebhardt03}
extract internal absorption-line velocity dispersions of 36
galaxies and add luminosities and surface brightness data from
$HST$ images to study the evolution of the Fundamental Plane from
redshifts $z \sim 0.2 - 1$.  In another companion paper, \citet[][: GSS10]{im02} identify a sample of 145 E-S0 candidates over a
wide redshift range ($0.1 < z \lesssim 1$) and brighter than $I_{AB} =
22.5$ to tackle the issue of the volume density evolution of E-S0's; 
this sample also includes galaxies with only
photometric redshifts. In a third related paper \citep{im01}, the likely descendants of 10
distant blue spheroidal  candidates are examined in more detail.

The present work isolates a {\it spectroscopic} redshift
sample of 86 candidate bulges at high redshifts $0.73 < z < 1.04$
with 
$r^{1/4}$ {\it light profiles} and brighter than $I  =  23.57$. This limit
ensures high completeness, which is important for studies of volume
densities, and good-quality photometry, which is needed for deriving
reliable structural parameters and colors. 
Unlike most surveys selected by the brightness of the total galaxy,
this sample is selected on the brightness of the bulge  alone. 
Note that our selection in
the $I$ passband corresponds roughly to selection in restframe $B$ at redshift $z
\sim 0.8$.

  Several key issues can be addressed by this sample that are
not part of the other two main companion papers (GSS9 and GSS10):

1) What are the colors of the bulge  without contamination
from the disk? 

2) How do these colors relate to other properties of the galaxies such
as disk and galaxy colors, bulge-to-total ratios ({\it B/T}), galaxy or bulge
luminosities, and bulge sizes or surface brightnesses?

3) What is  the total elliptical and bulge  luminosity density at high redshifts?

The paper is organized as follows.  Section 2 gives an overview of the
HST and Keck observations and reduction procedures and details the
determination of the selection function for the bulge candidate
sample. Section 3 describes the sample
characteristics and correlations among colors, luminosities, {\it  B/T}, 
sizes, and  surface brightnesses. This section also makes estimates of
the luminosity density of  distant, old bulge  stellar populations. Readers wishing to
bypass   the details may want to examine figures \ref {ub_mag}  to \ref{size_mag} and 
otherwise skip Sections 2 and 3 and jump directly to the
discussion.  The discussion in Section 4 starts with a summary of
the key results from Section 3 and then  
compares results to those of the other DEEP papers mentioned
above and to those of other bulge-related surveys. Section 5 makes
direct comparisons to predictions of several models of bulge formation
from Bouwens  and to semi-analytic models of
Kauffmann. Section 6 closes the discussion with a summary of our key
conclusions and implications for the formation of ellipticals  and
the bulges of S0's and spirals. The appendix includes further discussion of the sample
selection function, additional figures comparing the observations to theoretical
predictions, and detailed notes on individual objects.

We adopt a Hubble constant $H_o$ = 70~\kms Mpc$^{-1}$ and a flat
cosmology with $\Omega_m = 0.3$ and $\Omega_{\Lambda} = 0.7$.
At redshift $z \sim 1$, this cosmology yields a scale of
1 arcsec = 8.0 kpc, while 
$L^*$  of galaxies today at $B \sim -20.2$ appears at $I_{814} \sim 23.34$ for a very red
spectral type and at $I_{814} \sim 22.75$ for an actively star forming
galaxy with restframe $B-V < 0.6$. The lookback time is 7.7 Gyr
for a universe that is 13.5 Gyr old. 
Our photometry is in the Vega
system (see \citealt{fukugita95} for conversion factors and definitions) with
$V_{606} \sim V - X$,  where $X$ ranges from 0.2 to 1.0,
depending on the spectral shape and redshift; 
$I_{814} \sim I_C + 0.08$, 
where $I_C$ is that of Cousins.
For conversion to the $AB$ system:
$I_{AB,814} = I_{814} + 0.434$,
and 
$V_{AB,606} = V_{606} + 0.111$

Our sample limit of $I_{814} = 23.566$ is the same as $I_{AB,814} = 24$.
Throughout the paper, $V$ will refer to $V_{606}$ and $I$ to $I_{814}$
for data from $HST$. 

Colors for our sample are in the HST WFPC2 $V_{606} - I_{814}$ system,
corresponding roughly to restframe $U-B$ at redshifts $z \sim 0.8$
(see Fig. A11 in GSS9).  To quantify our color terminology, the
demarcation between ``red'' and ``blue'' is at restframe $U-B = 0$,
which is the average color of Sbc galaxies \citep{fukugita95}. For
passively evolving populations formed at high redshifts $ z > 1.5$,
the colors will be ``very red'', i.e., $U-B \geq 0.25$ since $z \sim
1$.  As needed, we will adopt finer binnings and clarify the divisions
between blue and red adopted by previous studies.








\section {OBSERVATIONS}

\subsection{Structural Measurements from HST Images} \label{GIM2D}

This section provides a brief summary of the overall survey as
detailed by \citet[][: GSS1]{vogt04} and the procedures to produce
the structural measurements as detailed by \citet[][: GSS2]{simard02}.

The {\it HST} data 
known as the ``Groth Strip Survey'' (GSS) consists of 28 overlapping WFPC2 subfields 
oriented NE to SW at roughly 14:17+52 at Galactic latitude $b \sim
60\deg$. 
All subfields have
exposures of 2800 s in the broad $V$ filter ($F606W$) and 4400 s in the
broad $I$ filter ($F814W$) that reach a detection limit of $I \sim 26$,
except for one subfield 
with total exposures of 24,400 s in $V$ and 25,200 s in $I$.  
Object catalogs were produced with SExtractor version 1.0a 
\citep{bertin96} 
while the surface brightness profiles of galaxies in the object catalog 
were fitted with a PSF-convolved 2D two-component model (GIM2D: GSS2;
\citealt*{simard98, marleau98}).
The best
fitting parameter values along with their confidence intervals  were found 
using Monte-Carlo sampling of parameter
space to maximize the likelihood function.

The first photometric component (which we term the ``photo-bulge,'' short  for 
photometric bulge) of the 2D 
surface brightness model  
is a S\'ersic profile of the form:

 \begin{equation}
 	\Sigma(r) = \Sigma_{e} {\rm exp} \{-k[(r/r_{e})^{1/n} - 1]\}
 	\label{sersic}
 \end{equation} 

\noindent 
where $\Sigma(r)$ is the surface brightness at $r$ {\it along the
semi-major axis} in linear flux units per unit area, $r_e$ is the
bulge effective radius and $\Sigma_{e}$ is the surface brightness at
this radius.  The parameter $k$ was set equal to 1.9992$n$ $-$ 0.3271
so that $r_{e}$ remained the projected major-axis radius enclosing
half of the light in this component.  Thus the effective radius measured 
with a circular aperture is $r_e \sqrt{b/a}$ where $a$ and $b$ are the
major and minor axis sizes, respectively.

Although we have the option of letting $n$ be another free parameter
in our fits, our data, relative to that used in fits of nearby
galaxies, are of much lower S/N and have poorer spatial sampling.  We
thus choose for this starting work on distant bulges to lock $n$ to a
constant for the current analysis, namely the classical de Vaucouleurs
profile value of 4.  This is certainly an oversimplification and
likely to be incorrect for bulges in general. Local, late-type spiral
galaxies with $B/T \leq 0.1$, for example, are better fit by $n$ = 1,
i.e., an exponential profile \citep*{dejong94}. Furthermore, to improve
convergence of the fitting, an exponential profile for bulges may be
justified even if it is not theoretically the best fit
\citep{dejong96b}. Our sample of bulges is, however, quite luminous,
roughly $M^{*}$ or brighter at redshift $z \sim 1$, and is thus still
quite massive even after allowing for one or two magnitudes of
possible luminosity evolution.  Since there is extensive evidence that
{\it bright} local ellipticals and the bulges of early-type spiral
galaxies generally follow such a profile more closely than an
exponential \citep{dejong94, andredakis95, courteau96}, our choice of
$n$ = 4 is justified as a reasonable starting assumption. Future work
with much deeper  data on distant galaxies, e.g., the Hubble Ultra Deep Field (UDF {\footnote{http://www.stsci.edu/hst/udf}}; PI. S. Beckwith),  should explore a wider range of
bulge profiles.

The second component (or  ``photo-disk'',  short for photometric disk) is a
simple exponential profile of the form:

\begin{equation}
	\Sigma(r) = \Sigma_{0} {\rm exp} (-r/r_d), 
	\label{disk}
\end{equation}

\noindent where $\Sigma_{0}$ is the face-on  central surface brightness, 
$r$ is the radius along the major axis, and $r_d$ is the disk scale length.


When referring
to the GIM2D {\it photometric parameters}, 
we adopt the terms photo-bulge ({\it pB}), photo-disk ({\it pD}), and
photo-bulge to total ratio ({\it pB/T}), since the true structure and
internal kinematics of the components from the photometric fits remain
uncertain.  The presence of an exponential component, e.g., does not
necessarily imply the presence of an actual disk, since dynamically
hot systems may also have exponential profiles \citep{lin83,
kormendy85}.  Likewise, an $r^{1/4}$ component may represent a central
starburst or an AGN rather than a genuine, dynamically-hot bulge.
Additional complications in interpretation arise when our simplifying
assumption of a smooth, symmetric, exponential disk is invalidated by
the presence of inner or outer truncations in the disks, of bars, of
spiral arms, of rings, of tidal distortions, etc. Finally, we note
that even bonafide pure ellipticals (e.g., cD's or dwarf ellipticals)
may yield photo-disk components in the fits if their true profiles do
not follow our assumed $n = 4$ de Vaucouleurs shape exactly and vice
versa, genuine, pure disk galaxies may masquerade as having both a
bulge and disk if the disk is not a pure exponential or if the
exponential disk has a color gradient when the simultaneous GIM2D fit
is adopted as described below.

\label{simul_fit_section}

For this work, we used the simultaneous fit option of GIM2D (GSS2). In
this case, the scale lengths, central positions, ellipticities, and
position angles of each component were made to be the same in both the
$V$ and $I$ images, and only the fluxes were free to vary to yield a
color (GSS2).  The underlying assumption in adopting this approach is
that any internal color gradients or color-dependent asymmetries in
the bulge and disk components can be neglected.  As an empirical check
of this assumption, we selected a sample limited to photo-bulges
brighter than $I \sim 24$ in our spectroscopic sample and compared the
photo-bulge sizes as measured from separate GIM2D fits to the
$I_{814}$ and $V_{606}$ images. Although the scatter was high,
typically a factor of two, we found no evidence for any systematic
color gradients among either the very red or blue photo-bulges that
dominated (90\%) the sample. The less red photo-bulges did appear to
have larger effective radii in the blue image, indicating a redder
center. But these comprised only a 10\% fraction of the total sample,
not enough to justify using separate GIM2D fits for the analysis,
especially given the improved precision in colors (typically a factor
of two) by adopting simultaneous GIM2D fits.  This check for color
gradients was repeated for the photo-disk scale lengths. Except for
perhaps a 20\% larger scale length in the bluer image for the few
(12\%) very large photo-disks, we again found no systematic color
gradients discernible within our random errors.

The simultaneous fits have 
three major advantages over the separate fits. First, simultaneous
fits ensure that derived colors represent flux ratios as measured over
the same spatial regions. There is no such assurance for colors
derived from separate GIM2D fits, in which measurements in the $V$ and
$I$ images may use different centers, ellipticities (or inclination
angles), and position angles.  Second, the analysis and discussion of
colors is greatly simplified when using the simultaneous fit
method
by having only one average color value for each photo-bulge and
photo-disk component. In contrast, the separate fit method results in
at least a mean color and color gradient for each component, and this
is meaningful only if the ellipticity (or inclination angle) and
position angles are measured to be the same in both bands. Third, the
simultaneous fit method dramatically reduces the number of free
parameters by locking the central positions, sizes, ellipticities or
inclination angles, and position angles to be the same for each of the
photo-bulge and photo-disk components in the two bands.  This
reduction of up to 8 free parameters when using the simultaneous fit
method yields colors that have significantly smaller and more reliable
random errors, typically a factor of two, than from separate fits. We
next address systematic errors in using the simultaneous fit method.

\noindent 

Following the approach described in Section 3.4 of \citet{marleau98},
GSS2 simulated 6000 model galaxies with variations in the luminosities for
the bulge and disk components to reflect the observed range of $V-I$
colors from 0.5 to 2.2 and to depths corresponding to $V_{606} = 26$;
sizes of each ranged  from 0$\arcsec$ to 0.7$\arcsec$; bulge eccentricities
from 0 to 0.7; and disk inclinations from $0\deg$ to $85\deg$. After adding
Poisson noise, simulated images were placed within actual sky frames
and then analyzed in exactly the same way as real galaxies. This
procedure allowed a good test of the reliability of the GSS parameter
values as measured with GIM2D.

The main purpose of using simultaneous fits is to improve the color
estimates.  Based on simulations of this mode of GIM2D, GSS2 conclude
that there are no significant systematic color errors for the galaxy,
bulge, or disk.  For bulges with $I_{814}$ from 22 to 23.5, near our
sample limit of 23.566, the average difference between the measured
and input $V-I$ colors is only 0.03, small compared to the amount of
color evolution expected (0.2 mag or more). Outliers were occasionally
found, especially in cases where the disk and bulge colors or their
relative sizes were at the extremes. In some cases, disks and bulges
were interchanged from the input values.  Thus a search was made in
the simulations for regions of bulge fraction, ratio of bulge and disk
size, and bulge to disk colors where bulges were mistaken for disks
and vice versa.  No regions were found with systematic errors, though
outliers did exist, especially when bulge/disk ratios were very small
or very large. These simulations also show that reliable photo-bulge
fluxes and $V-I$ colors (systematic errors of less than 0.04 mag and
typical random errors of 0.1 mag to 0.3 mag) can be expected for
half-light sizes greater than $\sim 0.3$ pixels (0.03 arcsec).  For a
given photo-bulge flux, both the systematic and random errors are
found to be actually smaller for smaller sizes until the 0.03 arcsec
limit. We will later adopt this threshold in defining the bulge
sample.

As explained in GSS2, systematic biases are expected in {\it B/T} at
the extremes, i.e., near $pB/T = 0$, where some measurements are
underestimates of the true values, and near 1, where they are
overestimates. Based on the systematic errors derived in GSS2, only 8
galaxies from our total sample of 86 objects are subject to systematic
errors in {\it pB/T} greater than 0.04, and all are overestimates.
Five are in the "quality sample" defined in section 3.4: 094\_1313
(0.07), 094\_6234(0.09), 103\_2074(0.09), 104\_6432(0.07), and
113\_3646(0.07) where the values in parentheses are the systematic
overestimates of {\it B/T}.  Three others not in the quality sample
are: 092\_6027(0.07), 153\_5853(0.14), and 313\_4845(0.11). Note that
the simulations show that random errors in {\it B/T} are roughly 0.1
to 0.15 over the range of our data, so the 8 objects above are the
only ones with systematic errors approaching random errors.

\begin{figure}
\plotone{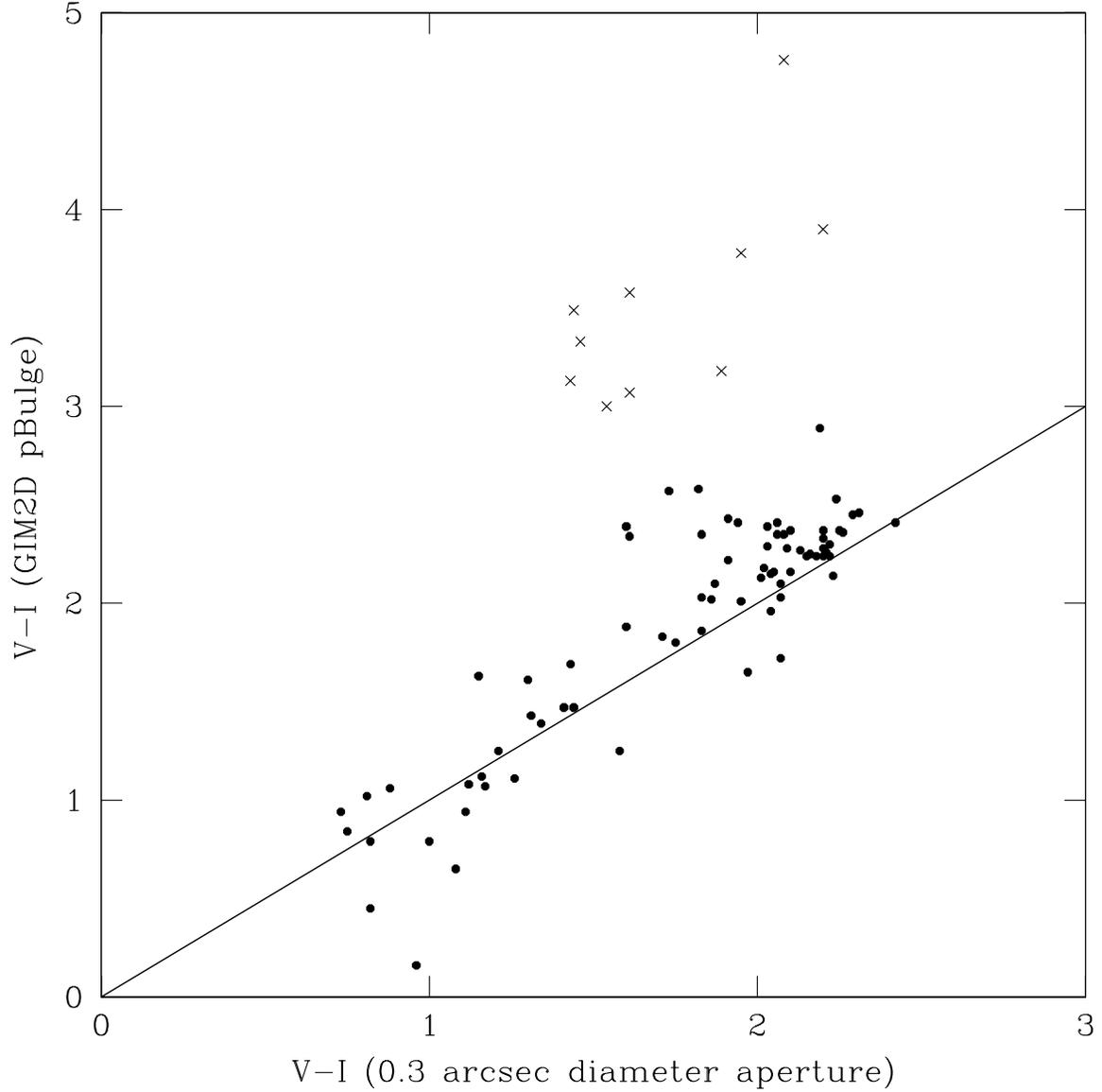}
\caption[aperture.eps] 
{GIM2D photo-bulge color versus  central aperture (circular 0.3 arcsec diameter) color for the full sample of 86 galaxies. Crosses mark objects
with $V-I$ color errors greater than 1.2 mag peak to peak at the 68\%
confidence limits (see Table 2). The solid line shows  the track for 
equal colors.   
\label{aperture}}
\end{figure}

Besides the simulations, we also compared colors measured using
circular apertures to those derived by GIM2D.  The $V-I$ colors from
seven aperture diameters of 0.3, 0.4, 0.5, 1, 1.5, 3, and 6 arcsecs
were compared among themselves and to the galaxy, photo-bulge, and
photo-disk colors from GIM2D for the bulge sample.  The most revealing
were the smallest aperture colors, so these central (0.3 arcsec
diameter) aperture colors have been included with the GIM2D photometry
in Table 1.  Assuming bulges have steeper light profiles than
exponential disks, the contamination from any disk light is expected
to be less for smaller apertures, and indeed we find that the central
aperture colors are almost always consistent with the GIM2D
photo-bulge colors (see Fig.~\ref{aperture}), with a median difference
of less than 0.12 in $V-I$ and in the expected sense that the GIM2D
colors are redder, since they should be less contaminated by any bluer
disk light than raw colors measured via apertures. The few exceptions showing
a blue nuclear color ($V-I < 1.6$) and a redder ($V-I > 1.8$) GIM2D
photo-bulge color included cases where GIM2D identified the whole
galaxy as being a very-red photo-bulge while a central blue component
was considered a photo-disk (e.g., 283\_5331).  In several cases, the
bulk of the galaxy is blue, the aperture color is redder, and the
GIM2D photo-bulge color is very red (e.g., 294\_2078).  The GIM2D
photo-bulge colors in these cases, however, have very large errors
($\sim 0.7$ mag in the example), so such discrepancies are not
statistically significant. Fig.~\ref{aperture} shows that {\it all}
the extremely red GIM2D colors for photo-bulges have large estimated
errors greater than 1.2 magnitudes peak to peak.

Finally, besides
using aperture photometry and simulations to check our GIM2D results,
we have also visually examined color images of the central 1 arcsec
regions of
the galaxies to check the photo-bulge colors. This sanity
check confirms that GIM2D is giving
reasonable photo-bulge results
for almost all objects, with the few illusory exceptions being
those
where GIM2D claims the presence of a tiny red photo-disk imbedded
within
a larger, bluer photo-bulge (e.g., 273\_7619).

\subsection {Keck/LRIS Spectroscopy} 

Spectra of GSS galaxies have been obtained with the Low Resolution
Imaging Spectrograph (LRIS, \citealt{oke95}) on the W.M. Keck II 10 m
Telescope.  DEEP/GSS galaxies were selected predominantly by the
magnitude criterion $(V+I)/2 \sim R < 24$ in 27 subfields and $< 25$ in the
Deep Field.  Spectra of over 600  galaxies were acquired between May 1995
and April 1999, and these are the spectra used in this paper.  Two
separate exposures with red and blue gratings covered a total spectral range of about
4500--9100\ \AA , depending on the exact position of
the target on the multi-slit masks. A 900 lines mm$^{-1}$ grating
gave a central wavelength $\simeq$ 5800 \AA, dispersion of 0.85\AA/pixel,  and
resolution $\sim$ 3--4 \AA\thinspace FWHM. A 600 lines mm$^{-1}$
grating gave a  central wavelength $\simeq$ 7700 \AA, dispersion of 1.26
\AA/pixel, and resolution $\sim$ 4--5 \AA\thinspace FWHM. Slit widths
ranged from 1.0\arcsec to 1.23\arcsec. Typical
total exposure times per target and per grating were 2700 seconds.
Rectified, wavelength-calibrated, sky-subtracted 2D spectra were
produced with a custom LRIS reduction package.  Details of the
spectral reduction are provided by Weiner \etal 
(2004: GSS3). 

\subsection {Selection Function for the Bulge Sample} \label{sel_fct}

Since selection effects may mimic real evolutionary
changes in the high-redshift galaxy population, it is important to
determine how they affect the DEEP/GSS sample in general and the present bulge
sample in particular.  Our approach has two parts.  The
first is to use simulations to determine the incompleteness of our
photometric catalog from which the spectroscopic sample is derived.
The second is to use a purely empirical determination of any
incompleteness of the final, spectroscopically-confirmed sample by
comparing it to the full photometric catalog.  In both cases,
simplifying assumptions as detailed in Appendix A are
adopted in the analysis of selection functions.

The selection function can be quantified by a weight, $W$, for each
bulge that is proportional to the inverse of the effective areal
coverage of the entire GSS sample (134 square arcmins) and which
combines the selection functions that depend on multiple parameters.
The total area covered by the spectroscopic survey is 90 square
arcmins, with the minimum value for $W \sim 1.5$.  For this work on
bulges, we restrict the dependencies of the weight to a small subset
of possible parameters that will be part of our analysis, namely,
apparent flux, size (or surface brightness), color of the photo-bulge,
and the photo-bulge to total ratio ({\it pB/T}).  A more detailed
discussion of selection functions, but for disks rather than bulges,
is provided in \citet{simard99}. Appendix A summarizes the main steps
adopted for this study of the high redshift bulges. In general, we
find that photo-bulge flux appears to dominate the dependencies and
there is no evidence for any significant dependencies of $W$ on {\it
pB/T}, size, or color at greater than 95\% confidence
limit.  Based
on the empirically derived ratio of the observed spectroscopic sample
to that of the entire photometric sample, we adopt a simple selection
function as follows:

	{\rm W(76/28)} = 2.7 for $I_{pB}$ between 20 and 21;

	{\rm W(57/26)} = 2.2 for $I_{pB}$ between 21 and 21.5; and 

	{\rm W(532/167)} = $2.0*I_{pB}$ - 41.5 for $I_{pB} > 21.5$, ~~~~~~~~~~~~(3)

\noindent
where $I_{pB}$ is the $I_{814}$ magnitude  of the photo-bulge
component as measured in the catalog using separate fits
to the $HST$ \  $I$ and $V$ images (see Table 1). The numbers
in parentheses show the total number of objects in the photometric 
catalog over the number of spectroscopic targets. The slightly greater
weight for the brightest interval reflects our selection bias against
the very brightest galaxies for the spectroscopic survey. Further 
explanation of the weights is given in Appendix A.

\subsection{K Corrections}

To compare our high redshift observations to local samples, we have
chosen $M_B$ for luminosity and $U-B$ for color, since the $V_{606}$
and $I_{814}$ filters coincide roughly with restframe $U$ and $B$ at
redshifts near $z \sim 0.8$. These choices reduce 
uncertainties in the K-corrections that result from variations in the
spectral energy distributions (SED) of galaxies. Slightly bluer
restframe bands would be better matched to our data near redshift $z
\sim 1$, but few local observations would then be available for
comparison.

To convert our observed $I_{814}$ magnitudes ($I$) and $V_{606}-I_{814}$
colors [($V-I$)] to restframe $M_B$ and $U-B$, we adopt the following parametric
conversions from GSS9:

\bigskip
$U-B = -0.8079-0.049752z-1.6232z^2+1.04067z^3+1.5294z^4-0.41190z^5-0.56986z^6
+(0.61591+1.07249z-2.2925z^2+1.3370z^3)(V-I)+(0.280481-0.387205z+0.043121z^2)(V-I)^2$,

and 

\bigskip
$M_B = I_{814} -DM(\Omega_m, \Omega_{\Lambda}, \Omega_K) + K_{IB}$,

\bigskip
\noindent
where DM is the distance modulus for the adopted cosmology and

\bigskip
$K_{IB} = 0.0496+0.46057z+1.40430z^2-0.19436z^3
 -0.2232z^4-0.36506z^5+0.17594z^6
 +(2.0532-2.8326z+1.05580z^2-0.67625z^3)(V-I)
 +(0.10826-0.68097z+0.61781z^2)(V-I)^2
$

\bigskip
\noindent
is the K-correction to convert from our $I$ band observations to
restframe $M_B$.

These transformations are valid in the redshift range $0.1 \leq z 
\leq 1.1$
and were derived from a subset (34 spectra) of an atlas  of 43 spectra of
local galaxies that extend far enough into the UV to match our filters
beyond redshifts $z \sim 1$ (see \citealt{kinney96} or GSS9 for
details). Two key advantages over the use of theoretical SED's from stellar
population synthesis \citep[such as][]{bruzual03} are
1) the empirical inclusion of other factors that affect the SED, such 
as internal
dust, variations in metallicity, and
emission lines and 2) the extraction of intrinsic dispersions to the 
fits
that yield estimates of the K-correction uncertainties.  We
find an RMS  in the $U-B$ conversion 
that varies from about 0.03 mag at redshifts $z \sim 0.8$
to about 0.08 mag at redshifts $z \sim 1$. The
$M_B$ conversion has an intrinsic dispersion of roughly $0.25*|z - 0.8|$
mag, i.e. about 0.05 mag at $z \sim 1$.  We avoided the use of the
popular set of empirical SEDs from \citet{coleman80}  
or other sources for K-corrections that depend on this
set \citep[e.g.,][]{fukugita95}, because they comprise a very limited
sample of only a few SEDs and, of more serious concern, 
are composites of spectra that do not
actually match the SEDs of individual galaxies, thus introducing
systematic errors.  The major downside of using
local SEDs rather than model SEDs is the possibility that evolution
may affect the K-corrections, but given how close our filters are to 
the
restframe bands of interest, any such  biases  are likely  to be small. 
The values for $M_B$ and $U-B$ in Table 3 derived from Table 2 are 
based
on the above relations.

\subsection {Data Tables and Appendices}

 
Table 1 provides the source identifications for the full sample of 86
candidate bulges; their J2000 coordinates; $I$ magnitudes, $V-I$
colors, and half-light radii of the whole galaxy and of the
photo-bulge subcomponent from the GIM2D catalog of the entire GSS {\it
using separate two-component fits} to each of $V$ and $I$.  The
redshift and redshift quality; weight $W$ for use in estimating number
densities; and notes that identify other publications on the galaxy
are also given.  Table 1 is ordered
by the source identification name
with the sequential number having
an asterisk (*) added for those
objects in the higher-quality subsample
discussed below. The
separate-fit catalog was used to select the starting sample in the
present work; to determine the selection function (which would not be
possible with the simultaneous catalog which included only the
spectroscopic sample); and possible dependencies of the selection
function on galaxy or photo-bulge flux, color, and size.

Appendix C provides comments for 66 galaxies, including more details
on possible problems in the GIM2D fits, on emission-line velocity
width data (see GSS3), and on identification of special subsamples
such as the most luminous galaxies in which both the photo-bulge and
photo-disks are very red (i.e., good S0 candidates).

Table 2, sorted by sequential numbers and source ID as in Table 1,
provides the {\it measured quantities} from GIM2D using {\it
simultaneous two-component fits in the two filters}. These are the
measurements we have previously argued to be more accurate and
reliable than from the separate-fit catalog.  Besides the $I$
magnitudes and $V-I$ colors for the whole galaxy, photo-bulge, and
photo-disk, the table gives {\it pB/T} as measured in the $I$ band;
the major-axis effective radius of the photo-bulge and face-on
scale-length of the photo-disk in arcsecs; the eccentricity of the
photo-bulge; the inclination of the disk in degrees; and the reduced
$\chi^2$ of the fits in each of the $V$ and $I$ images. Random errors
at the 68\% confidence level (roughly one sigma for normal
distributions) from GIM2D fits are provided for all measurements. No
corrections for any systematic or random errors as determined from
simulations have been included (see Section ~\ref{GIM2D}).  Note that
errors of 0.00 returned by GIM2D correspond to values below 0.01, but
all of these have been increased to 0.01 in the tables.

Table 3 provides {\it derived quantities} that depend on the choice of
cosmology (i.e., $h = 0.7, \Omega_m = 0.3, \Omega_{\Lambda} = 0.7$)
and K-corrections, including absolute magnitudes (and {\it pB/T}) in
rest-frame $B$ and rest-frame $U-B$ colors for the whole galaxy, the
photo-bulge, and photo-disk; the major-axis effective (half-light)
radius of the photo-bulge; the face-on exponential scale length of the
photo-disk in kpc; and the cosmology independent surface-brightness of
the photo-bulge in rest-frame $B$ per square arcsec as measured within
the effective radius. As in Table 2, the errors are at the 68\%
confidence level, though again, no corrections have been applied for
systematic or random errors (as determined from simulations and
discussed in Section ~\ref{GIM2D}).

Table 4 consolidates the various subsamples divided by color and
structure that will be relevant to the discussion of
results. Statistics are provided for both the total sample of 86
bulge candidates and the ``quality'' sample of 52 candidates as
defined below.

Table 5 consolidates the measurements of median $U-B$ colors of the
photo-bulges and integrated galaxies for various subsamples here and from other studies .

\section {RESULTS}

\subsection{Sample Characteristics}\label{samplechar}

We will work with two bulge samples.  The larger one of 86 galaxies
represents a magnitude-limited and thus a statistically complete
sample of bulges.  This sample is compared to the full galaxy redshift
sample, is used to address whether our photo-bulges are genuine
bulges, and provides the data for estimating the high-redshift,
red-bulge luminosity density. From the larger sample, a smaller
``Quality Sample'' of 52 objects is extracted.  This Quality Sample is
designed to have brighter bulges to yield better colors and photometry
and to have a more reliable sample of genuine bulges by removing
galaxies where GIM2D claims a very tiny disk embedded within a larger
bulge.

The larger bulge sample is extracted from the full GSS spectroscopic
redshift set with Keck redshifts (henceforth ``full GSS-SRS'') using
four selection criteria.

1) The redshift $z$ must be between 0.73 and 1.04.  The lower limit
was originally aimed to be at $z = 0.75$ to match that of the high
redshift bin of the CFRS sample and where the $HST$ $V$ filter matches
3500 \AA, close to restframe $U$.  But we found a small spike of
redshifts centered at $z = 0.75$ and thus decided to lower the limit
to include it.  The upper redshift limit was chosen so that [O II]
3727 \AA \ just enters into the deep atmospheric A-band absorption at
7600 \AA. The photo-bulge sample could be significantly expanded by
increasing the redshift range, but at the cost of increasing the
uncertainty in the K-corrections or in the homogeneity of the
colors. At $z = 1.04$, $HST$ $I$ is at restframe 4000 \AA, close to
the midpoint between $U$ and $B$, and $HST$ $V$ samples restframe 3000
\AA.

2) The spectroscopic redshift should be reliable, i.e., {\it Qz} in
column 12 of Table 1 must be 2.9 or greater. Individual objects were
examined in detail so that we are virtually certain that redshifts are
reliable.

3) The photo-bulge component alone must be brighter than $I_{814} =
23.566$ (i.e., $I_{AB} = 24$). This flux limit was chosen to be near
that yielding RMS errors of 0.5 mag for the photo-bulge in $I$. Random
$V-I$ color errors for very red photo-bulges are so large at this
limit that we will later reduce it by another 0.5 mag to improve the
quality of the photo-bulge colors.

4) The half-light size of the photo-bulge must be greater than 0.03 arcsec (0.3
pixels). This limit is chosen to exclude two  clear-cut cases of AGN
that masquerade as photo-bulges and to reduce the systematic and random errors of
the photo-bulge (Section 2.1). 

These four limitations reduce the full GSS-SRS set of 603 objects
(stars, galaxies, and AGNs) to 86 objects.  With the redshift cut 1)
alone, the full GSS-SRS sample would be reduced to 216 objects; with
the added redshift quality cut 2), the sample loses 5 to 211; with the
addition of the cut by bulge luminosity 3), the sample reduces to 88
galaxies; and finally, with the bulge size cut 4), two obvious AGN's
(142\_4838, a likely Seyfert 1, and 273\_4925, a likely QSO; 
see Sarajedini \etal 2004 [GSSXII]) are
eliminated to yield the  final sample of 86.  In principle, we could
increase the sample by including redshifts from the Canada France
Redshift Survey (CFRS: \citealt{lilly95}).  Out of 31 galaxies in
their 14h redshift sample that are not already in the GSS-SRS, 7 fall
within our high redshift range.  Of these, only one has a photo-bulge
brighter than $I_{814} = 23.566$, namely, CFRS ID 14.0411 (GSS ID 043\_3071),
but it was excluded to retain homogeneity in the spectral information.
The remaining paragraphs in this section compare this photo-bulge
sample to that of the full 556 GSS-SRS {\it galaxy}  sample 
with reliable redshifts. 

\clearpage

\begin{figure}
\plotone{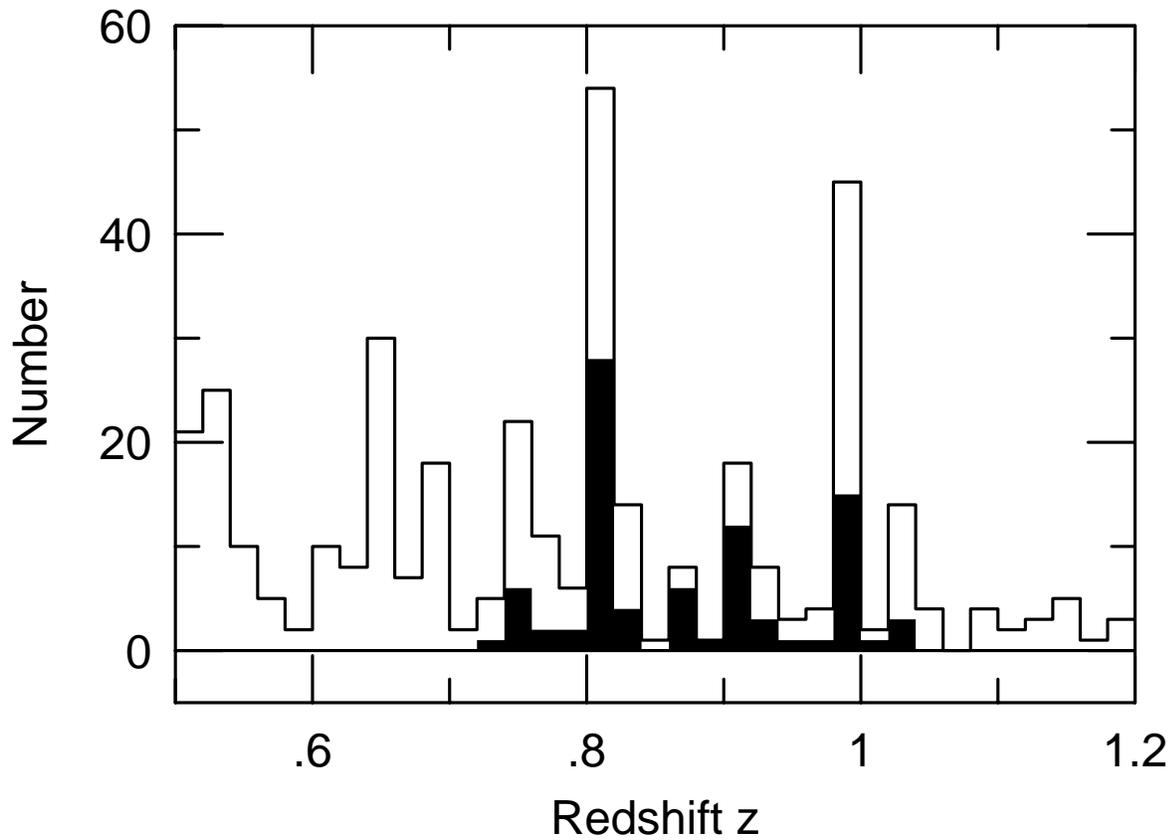}
\caption[histz.eps] 
{Histograms of the full, high-redshift GSS spectroscopic sample (open)
and the photo-bulge sample (filled); note 
the drop of objects in the full sample 
beyond $z \sim 1.03$, where [O II] 3727 \AA \ 
falls  into the 7600 \AA \ atmospheric A-band absorption. No 
comparable reduction in the number of 
objects is seen near $z \sim 0.91$ where the 4000\ \AA \ break is 
expected to be harder to discern when it enters the 7600\ \AA \ 
absorption feature.
\label{z_hist}}
\end{figure}

\clearpage

{\it Redshift Distribution:} As seen in Fig.~\ref{z_hist}, the
redshift distribution of the full GSS-SRS shows two strong spikes at
redshifts 0.81 and 0.99, with two weaker concentrations at 0.75 and
0.91. Both of the strong spikes extend across the full GSS field (42
arc minutes, or roughly 19 Mpc) and are thus likely parts of larger
superclusters rather than galaxies in the cores of rich clusters (see
further discussions of these features by \citealt{lefevre94, koo96},
and \citealt{weiner04}).  The photo-bulge sample appears
representative by showing redshift peaks at the same redshifts, though
the proportions drop significantly at redshifts near $z$ = 1.0, as
expected when criterion (3) (bulge luminosity) is taken into account.

{\it Spatial Distribution:} When compared to the spatial distribution
of the full, {\it high-redshift} GSS-SRS of 211 galaxies, the
photo-bulge sample shows an excess in field 9, i.e., those with source
ID of 09X\_YYYY.  More specifically, 30 of the 211 are in field 9
while the photo-bulge sample includes 19 of these. These 19/30 (63\%)
can be compared to the corresponding 67/181 (37\%) in the remaining
fields.  Somewhat surprising is a strong concentration of 9/19 targets
in the field-9 photo-bulge sample at redshifts between 0.900 to 0.905,
rather than at the stronger peaks at 0.81 and 0.99. This result
suggests the presence of a rich group of galaxies with luminous
bulges at $z$ = 0.90 within field 9. This field happens to be near
one of the clusters claimed by \citet{ostrander98}, but the DEEP
redshifts show their putative cluster to be a mixture of different
redshifts. None of our conclusions are changed if field-9 galaxies are
excluded from the analysis.

\begin{figure}
\plotone{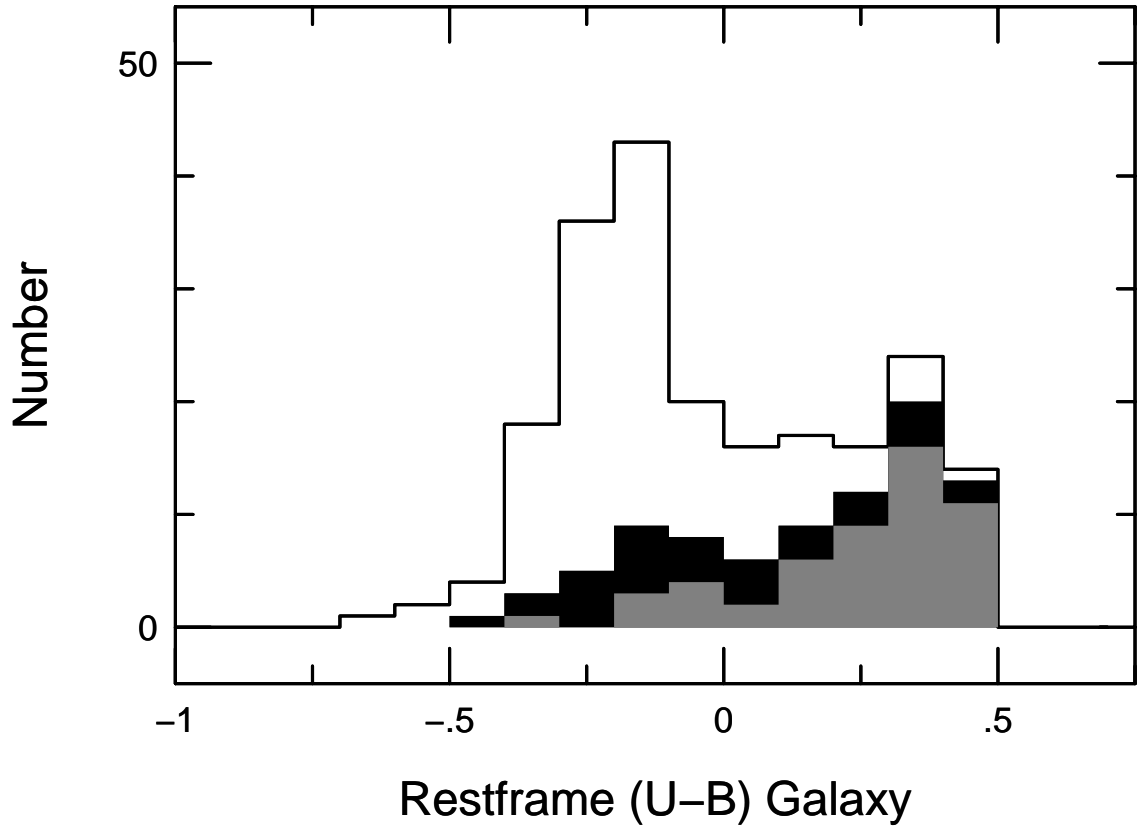}
\caption[histub] 
{Histograms 
in restframe $U-B$ of the whole galaxy 
for the full, high-redshift ($0.73 < z < 1.04$) 
GSS spectroscopic sample of 211 galaxies (open),
the photo-bulge sample of 86 galaxies (dark-filled), 
and the quality bulge sample of 52 galaxies
(light-filled). Note 
the bimodality in the colors of these high redshift galaxies.
\label{ub_hist}}
\end{figure}

{\it Magnitudes, Colors, Sizes, and Surface Brightnesses of Galaxies:}
Our selection provides a statistically complete sample of luminous,
high-redshift bulges. Indeed, the great bulk (70\%) of the galaxies
hosting the bulges are brighter than $I_{814} = 22.0$, the limit of
the CFRS \citep{lilly95}, while less than half (42\%) of the full
GSS-SRS in the same redshift range are so bright. Thus the bulge
sample is weighted to more luminous galaxies at high redshift.  As for
the colors of the galaxies, we find a relatively clean demarcation at
$V-I \sim 1.7$ between galaxies nearly as red as local E-S0's ($V-I
\sim 2$) and those with active star formation as seen in spirals or
later-type galaxies.  Fig.~\ref{ub_hist} shows the $U-B$ distribution
of the high redshift full GSS-SRS sample where we clearly see a
bimodal distribution with the red peak near $U-B \sim 0.35$.  This
bimodal color distribution can also be discerned in our GSS sample at
lower redshifts (Im \etal 2002; GSS3) and in other redshift samples
with high-precision colors (see e.g., Figs. 3 or 4 in the CNOC2 work
by \citealt{lin99}; the 21-22 magnitude subpanel of Fig. 9 of
\citealt{koo86}; the SDSS sample studied by \citealt{strateva01}; the
COMBO-17 sample studied by \citealt{bell04}).  We find that while
164/211 (77\%) of the full, high-redshift GSS-SRS galaxies have
integral colors bluer than $U-B \sim 0.25$, only roughly half (46/86)
of the photo-bulge sample is in this group (see Fig.~\ref{ub_hist}).

\clearpage

\clearpage

Finally, for the half-light sizes of the galaxies, we find
little evidence for any significant difference in the distribution
between that found for the high redshift full GSS-SRS sample and the
photo-bulge sample. However, when the surface brightness distributions are
compared, we find that,  while only 86/211 (40\%) of the full, 
high-redshift GSS-SRS sample have average surface brightnesses within the
half-light radius brighter than $\Sigma_{I} = 22$ mag per square
arcsec, about two-thirds  (57 or 66\%) of the photo-bulge sample
qualify. Thus any biases {\it against}  
detection due to surface brightness are
likely to be significantly less severe for the photo-bulge sample than for the full
GSS-SRS sample.

{\it Morphologies and Image Structure:}
The  photo-bulge sample was defined using only a simple
{\it photo-bulge apparent luminosity} cut ($I$ of photo-bulge  $< 23.566$)
with no {\it explicit} attempt to
restrict the sample by  image structure. 
%
Although we have made no attempt to select early-type galaxies,
such galaxies will be preferentially selected if they continue at high
redshifts to possess more luminous bulges than later-type galaxies.
Whether such early-types actually dominate the total sample depends on
whether the multiple pre-merger or single predecessors of today's
early-type galaxies had early-type morphologies in our redshift range.
Unless spirals with luminous bulges disappear at redshifts $z > 0.73$,
we expect that our sample will also include some spirals.  Indeed, we
do have a significant number of apparent spirals in our sample (see
Fig. \ref{vi_panel}), including some appearing to be of very late type
(e.g, 094\_2210), some even seen as very nearly edge-on (e.g.,
064\_4412, 094\_7063), and some with multiple components or internal
structures that resemble bars or arms (e.g., 064\_4813, 163\_4865),
bright H II regions (e.g., 094\_4767), interacting neighbors (e.g.,
073\_1809, 153\_2422), or tidal features (e.g., 093\_2327, 084\_1138).
Both 064\_4412 and 094\_2210 have
well-traced disk emission, with kinematics and masses as expected for 
disk systems \citep{vogt96}. As discussed later, the presence of star-forming
disks, presumably with significant gas and hence dust, may affect the
apparent colors of any genuine bulges.  But more importantly,
such disks  may have  concentrated regions of active
star formation that may masquerade as $r^{1/4}$ bulges in our 
two-component decomposition. 

Based on a visual examination by one of us (SMF), a 
rough division into three groups yields the largest to be E-S0's (35)
and slightly fewer but roughly equal numbers among 
normal spirals (25) and the catch-all remaining class of 
peculiars, compacts, and mergers (26). The diversity of morphologies
of the  disks and galaxies hosting high-redshift, luminous 
bulges  should serve as  a cautionary flag  that
bulge formation and evolution may include diverse histories and physical processes.

\subsection {Photo-Bulge to Total ({\it  pB/T}) Distribution}

\begin{figure}
\plotone{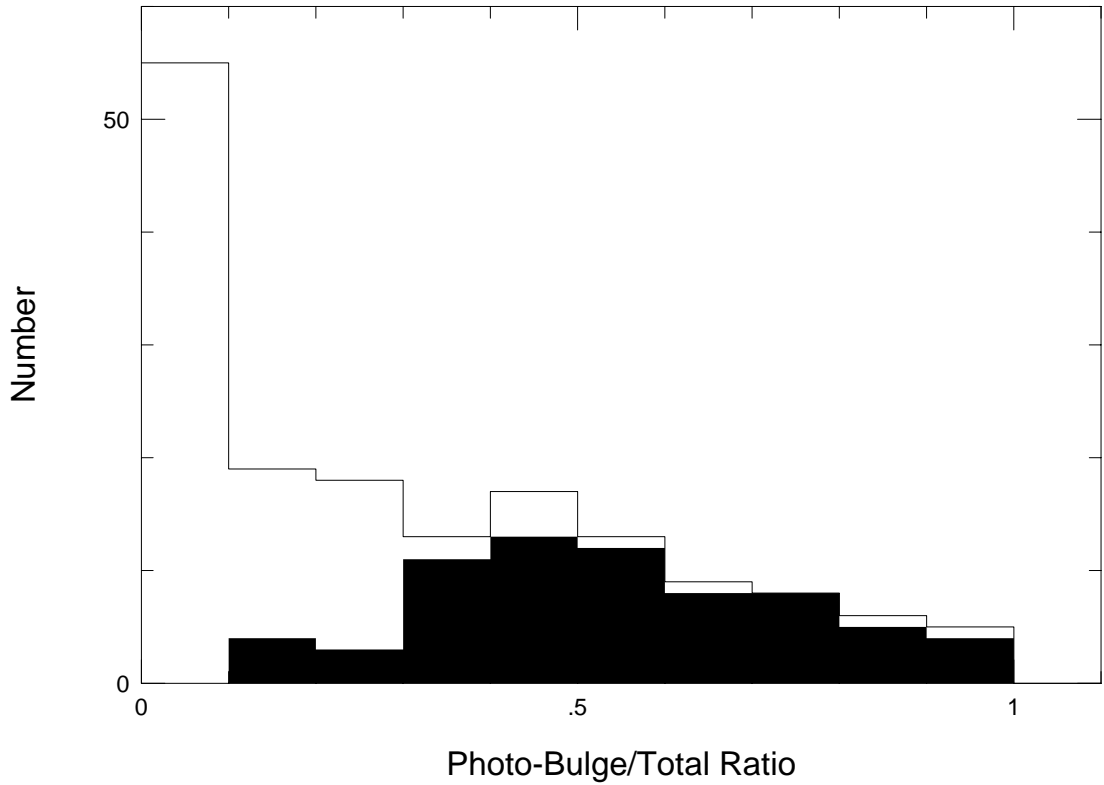}
\caption[bt_hist.eps] 
{Histogram of photo-bulge to total ratio ({\it pB/T}) in restframe
$B$. The open histogram is for the GSS spectroscopic redshift
subsample constrained to 192 galaxies with total $I_{814} < 23.566$
and redshifts $0.73< z < 1.04$. The filled histogram of 86 objects is
for the same constraints except that the photo-bulge luminosity has
$I_{814} < 23.566$. Note the dramatic loss of objects with low {\it
pB/T} with this additional constraint. A figure showing the quality
bulges can be found in Appendix B.
\label{bt_hist}}
\end{figure}

Fig.~\ref{bt_hist} compares the distribution of {\it pB/T} in our
sample versus a companion sample of GSS-SRS galaxies restricted to be
in the same high redshift range but with {\it total galaxy
brightnesses} chosen to be brighter than $I_{814} = 23.566$ mag. The
192 GSS-SRS galaxies in this category show a peak at the pure
photo-disk end in {\it pB/T} with a rapid drop at {\it pB/T} $\sim$
0.1, followed by a more gradual decline towards the pure photo-bulge
end.  Our sample of 86 photo-bulges shows only 5 galaxies with {\it
pB/T} $< 0.2$, a peak near $ {\it pB/T} \sim 0.45$, followed by a
decline towards the pure photo-bulge end that almost totally overlaps
the full high redshift sample.  The strong bias against low- {\it
pB/T} systems in the bulge sample can be understood as the direct
result of our choosing a brightness limit for the photo-bulge
component.  A further restriction to high redshift then forces the
photo-bulge sample to be intrinsically luminous ($M_{B} < -19$) and
thus understandably results in few, if any, very low {\it pB/T}
galaxies.  For example, a system with {\it pB/T} $\sim 0.1$ must be
accompanied by a very luminous disk ($M_{B} < -21$) to be within our
sample limits.  Our selection of only luminous photo-bulges thus
prevents us from placing strong constraints on the nature of lower
luminosity bulges in late-type galaxies at high redshifts. We will
compare the two observed {\it pB/T} distributions in Fig.~\ref{bt_hist} 
to model
predictions in the discussion section.

\clearpage

\clearpage

\subsection {Are Photo-Bulges Genuine Bulges?} \label{gen_bulges}

As already discussed in Section 2.1, we have intentionally used the
term photo-bulge and photo-disk throughout this work in order not to
prejudge the nature of the subcomponents extracted by the GIM2D
software package from the HST images.  As warned by the developers of
GIM2D \citep[GSS2;][]{simard99, marleau98}, the use of a two-component
decomposition does not assure that genuine bulges and disks are being
extracted.  In some cases, very blue central knots of active star
formation reside in an otherwise normal late-type disk.  These could
conceivably be genuine bulges in formation, as envisioned by secular
evolution theories. In other cases, GIM2D has chosen to fit the {\it
blue outer disk} of a galaxy with the $r^{1/4}$ profile and the inner
{\it very red}, true bulge with the exponential (more details of this
effect are discussed in GSS2).  Examples of such reversals of bulge
and disk are noted in Appendix C and include objects 104\_6432,
273\_7619, 303\_4538).  In such cases, the photo-bulge might
erroneously be regarded as a blue bulge.  Another potential problem in
identifying photo-bulges is that lower-luminosity bulges are observed
locally to have light profiles closer to that of an exponential than
an $r^{1/4}$ shape \citep{andredakis95}.  If such bulges were once
bright enough to enter our high redshift sample, they would be
measured here as photo-disks and thus lost from our photo-bulge
sample.

So,  what is the nature of our photo-bulges?  We find two major groups.
From the full 86-object sample, the dominant one (69, or 80\%) includes
red (restframe $U-B \ga 0$), high surface brightness ($\Sigma_e <$
20 mag per square arcsec), luminous ($M_B < -19.5$) photo-bulges.  Most of
these red and very-red photo-bulges (44/69, or 65\%) are accompanied by
red and very-red photo-disks and are likely counterparts of E-S0's
today, while others (25/69, or 35\%) are accompanied by blue and very
blue ($U-B < 0$) disks of varying proportions and are thus likely
counterparts of spiral bulges.

The minor (17, or 20\%), but more intriguing, group is associated with
blue and very-blue photo-bulges.  Some are merely the result of
misidentifying a blue disk as the photo-bulge component, as explained
above. A few, such as 092\_1339, appear to be blue $r^{1/4}$ bulges.
But as indicated in the notes (Appendix C), this particular galaxy has
strong emission lines with well measured velocity widths well under
100~\kms, and therefore is not a probable progenitor of luminous
bulges (which are expected to exhibit widths closer to 200~\kms).
Most, however, appear to be blue central regions lying within disks.
Several lines of evidence suggest that, unless stellar mass is added
through future star formation, mergers, or infall, such blue
subcomponents {\it are not the progenitors of luminous bulges today}:

1) Almost all of the bluest (16 of 17 with $U-B < 0$ in the full 86
sample) have restframe $B$-band surface brightnesses similar to or
dimmer than that of bulges of comparable size today. We find this
evidence alone to be compelling, since after their intense star
formation activity subsides, the resultant fading by several
magnitudes (depending on the fraction and size of an underlying older,
red stellar population) will reduce their average half-light surface
brightness to values below that seen among bulges today.

2) Their luminosities are fainter, rather than brighter, than most of
the redder photo-bulges; thus, after fading and evolving to redder
colors, they cannot be the counterparts of the luminous bulges of
today. In principle they may become lower-luminosity bulges, but
locally these are generally of smaller size and have profiles that are
exponential rather than $r^{1/4}$.

3) Roughly 80\% of the blue photo-bulges reside in photo-disks that
are more luminous than they are, i.e., {\it pB/T} $< 0.5$.  The
theoretical expectation is that a blue, and thus forming, bulge would
be so luminous that the bulge would dominate the total
light, i.e., yield a high {\it B/T} ratio.

4) Many have very blue colors ($U-B < -0.25$) that correspond to
intense star formation.  Since lifetimes are longer during the redder,
fading phase than during the active starburst phase, an even larger
proportion of bulges should have intermediate colors.  We see a dearth
of such bulges with intermediate colors.

5) Most (11/18, or 61\%) of the photo-bulges with $U-B < 0$ reside in
redder disks, presumably of older age or with less active star
formation.  This bluer-core color gradient would not be expected in
hierarchical formation scenarios for bulge formation, where the outer
disks form after the central bulges, and should thus appear younger
and bluer.  Secular evolution models, however, propose that bulges
form at the same time or after disks so that disks may then be
expected to be older and thus redder. The blue bulge colors are then
explained, but secular processes produce fainter bulges
\citep[e.g.,][]{macarthur03, carollo04}, not luminous bulges such as
we see in our sample.

Several of these points will be illustrated quantitatively in figures
based on a smaller, higher-quality sample to be defined in the next
section.

\subsection{Selection of ``Quality'' Bulges} \label{quality}

The primary selection criteria of the full bulge sample were
deliberately chosen to be relatively simple, well defined, and able to
yield a statistically complete sample.  As previously mentioned, GIM2D
is limited in this study to decomposing galaxies into two simple
subcomponents, whereas galaxies clearly span a wide range of
structural properties not necessarily well described by the adopted
model. Before continuing with the presentation and analysis of the
results, we have further refined the sample to what we consider to be
of higher photometric quality and reliability by adopting two
additional constraints:

a) We set the photo-bulge brightness limit to be a half magnitude
brighter, $I < 23.066$. This reduces the total sample of 86 to 64
bulges with more reliable structural and photometric measurements.

b) We have limited photo-bulges to have half-light major-axis radii
(effective radii) less than the half-light radius for an exponential
disk, i.e., 1.7 times the scale length of the photo-disk {\it unless}
{\it pB/T} is greater than 0.67 (from Table 3). This half-light radius
restriction is aimed to exclude likely cases of reversed bulge-disk
decomposition by GIM2D.  It ensures that galaxies that are dominated
by the photo-bulge (i.e., more luminous than twice the disk) are not
unintentionally eliminated due to the presence (or apparent
measurement) of very tiny, low-luminosity and thus poorly measured
photo-disks. The resultant sample is now reduced to 52 bulges, with
the excluded 13 photo-bulges roughly divided equally between blue
($U-B < 0$) and red ($U-B \geq 0$).

The final ``quality sample'' has 52 photo-bulges that should be
reliable and moderately well-measured bulge candidates.  We find that
only 4 (8\%) are in the broader blue category ($U-B < 0$) and all of
them belong to the subcolor class of being very blue with $U-B <
-0.25$ (see Figures \ref{ub_mag} and \ref{bt_ub} and Table 4). 
Other divisions by {\it pB},
{\it pD}, and {\it pB/T} are also provided in Table 4.

\subsection{Photo-Bulge Color-Magnitude  Relation}

\begin{figure}
\includegraphics[angle=0,scale=0.7] {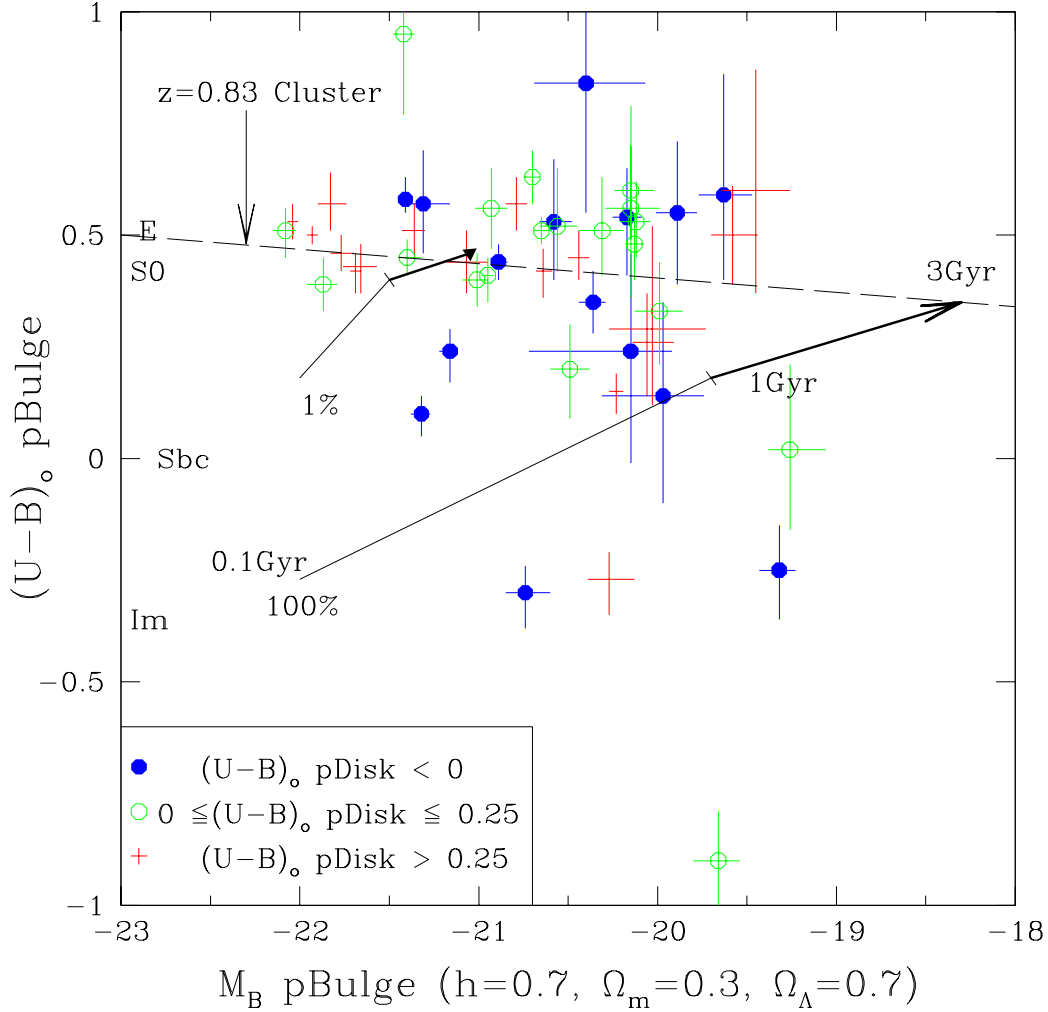}
\caption[24feb2004-ub-mag] 
{Restframe photo-bulge $U-B$ color vs photo-bulge $M_{B}$ for the
quality sample of 52 bulges along with their 68\% confidence
limits.  The dashed line with slope $\delta U-B/\delta M_B = -0.032$
is the total color-magnitude relation for early-type galaxies in
cluster MS1054-03 at $z \sim 0.83$ \citep{dokkum00}.  Symbols reveal
the {\it photo-disk colors} as indicated (colored in electronic
edition).  The E, S0, Sbc, and Im labels mark the colors of {\it local
galaxies} with different morphologies \citep[synthentic colors of
Table 2 from][]{fukugita95}.
Based on stellar population models assuming a Salpeter IMF and solar
metallicity \citep{bruzual03}, the long arrow shows the track from a
pure (100\%) single burst of star formation (duration $10^7$ years)
starting at $10^8$ years and after $3 \times 10^9$ years.  The shorter
arrow above shows the track from a similar burst of only 1\% by mass,
with the remaining 99\% of the galaxy having formed in a single burst
$3\times 10^9$ years before. Note that the 4 bluest photo-bulges
already have low luminosities and may fade by 3 mag or more within a
few Gyr.
\label{ub_mag}}
\end{figure}

Fig.~\ref{ub_mag} shows the color-magnitude relation for the quality
photo-bulges, with 68\% confidence error bars and symbols that
indicate the colors of the {\it disks}.  Very red photo-bulges span
the entire range of accessible $M_B$, while the bluest photo-bulges
are seen among less luminous galaxies.  Fig.~\ref{ub_mag} shows the
color-magnitude relation for early-type galaxies in a rich cluster, MS
1054-03, at redshift $z \sim 0.83$ \citep{dokkum00}. This relation is
almost identical to that found for local E-S0's
\citep{prugniel98,jansen00}, both using integral colors.

The key new result is that {\it red photo-bulges are nearly as red or
redder than the {\it integrated} colors of either local E-S0's or
distant cluster galaxies} (see also GSS9). Among the entire quality
sample, only 8 photo-bulges are bluer than the cluster or local E-S0
color-magnitude line at more than twice the 68\% confidence level
(roughly 2$\sigma$).  Thus 44/52, or 85\%, have colors that are
consistent with the very-red color-magnitude line.

\clearpage


\clearpage

That the absolute colors of our sample of bulges with a median
redshift of 0.81 (lookback time of 6.9 Gyr) are nearly the same as
that of E-S0's today, i.e. very red, is a surprise for which we have
no simple, compelling explanation {\it if only single-burst stellar
populations are considered}. For example, adopting a formation time at
redshift $z = 3$ (11.4 Gyr ago), Salpeter IMF, and solar metallicity,
a burst of duration 10 Myr would have $U-B \sim 0.56$ today and be
bluer by about 0.2 mag at a lookback time of 7.7 Gyr at redshift $z
\sim 1$ \citep{bruzual03}.  Thus we expect to observe a concentration
closer to $U-B \sim 0.35$, rather than 0.5 as observed.  Lower
redshifts of formation would result in larger color changes. Even with
formation at the Big Bang, the color change would be 0.1 in $U-B$;
coincidentally, the color at $z = 1$ (7.7 Gyr) for a simple burst is
close to the observed $U-B \sim 0.5$. We will return to this issue and
explore other options in Sec. 5.5.

The second surprise is that the colors of the field photo-bulges
appear as red or redder than the integrated colors of the cluster
galaxies at similar $z$. Part of the explanation may be that
subcomponent bulges are on average redder than the integrated colors
of galaxies, since galaxies may contain disks that are bluer on
average than the bulges (cf.  Fig.~\ref{bt_ubbd} below).  Galaxies in
rich clusters, however, are expected to have both bulge and disk
components to be older than field galaxies, so that even if bulges are
the oldest subcomponents of field galaxies, field bulges should still
have bluer colors than the integrated colors of cluster galaxies at
the same epoch.

Besides tracking the evolution of passively evolving populations, the
color-luminosity diagram serves as an important probe of bulge
formation itself. Fig.~\ref{ub_mag} shows how color and luminosity
evolve for a single starburst and for a 1\% starburst (by mass)
embedded within an older stellar population. Any stronger starburst
counterparts to luminous bulges or E-S0's ($M_B < -20$) would be
expected to be even brighter by perhaps several magnitudes, and with
very blue colors. Note, however, that the observed blue photo-bulges
1) are sometimes too blue to qualify as a minor (1\% or less)
starburst, 2) have luminosities too faint to become luminous bulges
after fading, and 3) are unusual in that several reside in disks that
are redder.  We will later return to these and other clues that
together suggest that blue photo-bulges are unlikely to be the
starbursting precursors of normal luminous bulges.

Fig.~\ref{ub_mag} also shows that the red photo-bulges, regardless
of their luminosity, reside within disks that span a wide range in
color.  The mere existence of very red disks at high redshift is another important
finding, with implications for the formation of S0's, the formation
epoch of spirals, the relative formation epochs of disks and bulges,
etc.  We will return to this issue in the next subsection.

Besides absolute colors, two other useful measurements are the slope
and intrinsic scatter of the color-magnitude relation.  As seen in
Fig.~\ref{ub_mag}, the colors of distant photo-bulges largely track
the slope seen among local E-S0's and among early-type galaxies in the
distant cluster at $z \sim 0.83$.  A major uncertainty is that the
slope among local bulges is not that well measured.  Using colors
from \citet{prugniel98} on early-type {\it field} galaxies, we obtain
a change of -0.09 mag in $U-B$ per magnitude change in $M_B$ while the
fits to RC3 by \citep{schweizer92} yield a shallower value of
-0.035. The steeper slope is also seen in the Coma cluster, which
yields a slope of -0.08 \citep{terlevich01}.  On the other hand, the
color-magnitude slopes for red galaxies are shallow in both the
distant $z = 0.83$ cluster (slope of -0.032 shown in
Fig.~\ref{ub_mag}) and the distant field galaxies in the HDF-N studied
by \cite{kodama99}. The latter sample uses the early-type galaxies
identified by \cite{franceschini98} via K-band surface brightness
profiles.

A biweight statistical measure \citep{beers90} yields for our quality
sample a slope of $-0.04 \pm 0.04$, where the errors are estimated via
Monte-Carlo bootstrap. Restricting the quality sample to include only
the very red photo-bulges yields a slope of $-0.02 \pm 0.02$.  Except
for being {\it redder} by 0.05 mag in $U-B$, the resultant high
redshift field bulge color-magnitude relation is closer to that
found for early-type galaxies in the cluster at $z = 0.83$ than to
that for local bulges or E-S0's.

An important diagnostic of the age spread of bulge formation is the
intrinsic scatter of the data about the color-magnitude relation
\citep[e.g.,][] {bower92}.  For cluster galaxies at high redshifts $z
\sim 0.8$, \citet{stanford98} and \citet{dokkum00} both find
small intrinsic scatter that supports a small age spread and an old
age for the early-type cluster galaxies, though morphological or
progenitor bias may artificially reduce the scatter
\citep{dokkum01}. Although our color measurement errors for the
bulges are typically larger than that for the entire
galaxy as measured in the cluster work, we can nevertheless place
useful constraints.  As seen by the proximity of the error bars in the
color-magnitude to the cluster line, and confirmed by a robust
estimate using the biweight statistical method \citep{beers90}, we
estimate the intrinsic $U-B$ color scatter to be $\sigma = 0 - 0.03 $
mag at the 68\% confidence level (CL), where the errors are estimated
via Monte-Carlo bootstrap. This small scatter is consistent with the
value of 0.03 found by van \citet{dokkum00} for the $z = 0.83$
cluster MS1054-03. We note that the morphological or progenitor bias
discussed by \citet{dokkum01} does not apply to our sample (we
include spirals), but that a similar type of bias may exist if the
bluer progenitors of genuine bulges do not possess the same
$r^{1/4}$-profile.

We will now examine the relationship between the colors of
photo-bulges and photo-disks.

\subsection{ {\it  pB/T} Ratio vs. Colors of Photo-Bulges and Photo-Disks}

\begin{figure}
\plotone{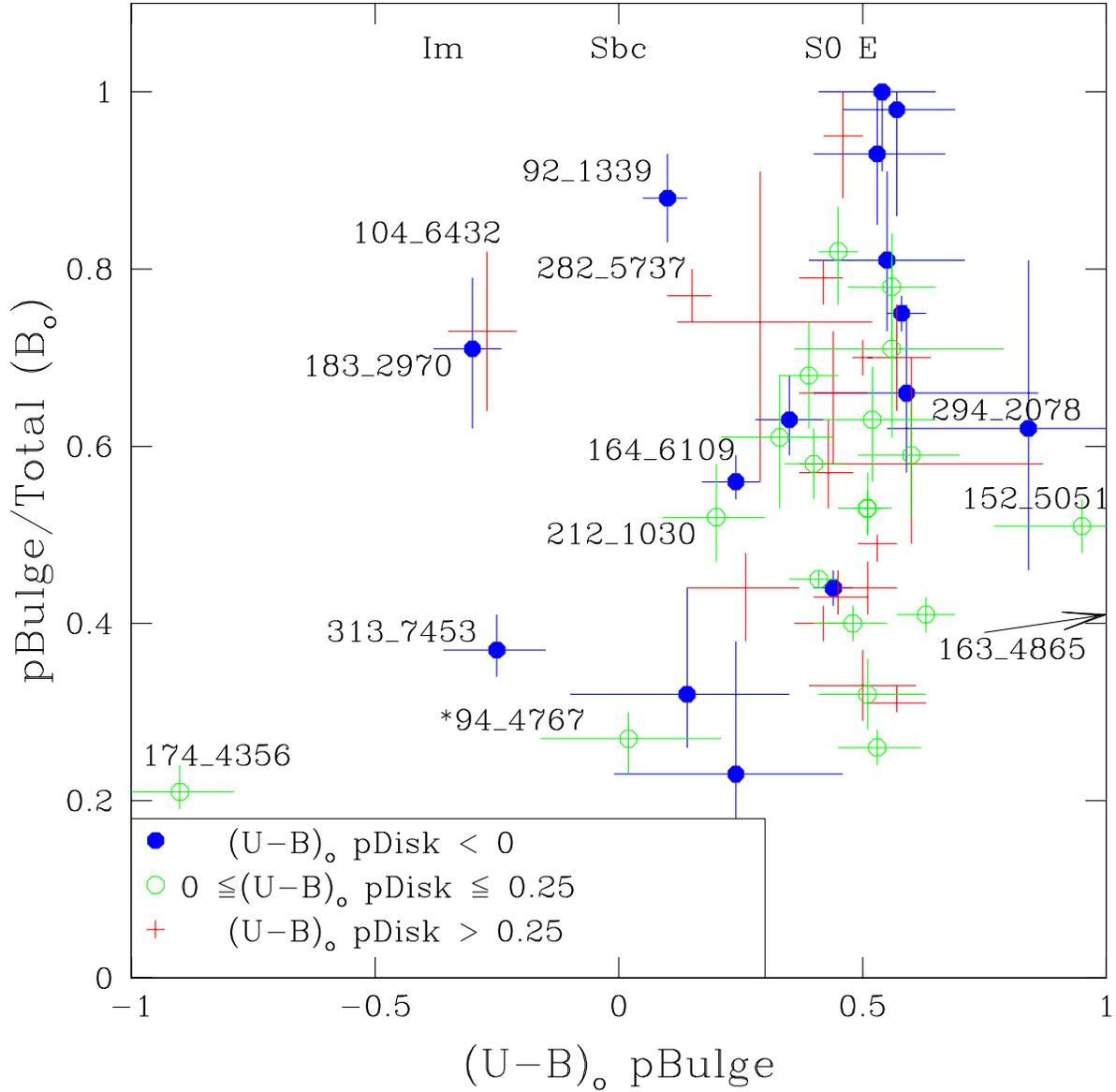}
\caption[24feb2004-bt-ub] 
{Photo-bulge to total ratio ({\it pB/T}) in rest-frame $B$ vs.
photo-bulge restframe $U-B$ colors for the 52 objects in the quality
photo-bulge sample.  The labels are the GSS-ID's of the outliers (see
Tables and Appendix C).  The data symbols indicate the restframe $U-B$
colors of the {\it photo-disk} component as indicated in the inset
(colored in electronic edition).  As a reference, the E, S0, Sbc, and
Im labels show the approximate {\it colors} for local galaxies with
the respective morphologies \citep{fukugita95}.  The figure shows that
the very-red photo-bulge colors appear to be independent of both {\it
pB/T} and the color of the associated photo-disk. No trend with
redshift is seen so we have not made further subdivision of the sample
in this figure.
\label{bt_ub}}
\end{figure}

Fig.~\ref{bt_ub} shows the photo-bulge to total ratio ({\it pB/T})
versus $U-B$ color of the photo-bulge, with different symbols
indicating the colors of the photo-disks.  The red clump of
photo-bulges is found to span the full range of observed {\it pB/T},
while the blue photo-bulges are shifted to systematically lower {\it
pB/T} systems.  If blue photo-bulges are classic bulges seen during
their active formation phase, we would expect instead to find that
blue bulges have {\it larger} {\it pB/T}.  Since no selections by
color of the galaxy, color of its subcomponents, or {\it pB/T} ratios
greater than 0.67 have been applied, blue bulges with high {\it B/T}
ratios, if they are common, are not missing in our sample\footnote{If
bulges in formation resemble point-source AGN's, then our size cut may
select against such objects. But only two objects were eliminated by
this selection, and thus they represent at best a rare population
(4\%). If bulges during formation have exponential rather than
$r^{1/4}$ light profiles, they may also be missing from our present
sample.}. Indeed, if bulges are passively evolving old populations
that fade with time while disks are more constant in luminosity, we
would also expect to find a higher proportion of large {\it B/T}
systems at higher redshifts.  Analysis beyond the scope of this work
is needed to assess whether this is true, but we see no gross evidence
for this in that the fraction of high {\it pB/T} $> 0.5$ systems
(Fig.~\ref{bt_hist}) in our distant galaxy sample is only about 25\%,
which is less than the roughly half of luminous galaxies being within
the red portion of the bimodal distribution of colors seen among local
galaxies \citep{strateva01}.

\clearpage


\clearpage

Another surprise in Fig.~\ref{bt_ub} is the lack of any strong correlation
between the colors of the disks of the red photo-bulges and {\it
pB/T}. Again, assuming that a bluer disk is at a brighter phase of its
life, we might expect bluer disks to reside among smaller {\it pB/T}
systems, but this is not seen.  Moreover, bluer disks might also be
associated with later Hubble types, which are roughly correlated with
{\it B/T} ratio so that blue disk systems might be expected to
dominate the low {\it B/T} regime.  This may be true for a complete
sample of galaxies but is not seen among the luminous, red,
photo-bulge systems.

The picture that emerges from these findings and those from the
previous subsection is one in which luminous bulges are universally
old, even at redshifts $z \sim 1$, and that disks form around them at
different epochs, with no strong correlation between the disk colors
(i.e., age) and bulge to disk ratio. This result on disk colors and
luminosities associatd with red photo-bulges serves as an important
constraint on the nature and history of luminous bulges.  As
previously discussed, the lack of very low {\it pB/T} galaxies within
our photo-bulge sample is a selection effect.  Thus we cannot directly
address the possibility that low-luminosity bulge systems have a
different formation history.

\clearpage


\begin{figure}
\plotone{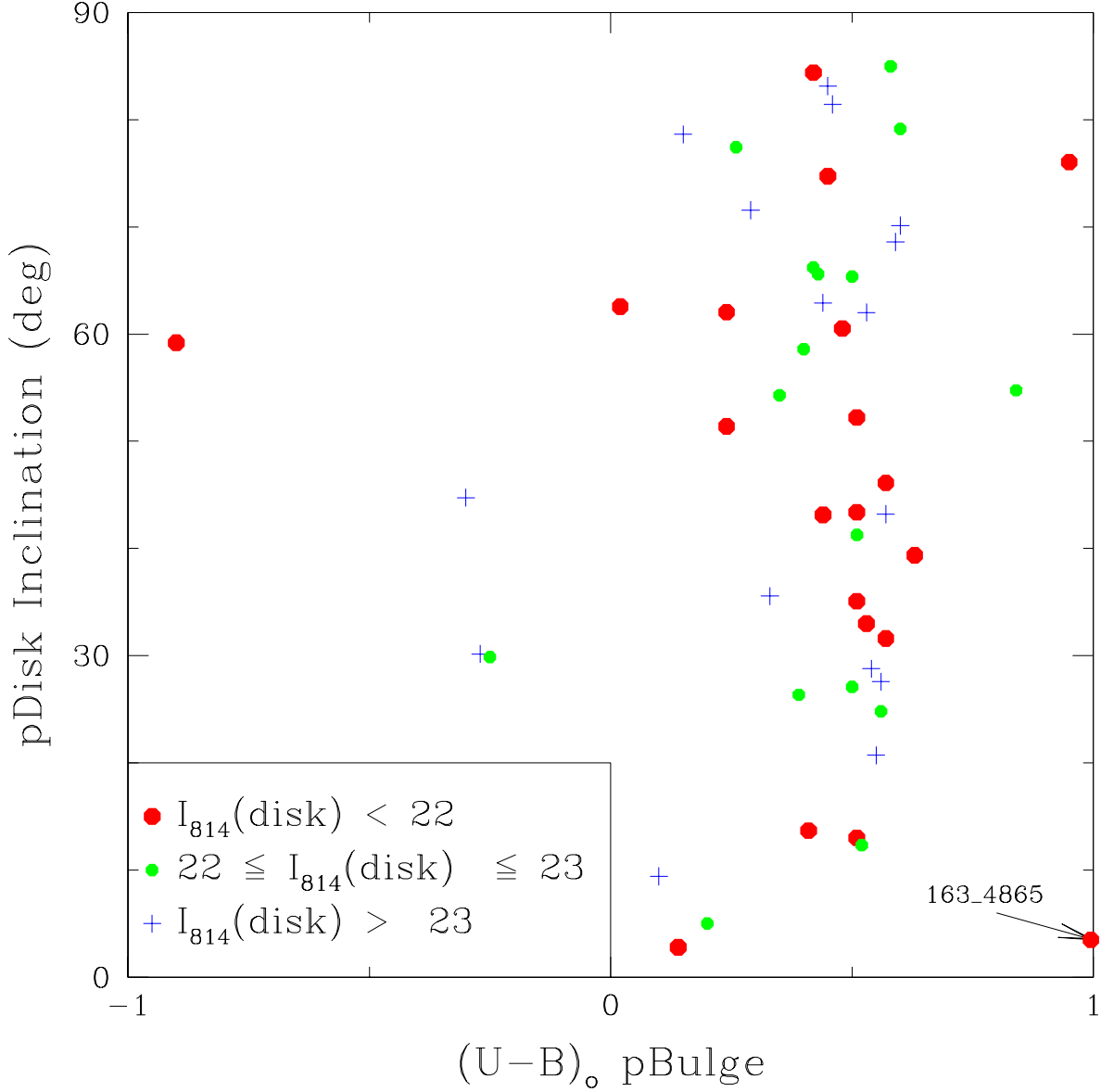}
\caption[16feb2004-inc-ub] {Photo-disk inclination angle vs.
photo-bulge restframe $U-B$ colors for the 52 objects in the quality
photo-bulge sample.  The data symbols indicate the $I_{814}$ mag of
the {\it photo-disk} component as indicated in the inset (colored in
electronic edition), with the expectation that the inclination angle
is more poorly determined for fainter disks. GSS ID 163\_4865 marked
in figure with arrow has pB color of $\ub = 1.53$.  Note that very red
($\ub > 0.25$) photo-bulge colors appear at all inclination angles of
the associated photo-disk, indicating that such red colors are likely
to be intrinsic to the photo-bulge component and not mainly caused by
dust reddening.
\label{inc_ub}}
\end{figure}

\clearpage

Fig.~\ref{inc_ub} shows the photo-disk inclination angle versus the
color of the photo-bulges in the quality sample. Again we see little
correlation. Since dust in high inclination disks might result in
redder bulges, the lack of correlation implies that any such effect is
not strong.  Although a few photo-bulges might be affected by dust,
e.g., GSS ID 152\_5051 and 163\_4865, the bulk of photo-bulges have
such uniformly red colors that, if dust were the major cause, its
effects must be nearly universal, i.e., it cannot vary much from
galaxy to galaxy. The uniformity, independence of the amount of disk
({\it pB/T}), and independence of photo-disk inclination angle
together suggest, but do not prove, that the very red colors of
photo-bulges are not due to dust obscuration.

\clearpage

\begin{figure}
\plotone{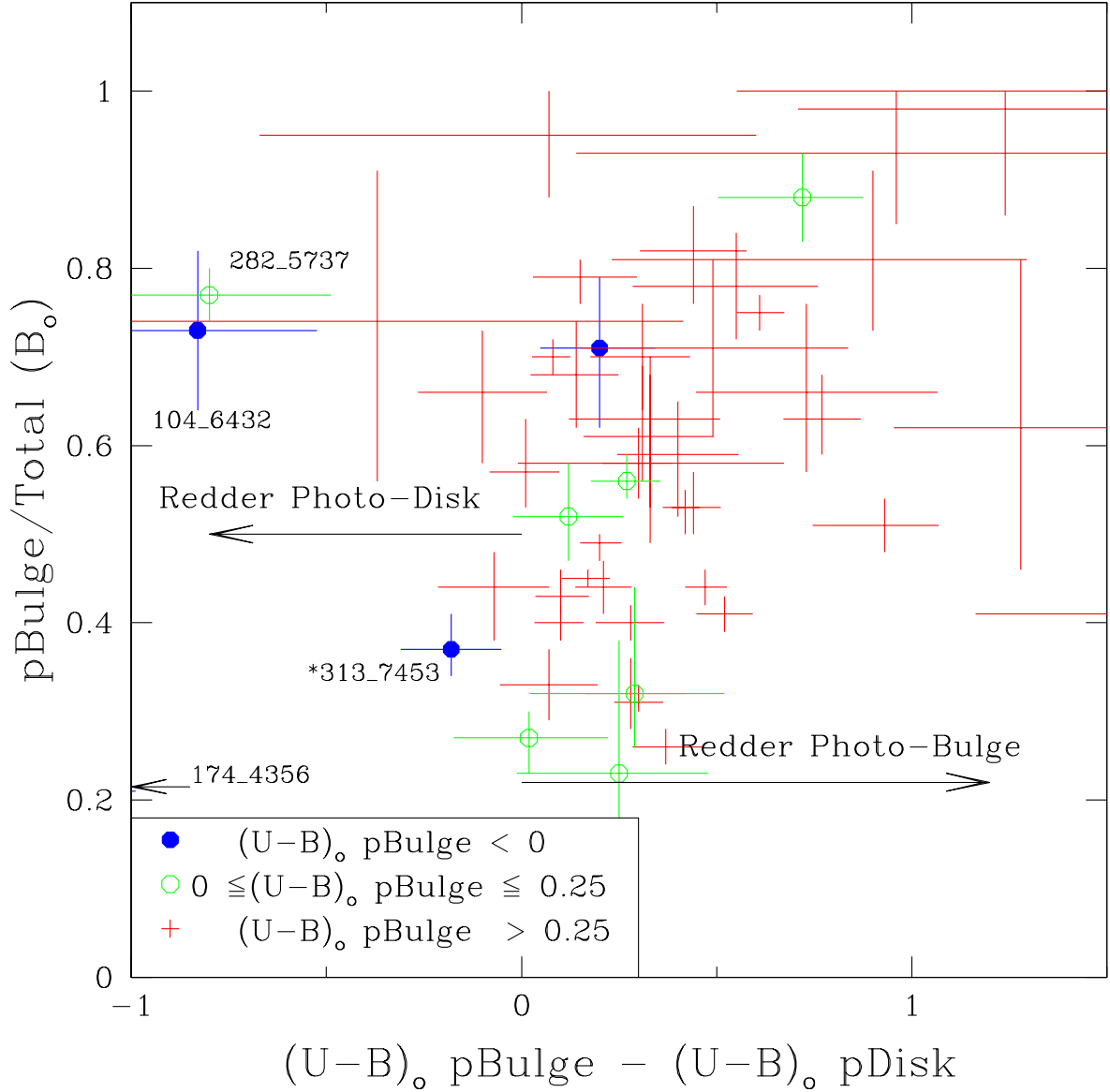}
\caption[15may2002-bt-ubbd.eps] 
{Photo-bulge to total ratio ({\it
pB/T}) in rest-frame $B$ vs.  the difference in the restframe $U-B$
color of the photo-bulges and photo-disks for the quality photo-bulge
sample of 52 objects. The data symbols show the {\it photo-bulge}
colors as indicated in the inset box (colored in electronic edition).
The vast bulk of very red ($U-B > 0.25$) photo-bulges lie on the
right-hand side, where photo-disks are bluer than photo-bulges.  In
contrast, most (3/4) of the blue photo-bulges ($U-B < 0$) reside in
photo-disks that are redder.
\label{bt_ubbd}}
\end{figure}

\clearpage

Fig.~\ref{bt_ubbd} shows {\it pB/T} versus the {\it color difference}
between the photo-bulges and photo-disks, with different symbols now
indicating the colors of the photo-bulges.  Almost the whole sample is
on the right hand side, where photo-bulges are redder than
photo-disks. A particularly interesting example is 094\_2210 (not in
the quality sample), which possesses a very red central bulge-like
component that is imbedded within a surrounding very-blue, disk-like
component that appears to be comprised of multiple blobs. The
structure might be a disk in its early formation phase, as originally
suggested by \citet{koo96}. On the other hand, while the red
photo-bulges reside with photo-disks of similar or bluer colors, three
of the four blue photo-bulges reside in redder photo-disks.

Figures \ref{bt_ub} and \ref{bt_ubbd} also show that some 
galaxies have quite red disks with
colors close to that of bulges. Such disks are important, for, if
not due to reddening by dust, they imply that at least some disks were
already quite old at redshifts $z \sim 1$. Given claims that S0's are
virtually absent in clusters of galaxies before redshift $z \sim 0.6$
\citep[e.g.,][]{dressler97, fasano00}, the existence, numbers, and
environments of {\it field} S0's whose disks and bulges are {\it
both red} at even higher redshifts $z > 0.7$ place important
constraints on plausible formation mechanisms.

In this regard, we note with interest that the most luminous galaxy in
our entire spectroscopic sample with $z$ between 0.73 and 1.04
(274\_5920) has a {\it pB/T} $\sim$ 0.5, i.e., equal light in the
photo-bulge and photo-disk components.  Visually, the galaxy appears
to be an ordinary elliptical galaxy.  The photo-disk in 274\_5920 is
also the most luminous photo-disk in the entire high redshift sample
of 205 galaxies, and it is very red ($U-B = 0.33$)\footnote {The
second most luminous galaxy is the quad-lens system 093\_2470, which
also has {\it pB/T} $\sim$ 0.5, but its disk has colors close to that
of Sbc galaxies ($U-B \sim 0$).}. We thus find that the most luminous
galaxy, photo-bulge, and photo-disk in our current sample are all very
red.  Our sample is presently too sparse to place good constraints on
the volume density of such very-red, bulge-disk systems (likely S0's),
but if photo-disks are genuine disks, their mere existence is
compelling evidence that very old disk systems (some very massive with
$M_B < -22$ in the disk alone) did exist side by side with very old
bulges in the field by redshift $z \sim 1$. Such massive, old field
disks are likely to be difficult to accommodate in current versions of
semi-analytic models. 

The quality sample includes 16 (31\%) such galaxies with photo-bulges
and photo-disks that are both very red and they are found to span the
full range of luminosities, $ {\it pB/T}$, and disk inclination
angles.  Such systems provide a unique sample to test for the possible
presence of residual star formation among ellipticals and bulges of
S0's and spirals with apparently old stellar populations, without the
confusion or ambiguity of emission lines arising from star-forming
blue disks.  Intriguingly, we do find emission lines even in these
galaxies which are red in both components.  In fact, while over 60\%
of these red photo-bulge and photo-disk systems show emission lines on
average, among the ten most luminous galaxies, we detect emission
lines from all but two, these being the most luminous (274\_5920) and
third most luminous (064\_3021).  The remaining eight
\footnote{brightest first: 074\_6044, 062\_2060, 094\_2660, 103\_7221,
203\_4339, 113\_3311, 103\_2074, 283\_6152}, or 80\%, all show
emission lines of O II.  The exact location (bulge, disk, halo, etc.)
and nature of these emission lines remain uncertain, but their high
frequency is a hint that star formation may be common within distant
galaxies, even those that appear quiescent by having very red colors
overall and separately in their photo-bulge and photo-disk
subcomponents. This last qualification is needed to avoid seeing
emission from bluer disks with active star formation.  Several
galaxies show relatively broad lines (but much narrower than from
typical AGN's), ranging from $\sigma =$ 100~\kms \ for, e.g.,
113\_3311, 150~\kms \ for 062\_2060, to around 200~\kms \ for
103\_2074 and 094\_2660, as might be expected for gas well-mixed
within a deep potential well. Indeed, these emission line values match
well the {\it absorption line} velocity dispersions measured for the
same galaxies in GSS9.
 
A rough estimate of the average star formation rate for these ten
luminous galaxies is about 0.5 to 1.5 M$_{\odot}$ yr$^{-1}$ per
10$^{10} $M$_{\odot}$ of stars\footnote{O II luminosities were derived
using the formula: log L(O II) = 31.97 - 0.4 M$_{3727}$ + logEW(O II),
where EW(O II) is the restframe equivalent width of the O II emission
line as given in Appendix C for each of the eight galaxies; an
estimate of the continuum luminosity at O II in AB magnitudes,
M$_{3727}$ = M$_B$ + 0.9($U-B$) + 0.628 ; M$_B$ and $U-B$ are for the
galaxy from Table 3; and the conversion from L(O II) to SFR adopted
the relation of SFR(M$_{\odot}$/yr) = 7.9x10$^{-42}$ L(H$\alpha$) from
\citet{kennicutt98} and L(O II) $\sim$ 0.4 L(H$\alpha$) from the
luminous portion of Fig. 1 of \citet*{jansen01}. The mass of stars
assumes the stellar populations are on average 1.5 mag brighter at the
observed redshifts and the local M/L$_B$ = 4.}.  The lower rate
assumes the gas has low sub-solar metallicity, while the higher value
assumes the solar to super-solar metallicity of luminous galaxies,
with no additional corrections for extinction.  Even the low rate
translates to significant mass accumulation -- roughly 5\% per Gyr or
a significant fraction of the entire galaxy after only a few Gyr.  As
discussed later, a total fraction of merely 4\% new stars, i.e., on
average only a fraction of a percent per Gyr, is needed to explain
constant colors.  These two estimates of accumulated new stars can be
reconciled by adding  a large fraction of the new stars to the disk rather than the bulge.

Regardless of the exact level of star formation activity, such star
formation among almost all very red, luminous, fading, stellar
populations is an important clue that virtually all field galaxies
probably experienced continual or episodic infusion of small amounts
of star formation at high redshifts.  This picture is qualitatively
consistent with hierarchical growth of galaxy via merging and provides
some additional support for a scenario, proposed later, to explain the
constancy of the very red colors of bulges from redshifts $z \sim
1$ to today while the galaxies are undergoing 1 to 2 mag of fading due
to passive evolution of the bulk of their old stars.

\subsection{Photo-Bulge Size-Luminosity Relation}

\begin{figure}
\includegraphics[angle=0,scale=0.7]{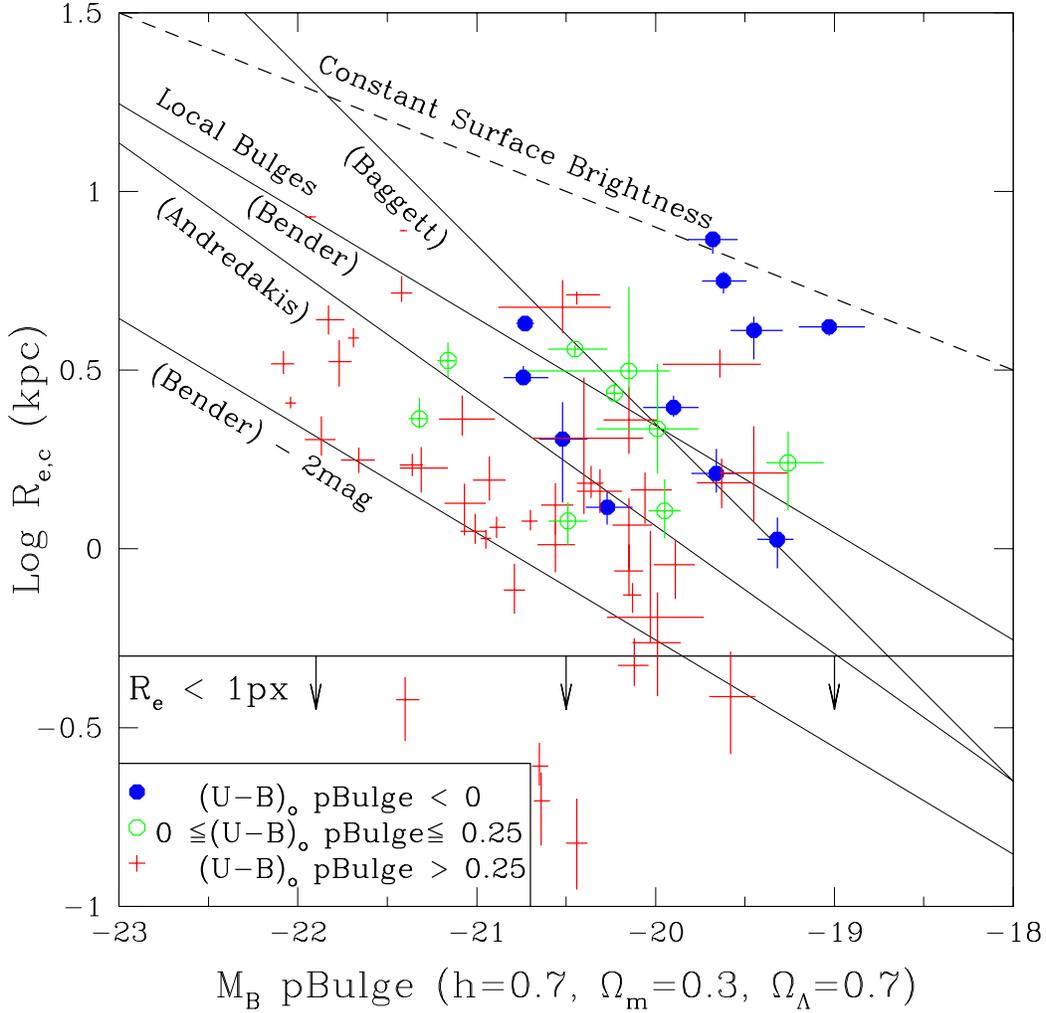}
\caption[24feb2004-size-mag.eps] 
{Photo-bulge circularized 
effective radius $R_{e,c}$ (see note for column 10 in  Table 3) in
kpc vs. the photo-bulge absolute luminosity ($M_{B}$) 
for the 64 objects with photo-bulges brighter than $I = 23.066$.
This restriction  is the same as that for the quality sample but without
any photo-bulge versus photo-disk size restrictions. 
Vertical error bars reflect only those for $R_e$, i.e., errors in the
eccentricity of the photo-bulges were ignored. 
The intrinsic colors of the photo-bulges are as indicated
in the inset box (colored in electronic edition). 
Photo-bulges with measured half-light sizes less than
0.1 arcsec (one pixel) are noted, for they are likely to  
be less reliable (see comments in
Appendix C for 092\_2023).   
Various solid diagonal lines
show  the locus for {\it local} bulges  from sources 
as marked \citep{bender92, andredakis95, baggett98} and the \citet{bender92}
line brightened  by 2 magnitudes, as marked. 
Lower surface brightness loci are parallel to the dashed diagonal line
and towards the upper right direction. 
The very red photo-bulges have 
a distribution that lies roughly  parallel to but $\sim 1$
mag brighter than the loci for local bulges. The bluer  photo-bulges
show a smaller  offset, with the bluest ($U-B < 0$) 
already close to the
local relation.  After any fading, most will
have lower surface brightnesses than that of bulges today.
\label{size_mag}}
\end{figure}

Fig.~\ref{size_mag} shows the sizes of photo-bulge effective radii
(kpc) vs.  photo-bulge luminosity ($M_B$), with different symbols
indicating the colors of the photo-bulges. Besides the quality sample
of 52 galaxies, the figure includes the 12 additional photo-bulges
(total 64) that meet the brightness limit of $I < 23.066$ for the
quality sample, but do not meet the criterion of the relative sizes of
the photo-bulges and photo-disks.  These 12 were excluded from the
quality sample to improve the reliability of the photo-bulge sample
and color measurements but have been added back in here to avoid a
strong selection by size.  The solid lines are the mean relations
found for local bulges \citep{andredakis95, baggett98, bender92}, all
showing a tilt towards higher surface brightness for lower-luminosity
bulges. The dashed line is one of constant surface brightness.  The
distant photo-bulges are found to have a correlation, albeit with
large scatter, that roughly follows the slopes of the local relations,
but with a shift to higher surface brightnesses.  If reliable, the
formal error bars imply that the large scatter is not primarily due to
data quality errors, but instead appears to be intrinsic to the
photo-bulge sample.  When separated by color, the bluer photo-bulges
(closed and open circles) lie towards the upper right,
lower-surface-brightness portion of the data distribution. After any
significant fading, these photo-bulges will lie well away from any of
the local relations for bulges.  This result, based on the brightest
64 photo-bulges, is only strengthened when the entire 86 photo-bulge
sample is examined.  This conclusion, that {\it blue photo-bulges are
actually of similar to or lower surface brightness than local bulges
of similar size}, is perhaps the strongest and most direct evidence
against their being genuine, pre-faded, young massive bulges
undergoing active star formation.

\clearpage

\clearpage

If we assume that bulges maintain stable structures since $z \sim 1$,
i.e., with no size evolution, we can use the average change in surface
brightness from the local relation as an estimate of any luminosity
evolution. Unfortunately, neither the slope nor the zero-point of the
local size-luminosity relation for bulges proves to be well
defined. For example, using the 17 bulge sample of \cite{bender92}, we
find a fit close to $M_B = -18.85 - 3.33$ log $R_e$, which is shown as
one of the solid lines in Fig.~\ref{size_mag}. The dispersion of the
local data around this relation is roughly $\pm 0.75$ mag at log $R_e
\sim 0$. Restricting our sample to just the 38 very red photo-bulges
and locking the slope to -3.33, we find the median offset to be -1.22
mag, i.e., $M_B = -20.07 - 3.33$ log $R_e$. Using a slighter steeper
slope of -2.8 as suggested by the work of \citet{andredakis95} (who
derive a fit of $M_B = -19.82 - 2.8$ log $R_e$) yields an intercept
close to the previous one of -20.37, but adopting a local intercept of
-19.82 then gives only a 0.55 mag offset, which is less than half the
estimate when compared to the \citet{bender92} sample. Finally, as
another independent check, we have compared our data to that from
\citet{baggett98}. To keep the measurements to the same $r^{1/4}$ plus
an exponential disk, we excluded all fits that required an inner
truncation radius. To allow conversion to $B$ and avoid the
uncertainties due to differences in the colors of the bulges and
disks, we included only galaxies which were on average quite red, with
$B-V > 0.8$ and which had T types earlier than 5. To avoid problems
with the Hubble flow, we used only galaxies beyond 20 Mpc. Finally, to
use only data with good fits, we included measurements with the $rms <
0.2$.  This reduced the sample of 620 objects to 94, yielding the fit
$M_B = -19.30 - 2.0$~log $R_e$. Adopting this steep slope, our sample
yielded an intercept of -20.42, implying a luminosity brightening of
1.12 mag.

In summary, by comparing the size-luminosity relation of our distant
sample to local bulges, we find strong evidence for luminosity
evolution.  The best estimate of the brightening is probably between
1.1 to 1.2 mag when using the Bender \etal or Baggett \etal samples,
but may be as low as 0.55 mag if we adopt instead the Andredakis \etal
sample.  We searched for, but did not see, any systematic trend with
redshift and so our estimates of luminosity evolution apply on average
to the full sample.

\subsection {Very-Red Bulge Luminosity Density }

Having derived the selection function and weights, we can, in
principle, derive the luminosity function of bulges. Our sample is,
however, too small for reliable results and is also subject to
significant fluctuations from large-scale structure. But given the
importance of the luminosity function evolution of bulges in our
understanding of galaxy formation and the role of bulges in hosting
AGN's \citep{gebhardt00}, we have obtained instead a related but more
robust measure of the integrated luminosity density.

Since the very-red photo-bulges are most likely to be genuine
bulges, we restrict this analysis to the 58 very-red ones from the
full sample of 86.  To our depth of $I \sim 23.6$, we reach bulges
as faint as $M_B \sim -19$ at $z \sim 1$.  We adopt the $1/V_{max}$
method, even though our data clearly show that the redshift
distribution is not uniform. This simple exercise yields an {\it
averaged} luminosity density in $B$ of $7.1 \times 10^7 L_{\odot}$
Mpc$^{-3}$ to our observed depth. The formal errors of $\pm$ 6\% based
on 500 Monte-Carlo bootstrap resamplings do not reflect cosmic scatter
due to large scale clustering, which we estimate to be roughly
$\pm$25\% \citep{im02}.  Note that our luminosity density includes the
light at redshifts $z \sim 0.9$ from only very-red bulges, i.e. no
photo-disks, even red ones or those that physically belong to the
bulge population but happen to be excluded because they do not have
$r^{1/4}$ light profiles.

The luminosity density in $B$ is measured to be 
7.0 $\times 10^7 L_{\odot}$ Mpc$^{-3}$ for very-red, high-redshift bulges.
This is $\sim$36\% of
the luminosity density of  19.6 $\times 10^7 L_{\odot}$ Mpc$^{-3}$, as measured
from  the light of the  whole galaxy for the full GSS  high-redshift 
sample \citep{willmer04}.
A recent local estimate of the total $B$
luminosity density by \citet{liske03} is  13.9
$\times 10^7 L_{\odot}$ Mpc$^{-3} (h=0.7)$ or roughly 70\% of that
observed at high redshift by \citet{willmer04}.  
Thus galaxies, when added together, were  brighter in the past. 
A recent estimate for local bulges \footnote{We used $h=0.7$
and converted from $I$ to $B$ assuming disks have the colors of Scd
galaxies and bulges have S0 colors from \citet{fukugita95}} is 0.4
$\times 10^7 L_{\odot}$ Mpc$^{-3}$ \citep*{benson02}.  This is over
10$\times$ less than what we find at redshift $z \sim 0.8$ and only 3\% of
the local total from \citet{liske03}.

We  suspect the \citet{benson02}
value  to be too low, since prior estimates of the bulge
fraction in $B$ (converted from $V$) 
range from about 25\% from \citet{schechter87}
to about 39\% as estimated by \citet{fukugita98}.  This range for
local galaxies is supported by a more recent estimate from the SDSS of
30\% in the $r$ and $i$ \citep{tasca03}. Their method also decomposes each
galaxy into a bulge and disk component.  In summary, the bulge
luminosities of both the local and our distant samples are presently
quite uncertain.  We find that roughly a third of the total luminosity
density of distant luminous galaxies reside in bulges, comparable
to some local estimates.  Larger samples are needed before we can have
a reliable check of the recent claim for a factor of two buildup of
the integrated stellar mass in spheroidal galaxies as a whole by
\citet{bell04}.

\subsection{Visual Morphologies of the Quality Sample}

The visual morphologies of the galaxies that belong to the very-red,
quality sample of 41 photo-bulges appear overwhelmingly normal, with
only three objects (84\_1138, 93\_2327, and 94\_6234) showing low
surface brightness features or very close neighbors suggestive of
interactions and mergers (see Fig.\ref{vi_panel}). In contrast, 7 of
the remaining 11 bluer (i.e., not very-red with $U-B < 0.25$)
photo-bulges show visual morphologies that are unusual, by having
double nuclei, distortions, or very close neighbors that are
suggestive of interactions and mergers. Such complex morphological
structures will affect our color measurements from GIM2D at some
level, but probably more those of the of the larger and more distorted
photo-disks than the more centrally concentrated photo-bulges. A more
detailed study from a much larger sample will be needed to assess the
impact of this correlation between morphology and color on the colors
of photo-bulges.  The key result from this work is that our sample
suggests a high correlation between the colors of photo-bulges and
whether they belong to galaxies that have unusual morphologies: less
than 10\% of the very red photo-bulge sample show such morphologies
while the remaining bluer sample is dominated (64\%) by them.

\section {COMPARISON TO PREVIOUS WORK}

\subsection{Summary of Key Results}

  We emphasize again that we have a statistically complete,
magnitude-limited sample of high-redshift, {\it luminous},
$r^{1/4}$-profile bulges that should include bonafide ellipticals,
bulges of S0's, and bulges of spirals. We caution the reader that our
selection and structure-extraction procedures may, however, also
contaminate the sample with non-bulges such as nuclear/central
star-forming regions of late-type galaxies or any subcomponent that is
not well fit simply by an exponential with one scale length.

  Before proceeding, we summarize the key results found in the
previous section:

1) The vast fraction (over 80\%) of luminous field photo-bulges at
redshifts $0.73 < z < 1.04$ are very red, independent of the observed
B/T, disk color, and disk inclination. Almost all reside in morphologically
normal early-type galaxy or spiral.  Moreover, the color-magnitude
($U-B$ vs. $M_B$) relation is similar to that of bulges today with a
shallow slope and small scatter. The bulge size-luminosity relation
indicates about 1 mag of fading since $z \sim 1$. 

2) The small remaining fraction of blue photo-bulges, compared to the
dominant very red photo-bulges, have on average lower surface
brightnesses, lower luminosities, and redder photo-disk colors that
argue against most of them being genuine proto-bulges. Many appear to
reside in morphologically peculiar galaxies.

\subsection {Comparison to Prior Studies}

  We divide the following discussion into three high redshift groups:
ellipticals (diskless bulges); early-type galaxies (E-S0's); and
spiral bulges. Since several of our conclusions differ from those of
other studies, we start by summarizing the major advantages of our
survey.  First, our survey sample size is substantial, with 86 objects
at high redshifts ($z > 0.73$), while some other surveys have fewer
than five objects. Second, we try to separate the bulge colors from
disk colors using 2-D decomposition. In contrast, others use
integrated colors and assume their galaxies are disk-free $r^{1/4}$
ellipticals or use small central aperture colors and assume that disk
contamination is negligible. Third, our sample is spectroscopically
confirmed. The spectra provide more reliable redshifts than
photometric redshifts and other useful diagnostics such as star
formation rates and internal kinematics.

\subsubsection{Integrated Colors of Distant Field Ellipticals}

\citet{schade99} studied the properties of 46 {\it field} ellipticals
at redshifts $0.2 < z < 1.0$ and found much bluer $U-V$ colors at
higher redshift.  Besides a brightening of 0.97 mag by $z \sim 0.92$,
they also find strong [O II]3727 emission lines in roughly one third of
these ellipticals.  

While we agree with the last two conclusions, we disagree with the
first.  To track the differences in more detail, we have examined the
7 galaxies in common between our two surveys (as indicated by comment
``c'' in the Notes column of Table 1 ). 

Overall, we find good agreement in $I_{814}$ magnitudes, but
relatively poor agreement on whether {\it B/T} is indeed
indistinguishable from 1, i.e., pure $r^{1/4}$ or elliptical by the
Schade \etal definition, namely, galaxies that are well described by
$r^{1/4}$ light profiles as derived from 2-D surface photometry of
$I_{814}$ images from HST. Our  {\it pB/T} values  for
6 galaxies lie  more than 7x
the 68\% confidence limits (i.e., roughly 7$\sigma$ for Gaussian error
distributions) away from {\it pB/T} = 1.  When systematic errors are
taken into account (see section ~\ref{GIM2D}), these galaxies are even
less likely to have {\it B/T} = 1\footnote{The object closest to a
pure $r^{1/4}$ profile is 092\_1339, which has a {\it pB/T} value of
0.85 and 68\% confidence limits of 0.03; this galaxy, however, is also
the best candidate for being a {\it blue} bulge, and, as detailed in
Appendix C and \cite{im01}, this galaxy has strong emission lines that
have a small velocity width
($\sigma$) of only 85~\kms . It is thus
unlikely to be a genuine,
young, massive E-S0.}.

Why the difference?  To identify ellipticals, \citet{schade99} use
visual inspections of the $r^{1/4}$ fits to the galaxy profiles in the
$I_{814}$ HST image.  Based on our own tests, we find that this
procedure can be deceptive for two reasons.  First, exponential
components (photo-disks) can easily hide as merely slight systematic
deviations from an $r^{1/4}$ fit, not easily discernible by eye, but
whose statistical significance is strongly supported.  Second, we find
our $V_{606}$ image, not used by \citet{schade99}, to be an important
additional and independent source of information to confirm the
presence of a disk, especially those that are blue.  The uncertainty
of their identifications is confirmed by their own visual
classifications, which sometimes assign Sab or later types to their
sample. Thus, while we can understand how \citet{schade99} might be
deceived into believing their sample consists of pure $r^{1/4}$
ellipticals, we believe that our measurements of {\it pB/T} with error
bars show that such galaxies are actually relatively scarce (5/52, or
10\%).

Besides finding poor agreement on type, we also find poor agreement on
the integrated colors of the galaxies. For example, \citet{schade99}
find that none of the 7 in common with our sample has total galaxy
colors matching those of unevolved early-type galaxies (i.e., redder
than Sab, $U-V_{0,AB} \sim 1.8$ or $U-B \sim 0.33$).  We find three
that do (062\_2060, 062\_6859, and 064\_3021). Except for 092\_1339
mentioned in the previous footnote, the remaining 6 all have very red
photo-bulges, while \citet{schade99} claim that they are {\it all}
blue, pure ellipticals.

To explain the large differences of colors, we suspect field-to-field
zero-point differences in the \citet{schade99} colors.  Among the 19
high redshift ($z > 0.75$) galaxies in their sample, over half (10)
are in the GSS and yet {\it none} have colors redder than
($(U-V)_{0,AB} = 1.83$), roughly the average color of an Sab galaxy.
Of the remaining 9 high redshift ellipticals outside of GSS and in the
other three fields in the Schade \etal sample, 6 have very red colors
($(U-V)_{0,AB} >= 2.0$). Without any variations in the color
zero-points, the probability of finding by chance that none of the 6
reddest objects are among 10 from a sample of 19 is about 0.3\%.  In
contrast to \citet{schade99}, we find many very red galaxies ($U-B >
0.25$) in GSS.

In comparing the two surveys, note that our measurements are of high
precision with reliable zero-points (HST $V$ and $I$) and that we have
derived colors for the bulge and disk separately.  In comparison,
\citet{schade99} used {\it ground-based} photometry in $V$ and $I$ for
their colors (but HST $I$ for the elliptical identifications) and assumed
that their ellipticals are diskless.  Thus when a bluer disk is
present, Schade \etal would conclude that they had found a blue
elliptical, i.e., blue integrated colors, while we might find instead
that the bulge is indeed very red, but the disk is blue.  Two good
examples of such objects are 84\_1138 and 93\_3251 (compare colors of
the whole galaxy to that of the photo-bulge and photo-disk in Tables
1-3).

In summary, we agree with \citet{schade99} that early-type galaxies
exhibit luminosity evolution at the $\sim 1$ mag level, along with the
frequent presence of [O II] emission lines. The work of \citet{im02}
(GSS10) also agrees with the claim by \citet{schade99} for little
volume density change of early-type galaxies since redshifts $z \sim
1$. But we question the claim for evolution towards much bluer colors
among ellipticals at high redshift, since 1) their sample appears to
include galaxies that are not pure ellipticals and 2) their colors are
measured to be too blue, perhaps due to photometric zero-point
problems, at least in the GSS, and to the use of integrated colors for
objects that may contain blue disks.

\subsubsection {Internal Color Dispersions of Bulges}

\citep{abraham99} undertook two studies in the Hubble Deep Field
(HDF-N) directly related to this work, one on the uniformity of the
star formation history of 11 E-S0's and another on the relative ages
of the bulges and disks of 13 spirals (discussed in the next
subsection), all with redshifts $0.3 < z < 1.1$.  These samples were
taken from the \citet*{bouwens97} sample of galaxies with $I_{814} <
21.9$ and with üspectroscopic redshifts.  The morphologies of
\citet{bouwens97} were replaced by a visual reclassification by one of
the co-authors.

In the first study, the inferred ages of the stellar populations in
{\it each pixel} were derived from the 4-band photometry available in
the HDF-N. The dispersion or distribution of these pixel-ages were
then used to divide the E-S0's  into those which did and did not
have 5\% or more of the pixels with ages younger than the most recent
third of the age of the oldest pixel.  

Five of the 11 E-S0's are in the same high redshift regime as our
sample, with 3 (2 E's and one S0) showing largely old coeval stellar
populations while 2 (both E's) show evidence for younger
populations. This would suggest that 40\% of the high redshift E-S0's
have a young component. 

In a follow-up study, \citet{menanteau01} studied 79 field E-S0's (24
with spectroscopic redshifts) to $I_{814} = 24$ and made a comparison
to galaxies to $I_{814} = 22$ in five distant clusters analyzed in the
same manner. They claim to ``provide strong evidence for the continued
formation of field E-S0's over $0 < z < 1$.'' This was based on
finding ``that a remarkably large fraction ($\gtrsim$ 30\%) of the
morphologically-classified E-S0's with $I_{814W} < 24$ show strong
variations in internal colour, which we take as evidence for recent
episodes of star-formation,'' with most showing bluer cores.  They
find significantly smaller color dispersions in the cluster galaxy
sample and estimate from modeling the star formation history ``that at
$z \sim 1$ about half the field E-S0's must be undergoing recent
episodes of star-formation.''

A direct comparison to our luminous bulge sample is not
straighforward, since we have not made any morphological
classifications.  GSS10, however, identifies 18 galaxies in the
present paper as being E-S0's on the basis of actual measurements of
high {\it pB/T} and low levels of asymmetries (see Table 1, comment
f). Of these, only 092\_1339 and 294\_2078, or 11\%, have blue {\it
total} colors (see column 8 of Table 3) while the remaining are all
very red by our criterion of $U-B \ga 0.25$.  The fraction of blue
early-type galaxies among these 18 may actually be lower to only 6\%,
since 294\_2078 appears visually to be a spiral (see
Fig.~\ref{vi_panel}) with a very blue disk; shows a rotation curve in
its spectrum \citep{im01}; and most importantly, possesses a central,
very-red bulge (see Fig.~\ref{vi_panel} and Table 3).

As previously noted, our photo-bulges are as red as local E-S0's or
the early-type cluster galaxies at $z \sim 0.83$.  Since nearly all
photo-disks are bluer than photo-bulges, the {\it integrated} galaxy
colors are usually bluer than that of photo-bulges. For the 18
galaxies identified by \citet{im02} as early-type, the median color of
the photo-bulges is $U-B = 0.51$, while the median for the total
colors of these same galaxies is $U-B = 0.39$ (the two values differ
at more than the 95\% confidence limit when the errors on the median
values are accounted for).
Compared to $U-B \sim 0.45$ for early-type galaxies in the $z \sim
0.83$ cluster \citep{dokkum00}, we find that the field early-type
galaxies are indeed bluer in $U-B$, by $\sim 0.06$.  This result is
expected in scenarios where early-type galaxies in clusters formed
earlier than those in the field.  Whether the {\it bulges} 
of cluster galaxies are also redder than that of field
galaxies needs to be checked (Koo \etal, in preparation).

In summary, we find 2/18 (11\%) early-type galaxies to have blue
overall colors, and only one of these (6\%) has  a blue bulge. These
fractions are smaller than the 30\% to 50\% fractions of blue E-S0's
claimed by \citet{menanteau01} and others \citep[e.g.,][]
{franceschini98, abraham99, stanford04}.  While the 50\% fraction can
be excluded by our sample of 18 at the 99\% confidence limit, the
disagreement is only at the 90\% confidence limit for the 30\% figure.
A larger sample is needed to improve these statistics.

\subsubsection{Bulge Colors  in Distant Spirals}

In the second study by \citet{abraham99} of 13 field spirals, 
bulges and disks were defined by the light within and outside, respectively,
an aperture of 1 arcsec (10 pixels)  diameter.
Deriving ages from colors, \citet{abraham99}
find that 8 out of 9 {\it normal} spirals have bulges that are older
than the disks, and thus they conclude that ``for morphologically
normal systems, bulges are indeed always the oldest parts of
galaxies.''  They note that even the oldest bulges do not appear to be
as ``uniformly red and old as the oldest ellipticals'' in the first
study. Only two of these spirals are in the high redshift range of our
sample.  In contrast, among the 4 {\it peculiar} systems, only one has
an older bulge, and two have clearly younger bulges. All of these are
at redshifts lower than the range in this paper.

In a follow-up study, \citet*{ellis01} compared the colors of the
bulges of 95 spirals to the integral colors of 60 early-type galaxies
using data from HDF-N and HDF-S. Most of the sample relies on
photometric redshifts, with only 20 in the sample having spectroscopic
redshifts that overlap our high-redshift range. The bulge colors for
galaxies down to integrated $I_{814} \sim 24$ were estimated from
aperture photometry within the inner 5\% radius using $V_{606} -
I_{814}$ colors from WFPC2 and $J_{120} - H_{160}$ colors from NICMOS.

Our results are fully consistent with theirs that central (bulge)
colors are generally redder than the outer disks (see
Fig.\ref{bt_ubbd}). However, our results disagree with their second
conclusion ``that bulges are, statistically, optically bluer than the
reddest ellipticals and show a large dispersion in their rest-frame
colors.''  Note that the Ellis \etal sample is selected by total
galaxy luminosities rather than by the luminosities of the bulges
as in our study. We speculate that the differences
between our study and those of Abraham \etal and Ellis \etal can be
understood as the result of the following factors: 1) our survey is
restricted to very luminous bulges while theirs includes galaxies with
very low
luminosity bulges; such low luminosity bulges are expected
to be bluer than luminous
ellipticals; 2) their bulge measurements
have contamination of central aperture colors by bluer disks; and 3)
their visually-selected bulges are sometimes misclassified and are
instead central, star-forming regions of late-type galaxies.

\subsubsection {Summary of Comparisons to Other Surveys}

No other survey of high redshift galaxies has separated the bulge
from the disk for studies of colors, sizes, and luminosities.  The
closest in spirit are the \citep{abraham99} and \citet{ellis01}
studies of the bulges of spirals using a small central aperture to
derive colors; they find the disks are generally bluer than the bulge.
We agree. They also find, however, a large dispersion in the colors of
the bulges and that they are bluer than the integrated color of the
reddest cluster ellipticals. Here we disagree. The vast majority of
the bulges in our sample (85\%) are very red and are not detectably
bluer than {\it even} the integrated colors of {\it local}
E-S0's. Although we also find a few blue bulges, their surface
brightnesses are too low to qualify them as precursors or pre-faded
counterparts of small, high surface brightness, redder bulges. As
discussed previously, the differing results may reflect the choice of
samples. Ours is confined to luminous bulges while others may have
included fainter bulges whose colors may be bluer, or which may be
confused with very bright central star formation complexes in spirals
and irregulars.

Another major issue is the fraction of blue E-S0's (not bulges) at high redshift.  A
key difficulty is the definition of E-S0's (especially if selected by
eye) and the level of sample contamination by bluer spirals and AGN's.
Those studies based on the small handful of
spectroscopically confirmed high-redshift E-S0's in HDF-N 
\citep{franceschini98, kodama99, tamura00,dokkum03}
include a known AGN and radio source 
\footnote{Hdf2-251.0 with redshift $z = 0.960$ located at J2000
12:36:46.3 +62:14:05.7;  see \citealt{phillips97} for spectroscopic
confirmation of AGN nature} as well as a galaxy with a very small {\it  B/T}
ratio (0.17) and highly distorted residuals\footnote{Hdf4-565.0 with
redshift $z = 0.751$ located at J2000 12:36:43.6 +62:12:18.3; see
Fig. 1 at $x=720$ and $y=120$ in \citealt{marleau98} }.  Overall, the
bulk of published works claim high blue fractions between 30\% at
moderate redshifts $z \sim 0.4$ to 50\% by redshift $z \sim 1$. An
exception to such claims comes from the work of \citet{im02}.
Indeed, when we adopt the same definition of E-S0's using {\it  B/T} and
asymmetry, we find 18 E-S0 candidates  in the present sample, but only 2, 
or 11\% are blue, and of these, one is a spiral and one is anomalous \citep{im01}.
On the other hand, we do confirm the claim by \citet{schade99}  for the
frequent presence of emission lines,  a finding that supports
scenarios that include continued star formation in otherwise quiescent
galaxies via infall or mergers, albeit at a low level. 

\section {Models of Elliptical and Bulge Formation}

\subsection {Introduction}
Three major classes of bulge formation mechanisms have been
proposed over the years, ranging from 1) the monolithic formation models
of \citet*{eggen62}; 2) 
major mergers of disks into ellipticals \citep{toomre72} or
mass accretion of dwarf satellites into a bulge; and 3) secular dynamical
evolution models where instabilities, resonances, and dynamical
interactions among disk, halo, and bars contribute to the formation of
bulges \citep[see reviews by][]{wyse97,combes00, carollo04}. 
Within the dominant paradigm of hierarchical formation of galaxies,
each of these mechanisms is likely to play some role.
As the reviewers  emphasize, bulge formation is unlikely to be a simple,
homogeneous process. To decipher the relative importance of these and
other mechanisms of bulge formation, observers need to
measure the mass function of bulges, their stellar populations
(ages, colors, and metallicity distribution), all as a function of
time or redshift as well as of environment. Theorists need to make
realistic simulations that can be compared to the observations.  We
are today far from reaching either ideal.

A comprehensive  discussion of models and theories of bulge formation is
beyond the scope of this work.  We will instead focus on comparing our
new data to a subset of models that make explicit predictions of
the luminosities, disk and bulge colors, and {\it  B/T} of field galaxies
at large lookback times. This comparison is strongly motivated by the
lack of discrimination among different models when comparisons were
made with high redshift data that existed  a few years back \citep{bouwens99}.
Our sample has  substantially improved the available
data and the following demonstrates the high level of discrimination
now possible. An important caveat is that our data apply only to
luminous ellipticals and the bulges of other galaxies 
in the field at high redshifts.

\subsection {Analytic Models of Bouwens \etal}

We compare our observations to a {\it modified} version of the
analytic bulge-formation models originally presented by \citet{bouwens99}.
These models adopted  various analytic  prescriptions for the formation epochs and 
evolution of bulges and disks and translated these into predicted 
luminosities, colors, and {\it B/T}.  The modifications are introduced to better
match the properties of the high-redshift disks observed in the DEEP
survey and to incorporate the influence of dust.

\citet{bouwens99} did not correct disk properties for inclination
though they noted that inclination-dependent biases could be important
in reconciling the results of the \citet{peletier96} sample with that
from the \citet{dejong96a} sample.  Here we use the \citet{tully85}
prescription to make corrections to the luminosity and color of the
disks as a function of inclination.  To improve the fits to the local
and redshifted $z \sim 1$ disk colors, we increased the total opacity
given by this prescription by 30\%.  We assume the \cite{bouchet85}
extinction curve, where $A_R = 0.53 A_B$.  For local comparisons, we
correct the disk $B$ and $R$ luminosities of the \citet{peletier96}
sample (composed of edge-on galaxies with inclinations greater than 50
deg) and of the \citet{dejong96a} sample (composed of face-on galaxies
with inclinations less than 51 deg) to reflect an average inclination
of 34 deg.

Unlike the previous work by \citet{bouwens99}, we assume that disks
form when 50\% of their final halo mass is assembled; that $\Omega_m =
0.2$ \ and $\Omega_{\Lambda}=0$; and that the fiducial mass of all
disks is $\sim 3\times 10^{11} M_{\odot}$, independent of their
luminosity (this ignores the mass dependence of the halo formation).
In modelling the disks, we adopt a Hubble parameter ratio of $h = 0.7$
instead of the $h = 0.5$ used in the original models.

Another modification is that we assume a distribution of e-folding
times for the disk star formation rates instead of adopting a single
value as in \citet{bouwens99}. We here assume 5\% of disks to have a
7$\times$ shorter e-folding time than adopted by \citet{bouwens99};
5\% with 4$\times$ shorter; 20\% with 2.5$\times$ shorter; 20\% with
1.7$\times$ shorter; and 20\% with 1.5$\times$ {\it longer} e-folding
times.  We set these values by attempting to fit both the $z=0$ and
$z=1$ disk color distributions simultaneously.

\subsection{ Semi-Analytic Models}

Unlike the \citet{bouwens99} approach, in which the formation epochs
of the disks and bulges are manually adjusted to match local
observations, several other groups have instead adopted the results of
N-body simulations or the Press-Schechter formalism to model the
formation of structure over time.  As reviewed by each of three major
groups working with this semi-analytic model (SAM) approach
\citep{kauffmann99, somerville99, cole00}, SAMs follow the merging
evolution of dark matter halos and, via constraints from local galaxy
properties, adjust a set of relatively simple parameters that relate
mainly to star formation, gas cooling, satellite mergers, and
supernovae feedback.  After adding stellar populations, and as
described in the reviews, the SAM approach from all three groups has
enjoyed a number of successes.  These include matching the fraction of
early-type galaxies to spirals; the luminosity functions of local
galaxies from the optical to near-infrared; the Tully-Fisher relation;
the amounts of neutral hydrogen in different galaxy types; and the
sizes of galaxies and their subcomponents.

We will concentrate here on the general trends to be expected from
SAMs regarding the relative ages (colors) of bulges in clusters
versus field galaxies and among bulges, ellipticals, and S0 galaxies.
As previously claimed \citep{kauffmann96a, baugh96}, hierarchical
models predict that the mean stellar ages in field ellipticals should
be several Gyr younger than cluster ellipticals. Moreover, since disks
take additional time to initiate and grow after the formation of their
central bulges from an earlier strong merger event, bulges in higher
{\it B/T} systems should be younger than those in lower {\it B/T}
systems.  Thus bulges in late-type spirals with lower {\it B/T} are
predicted to be older and thus redder than the bulges of early-type
spirals with higher {\it B/T}.  As seen in Fig. 6 in the work by
\cite{ellis01}, the color-magnitude relation from the unpublished
$\Lambda$CDM SAM predictions of \citet{baugh96} predict that at our
redshifts of interest ($0.7 < z < 1.1$), spiral bulges should be
redder by 0.1 to 0.2 mag in $V-I$ than ellipticals, which in turn are
redder or older on average than entire S0's by about 0.3-0.4 mag in
$V-I$.  The S0's are bluer than ellipticals presumably because their
colors include the light from not only an elliptical-like old
bulge, but also a younger (bluer) disk.  And finally, as emphasized
by \citet{kauffmann96b}, ellipticals should be forming over time, and
thus appear to be decreasing in volume density towards higher
redshifts, though admittedly, the amount of decrease is dependent on
the choice of cosmology.

To make the comparisons somewhat more concrete, we compare the
bulge data against the predictions of the SAM of
\citet{kauffmann99} for a given epoch near redshift $z = 1$. These
predictions are made in the form of Monte Carlo realizations of the
SAM, with rough inclusion of the selection function.  The original SAM
(referred to as SAM-B in the figures) adopted a prescription whereby
all satellites with mass ratios 1/3 or larger were incorporated into
the bulge of the primary galaxy. But we found that the observed {\it
B/T} distribution prior to bulge selection (see right hand two
panels of Fig. \ref{model_bt_hist} in Appendix B) yielded a much higher fraction of
low {\it B/T} galaxies than that predicted by this SAM.  This is
likely related to a well-known angular-momentum problem faced by all
CDM models, which are unable to make a significant population of
bulge-less disks as observed \citep[e.g.,][]{navarro94}. We thus worked also with a
second SAM model, in which satellites were added to the disk instead
of the bulge of the primary (referred to as SAM-D in the
figures). Overall and qualitatively, we find a reasonable match
between the observations and the SAM-D predictions, though there are
visible differences when examined in detail.

\subsection {Results of Model Comparisons}

We reference the three scenarios of \citet{bouwens99} based on the
relative ages of bulges and disks, namely {\it Early} for the
monolithic collapse formation model in which bulges form before disks;
{\it Simultaneous} for the simultaneous formation model; and {\it
Late} for the secular evolution models in which bulges are formed
after disks.  SAM-B will refer to the original Kauffmann semi-analytic
models in which 1:3 or larger satellites are all placed into the bulge
of the primary.  The revised models with satellites going into the
disk are designated SAM-D.  To improve the realism of the comparisons,
the Monte-Carlo realizations of the models include the same luminosity
selection factors as our quality observations (Sections ~\ref{sel_fct}
and 3.4).  The additional constraints based on size were not applied,
since the models did not include such parameters for the bulges and
disks. Since the restrictions based on size eliminated only a small
fraction of the data, in practice the comparisons should be reliable
enough to be illustrative.

To compare data to models, many diagnostics are possible given the
large number of parameters common to both the observations and the
models. 
To limit the discussion, we focus on the  {\it  B/T} vs. bulge color
distribution (see Fig.~\ref{bt_ub}). Model predictions of the {\it  B/T}
distribution, color-magnitude relation, and {\it  B/T} vs. color difference
between the bulge and disk, are provided in Appendix B.

As seen in Fig.~\ref{model_bt_ub}, the most striking result is the
poor fits of both the {\it Simultaneous} and {\it Late} Bouwens models
to the data.  \citet{bouwens99} were unable to eliminate any of their
three basic bulge formation scenarios with the very limited
observational data available at that time.  In contrast, our new data
illuminate significant differences between the models and data, even
after having added the aforementioned improvements to the original
models of \citet{bouwens99}. 

Thus one solid result is that neither the {\it Simultaneous} nor {\it
Late} models of \citet{bouwens99} are viable now.  Both models have
bulges forming at the same time or later than disks, and thus both
predict large fractions of distant high {\it B/T} galaxies with
very-blue, luminous bulges.  The vast fraction of observed luminous
photo-bulges , and thus presumably any subset of genuine luminous
bulges, is found to be very red.  The predominance of very-red bulges
is seen even at lookback times corresponding to the epoch of major
disk and bulge formation at redshifts $z \sim 1$.  In contrast, the
SAM-B model is not a bad rendition of the data, but both the {\it
Early} and SAM-D models yield distributions that are far better
matches to the data.

One clear feature of all models is that blue ($U-B \lesssim 0$) bulges
almost always have large $B/T > 0.6$.  This is true even for the
relatively few blue bulges seen in the {\it Early, SAM-B,} and {\it
SAM-D} models.  This result reflects the difficulty of building up a
significant disk very soon after the blue, early-formation phase of
the bulge.  This near-universal property of blue bulges having high
{\it B/T} ratios in all the models further supports our claim in
Section~\ref{gen_bulges} that the few photo-bulges in our observations
bluer than $U-B \sim 0$ and with low {\it B/T} (large disk fractions)
are more likely to be centralized regions of active star formation
rather than genuine massive bulges in early formation as envisioned by
theorists. The high luminosities predicted for blue bulges in the
models (see Fig. \ref{model_ub_mag} in Appendix B) also support our
contention that we have not actually found a significant population of
massive bulges in the early phase of active star formation.

\clearpage


\begin{figure}
\plotone{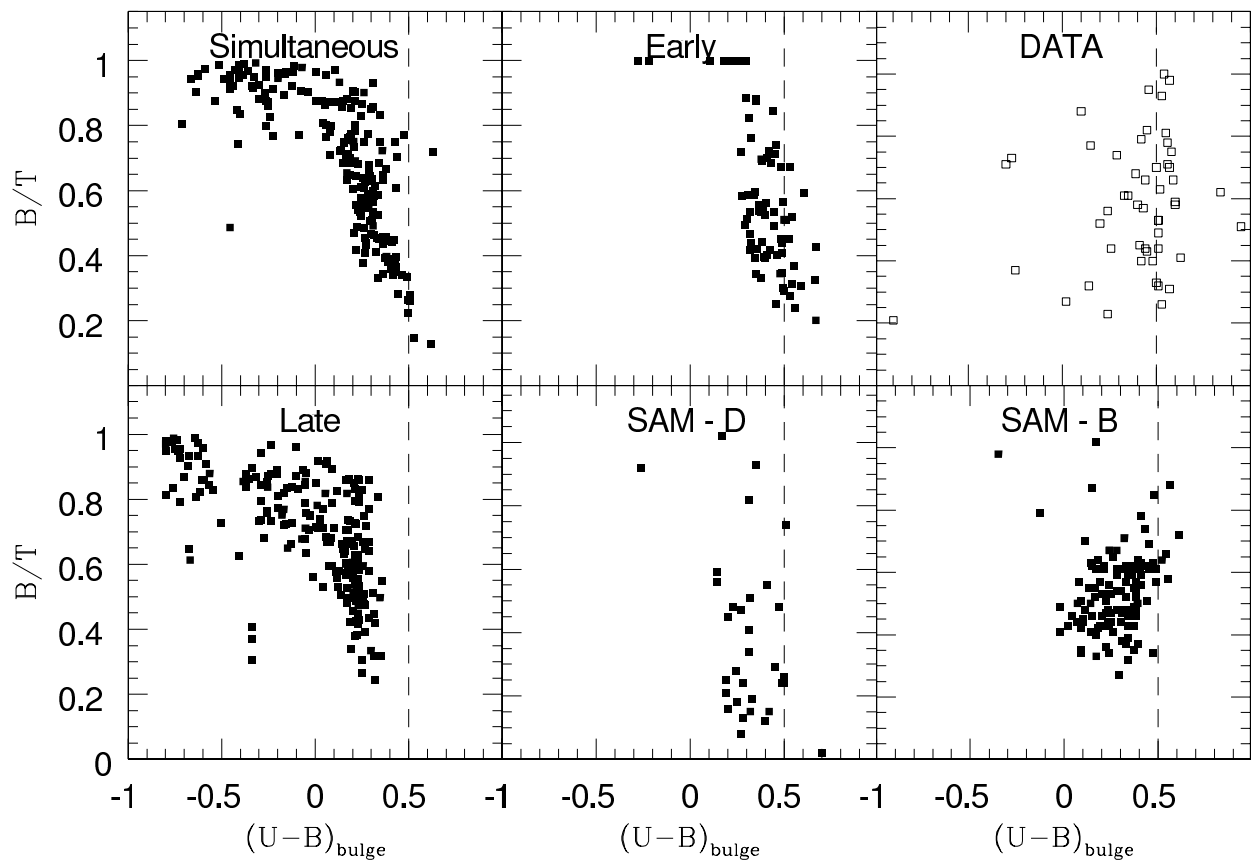}
\caption[model_btcolor.eps] 
{B/T in restframe $B$ vs restframe $U-B$
color for models (as labeled in each subpanel) 
vs. observations from the quality  bulge sample.
Each model is based on one Monte Carlo realization with known selection
biases included.  Note that the large
number of very blue bulges predicted by both the Simultaneous and Late
models are expected to have large {\it  B/T} ratios. The relatively few blue
bulges predicted by the Early and SAM models are also expected only
among the high {\it  B/T} systems. 
\label{model_bt_ub}}
\end{figure}


\clearpage

The {\it Early} and {\it SAM} models match the data fairly well, not
only in the {\it B/T} distribution, but also in the other diagrams
shown in Appendix B.  When examined more closely, some of the more
notable differences include slightly bluer predicted colors on
average, a higher abundance of large {\it B/T} $> 0.6$ in the
observations compared to either of the SAM models.  Also, although the
numbers are small, 4 of the 5 blue model bulges are among the most
luminous of all predicted bulges, whereas the bulk of the bluest
observed photo-bulges are among the fainter half.  Other differences
between the data and models are discussed further in Appendix B.
Assuming that the SAM-B models represent the state of the art among
SAMs,
have been adjusted  to match local observations
or less colors. 
the better match of SAM-D  to our 
data is a vindication of the power and potential of  high-redshift data
to help theorists improve their SAM parameters. 

Making the simplistic assumption that colors are correlated mainly
with age (i.e., that  metallicity variations are secondary), the
comparison of models to data suggests:

\begin{itemize}

\item Most {\it luminous}  bulges (as in the present sample)   appear to be 
very old,  with very red colors\footnote{To reach $U-B \sim 0.5$,
an instantaneous burst must fade for 6-8 Gyr if of solar metallicity. If
of 2.5$\times$ solar, only 2.5  Gyr are needed \citep{bruzual03}.}
 indicating  
formation redshifts at least 1-2 Gyr 
before $z \sim 1$.

\item {\it Luminous} bulges formed almost always before or
simultaneously with disks, rather than afterwards.

\item Bulge colors (and therefore ages) appear to be independent of
{\it B/T}, so that the ages of ellipticals (pure bulge) and the
bulges of S0's and spirals were all similarly old at $z \sim
1$.

\item Few luminous bulges are being formed outright at high
redshifts $z \sim 1$, since the observed blue bulges are generally
not very bright and do not have high {\it B/T}, and thus do not match
the predictions by any models.

\end{itemize}

\subsection {Why are High Redshift Bulges So Very Red?}

The main result of this work is that luminous bulges ($M_B < -19$)
of {\it field galaxies} at high redshifts ($0.73 < z < 1.04$) appear
to be predominantly (81\%) very red ($U-B \sim 0.5$) with relatively
small intrinsic scatter ($\delta (U-B) < 0.03$). This finding appears
to be independent of the relative amount of disk light (i.e., {\it
B/T}), the color of the disk, the luminosity of the galaxy, and even
the presence of emission lines in the integral spectra. The puzzle is
that the color-luminosity relation for luminous high-redshift
bulges appears virtually identical to that found for bulges
today.  Any luminosity evolution only deepens the puzzle, since fading
bulges need to become {\it bluer} with time to stay on the
color-magnitude relation. Such behavior is diametrically opposite to
expectations for any plausible set of models of an isolated, passively
evolving population of stars!

Let us address possible explanations for our results.

{\bf Cosmology:} Cosmology does not affect our basic findings, since
both color and surface brightness are measurements that depend only on
redshift and not on the geometry of the universe.

{\bf K-corrections:} The conversion of our HST WFPC2 $V_{606} -
I_{814}$ to restframe $U-B$ is relatively robust, since at redshifts
near $z \sim 0.8$, the observed bands correspond closely to restframe
$U$ and $B$ \citep[see][]{gebhardt03}. As an independent check, our
K-corrections for $z = 0.83$ match to within 0.02 mag to that derived
by \citet{dokkum00} in their equation B4.

{\bf Photometric zero points:} HST data have much more reliable and
stable zero-points than usually possible from ground measurements.  As
an independent check of our zero point, we can compare whole galaxy,
instead of subcomponent, colors measured by our team \citep[see Figure
8 in] []{im02} to those of other field \citep[e.g.,][]{ellis01} or
cluster \citep{dokkum00} galaxies, all at the same redshift range. All
three samples show $V_{606}-I_{814} \sim 2.0$.

{\bf Systematic errors in GIM2D:} One possible concern is that the
photo-bulge component is made artificially redder by GIM2D. To test
this bias, we have examined GIM2D {\it simultaneous-color} extractions
of a set of simulations where the colors of the photo-bulge and
photo-disks were nearly identical for a full range of S/N (galaxy
brightness), eccentricities, inclination, {\it B/T}, relative sizes,
disk inclinations, etc. We find no evidence for any systematic offset
to redder colors for the photo-bulge component.  As another simple
check, we can compare central colors (column 9 of Table 1) 
to that of the whole galaxy (column 8 of Table 1 or column 7 of Table 2),
\citep[cf. work of][]{abraham99, ellis01} -- the vast fraction
shows a redder center. Thus if 
photo-bulges  dominate  the central light, the photo-bulge colors should
be redder. A key assumption in
adopting  the simultaneous-fit  rather than separate-fit color measurements is
that neither the photo-bulge nor photo-disk has a
color gradient. Another assumption is that disks are
purely exponential to the galaxy center. If instead, disks are
truncated in their inner parts (see \citealt{kormendy77} or
\citealt{baggett98}), and {\it if bulges are redder than disks}, then
the measured colors for the bulge will be biased redder. Although such
inner truncated disks may explain the very red colors for some
photo-bulges, the low dispersion of the very red colors becomes a
serious challenge.

{\bf Local comparison samples:} The $U-B$ colors of local bulges
remain uncertain. The most direct comparison of our sample can be made
with the $U-B$ measurements of the {\it bulges} of 45
early-type (S0-Sbc) galaxies \citep{balcells94}.  After we shift the
distant bulges in the color-magnitude plot to account for about 1
mag of luminosity evolution, we find the $U-B$ colors to be
indistinguishable, either in average value or in scatter, to the
colors of bulges seen locally.  If we instead adopt the {\it
integrated} colors of {\it field} E-S0's, we find the $U-B$ surface
photometry of the field sample of \citet{jansen00} to yield a tight
color-magnitude relation for bright ($M_B < -19$) galaxies (along with
a shallow color-magnitude slope). In this case, our high redshift
sample appears to be slightly {\it bluer} by $\delta U-B \lesssim
0.05$ mag, but after again accounting for about 1 mag of luminosity
evolution, we find the mean colors to be indistinguishable.

Clearly the assumed slope of the CM-relation is important when
comparing distant vs. local colors, as a corection must be applied for
luminosity evolution to match the same physical objects, A shallow
slope means that the color correction due to this effect is less.

These findings of a shallow slope for local bulges are reaffirmed
by recent measures from the Sloan Digital Sky Survey (SDSS)
\citep{bernardi03b}. The {\it integral} $U-B$ photometry of E's from
the 7 Samurai \citep{burstein87} or of E-S0's from \citet{prugniel98}
have larger scatter, steeper color-mag slope, and slightly bluer
colors. In these comparisons, we find the color scatter of the high
redshift sample to be comparable to that actually observed in the
local sample, rather than to the inferred intrinsic scatter.  A full
discussion of this issue is beyond the scope of this work and we
remain cautious of the exact amount of color evolution, but the
predominance of the evidence suggests little, if any, change in the
intrinsic $U-B$ colors of bulges from high redshifts to today,
especially after accounting for a shift in the color-mag relation to 1
mag of luminosity evolution.

{\bf Dust:} Dust is a complication that is difficult to address well,
especially within the limits of our existing optical data for the high
redshift bulges in our sample. At some level, dust must be
present. Indeed, in cases where we see a highly inclined blue disk
through which the bulge appears to be plausibly obscured (e.g.,
064\_4412, 094\_7063, 152\_5051), we find implausibly red colors for
the photo-bulges if dust is not included.  Moreover, as
revealed in HST NICMOS observations of local bulges in early-type
spirals, where much more detailed and careful photometry is possible,
strong color gradients suggest the presence of dust at the significant
level of $A_v \sim \ $ 0.6--1.0 mag, but mainly in the very central
100-200pc \citet{peletier99}.  On larger scales more comparable to the
measurements we can make at high redshifts, the same authors find the
optical to near-IR colors to be so tight among the S0-Sb bulges that
the inferred age spread is no more than 2 Gyr. They also find similar
colors between Coma cluster early-type galaxies and local field galaxy
bulges.  Even if dust plagues the colors of bulges both locally and at
high redshift, a dust explanation of similar colors at both epochs
would imply some {\it evolution} in the relative effects of dust,
since the underlying stellar population is expected to be bluer in the
past.  Though our sample size and color precision are not high, we
find no clear evidence for any dependence of the very red photo-bulge
colors or of their scatter on the inclination angle of the photo-disk
or on the {\it B/T} ratio.  A naive expectation is either for a larger
color scatter among photo-bulges residing in disks (some of which may
be dusty) or for disks that are measured to be inclined. We surmise
that dust is present at some level, but we find no evidence that it is
a dominant source of our red colors.

{\bf Stellar population models:} Even after accounting for the high
level of degeneracy between metallicity and age among most broadband
colors, models from different authors still yield significantly
different (35\%) predictions for ages versus colors
\citep*{charlot96}.  However, regardless of uncertainties in the input
stellar evolution components, no models to our knowledge predict
constant very-red colors for passive evolution.  Though non-linear,
typical changes are 0.2 mag in $U-B$ for each magnitude of luminosity
change.  We remain open to the possibility that a significant effect
or component has been overlooked in all these models that would
salvage the pure passive evolution in explaining the constancy of very
red colors in $U-B$, while the luminosity has brightened by 1 mag. One
suggestion along these lines is that the IMF is truncated above 2
M$\odot$, in which case red and roughly constant colors would result,
even during the ``young'' stage of the first few gigayears after
formation \citep{broadhurst00}. The [O II] emission seen in distant
spheroidals would then need a separate explanation.

{\bf Revision to pure passive evolution:} As discussed in more detail
by \citet[][: GSS9]{gebhardt03}, a plausible scenario to explain our
results requires a more complicated history than pure passive
evolution after an initial, brief burst of star formation. We propose
a post-burst infusion of blue light from small amounts of additional
star formation or from metal poor stars over an extended period.
While GSS9 needed to explain 2.4 mag of luminosity evolution while
keeping {\it integral} colors of galaxies to $U-B \sim 0.4$, our
bulges are slightly redder at $U-B \sim 0.50$ and luminosity
evolution is somewhat milder at $\sim 1$ mag. In this case, the
typical scenario needs less (4\% instead of 7\% by mass) additional
star formation integrated over the lifetime of the bulge.  Thus a
viable scenario to yield 1 or more magnitudes of luminosity evolution
accompanied by a nearly constant $U-B \sim 0.5$ is found to be
achievable.

This suggestion of an additional phase of continued star formation is
compatible in spirit with the claims by, e.g., \citet{trager00}, that
the observed correlations among Mg, Fe, velocity dispersion, and ages
measured from high-quality, local early-type galaxy spectra can be
explained by adding a ``frosting'' of younger (but more metal-rich)
stars to older, solar-metallicity, single stellar populations.  While
\citet{trager00} examined their scenarios using two bursts,early and
late, our scenario includes a more continual infusion of star
formation. As we previously noted, the frequent presence of [O II]
emission lines provides strong evidence for continued star formation
in early-type galaxies, even those that appear to be very old (i.e.,
very red) in both the disk and bulge. Other authors have
also noted the common presence of [O II] emission lines among
early-type galaxies at intermediate redshifts $z \sim 0.4$ at the 25\%
level \citep{willis02, treu02} and even higher fractions ($\sim 33\%$)
and up to higher redshifts $z \sim 1$ \citep[e.g.,][]{schade99,
dokkum03}.  These results favor active star formation as the
``frosting'' component rather than hot, old metal-poor stars. Whether
the star formation arises from infalling gas from the halo or satellites, 
cooling
flows \citep{mathews99}, or cooling of internal residual gas left over
from feedback processes such as prior episodes of supernovae heating
\citep{ferreras02} is not easily discriminated from our data, but we
note that the small scatter observed in the color-magnitude plots
precludes episodic star formation that occurs mainly in strong bursts,
since otherwise significant color dispersions would be expected.
Moreover, whether the new star formation contributes mainly to the
disk rather than bulge is also a key uncertainty.

\section {SUMMARY and CONCLUSIONS}

We present a candidate sample of luminous, high-redshift spheroids
(ellipticals and the bulges of S0's and spirals) found within the
Groth Strip Survey (GSS), one of the early-phase DEEP surveys with
redshifts and spectra from the Keck Telescope and photometry from the
Hubble Space Telescope ({\it HST}).  A framework is adopted in which
the structure of each galaxy is decomposed into two simple
subcomponents, one with an $r^{1/4}$ profile which we dub a
photo-bulge, and another with an exponential profile we dub a
photo-disk.  We caution the reader that our selection and
structure-extraction procedures may, however, also contaminate the
sample with non-bulges such as nuclear/central star-forming regions of
late-type galaxies or any subcomponent that is not well fit simply by
an exponential with one scale length.  We define a statistically
complete sample of 86 galaxies that is constrained to have
photo-bulges brighter than $I_{AB,814} = 24$ and to have good-quality,
spectroscopically-confirmed redshifts in the range $0.73< z < 1.04$.
This sample is extracted from a larger and fainter redshift sample of
about 600 field galaxies within the GSS and comprises about 40\% \ of
the full sample in the same redshift range.  This photo-bulge sample
is the most extensive, faintest, and homogeneous sample of candidate
bulges with solid spectroscopic redshifts at $z \sim 1$ and should be
a statistically complete sample of high-redshift, {\it luminous},
$r^{1/4}$-profile bulges that include bonafide ellipticals, bulges of
S0's, and bulges of spirals.  We largely avoid the common problems in
prior studies of identifying pure ellipticals or early-type galaxies
at high redshift, of mixing galaxies with and without disks into the
analysis of bulge evolution, and of contamination of bulge colors by
bluer disks.

After further pruning the sample to exclude the faintest 0.5 mag and
possibly unreliable measurements, we retain a sample of 52 bulges
with which we analyze the photo-bulge luminosities, sizes, colors, and
volume densities. With the caveat of having adopted several key
assumptions for this work, namely, that galaxies can be decomposed
into $r^{1/4}$ bulges and exponential disks; that color gradients for
either subcomponent are negligible; that colors largely track age
rather than metallicity or dust, we find the following results:

1) The main conclusion is that the vast majority (85\%) of $I_{814}<
23.1$ luminous photo-bulges at redshift $z \sim 1$ are very red, with
median restframe $U-B = 0.46$ and bounded by the 25 percentile at 0.31
and 0.55.  This color matches that found for local E-S0 today ($U-B \sim
0.42-0.64$) and is {\it redder} than the observed {\it integrated U-B colors} of
luminous cluster or field early-type galaxies at similarly high
redshifts.

2) The very red colors of the {\it luminous} photo-bulges are found to
be independent of their fraction of the total light ({\it pB/T}),
i.e., we find no difference between the colors of bulges found in 
disk-dominated spirals and  in early-type, bulge-dominated 
E-S0's.  These red photo-bulges almost always ($\sim 90\%$) co-exist
with photo-disks at the 10\% or greater level ($B/T \lesssim 0.9$),
i.e., pure $r^{1/4}$ ellipticals are rare in our sample
\citep[c.f.,][]{schade99}. Likely systematic errors in our {\it pB/T}
values strengthen this conclusion.  We also find that almost all
(90\%) of the galaxies harboring such very-red photo-bulges appear as
normal early-type or spiral galaxies. In contrast, the galaxies
hosting bluer photo-bulges are dominated (over 60\%) by morphologies
(double nuclei, distortions, close neighbors) highly suggestive of
interactions and mergers.

3) The very-red photo-bulge colors are also independent of the color
of the associated disk and of the inclination of the disk. If dust is
playing a role, we have not been able to discern its effects directly,
except as seen in a few edge-on objects.

4) For the very red photo-bulges, the slope of the color-magnitude
relation is found to be shallow ($\sim -0.02 \pm0.02$) and the
intrinsic scatter about the color-magnitude relation is small, $\sigma
< 0.03$ mag.  These match well to the slope of -0.03 and scatter of
$\sigma \sim 0.024$ mag observed among early-type cluster galaxies at
$z \sim 0.83$ \citep{dokkum00}.  We note that the persistence from
high redshift $z \sim 1$ to today of a shallow slope in the
color-magnitude relation might naively imply that metallicity, rather
than age, is the more dominant cause of the color-magnitude relation
\citep[e.g.,][]{tamura00}. But, as mentioned next, a pure, simple,
passive evolution model is unlikely to be applicable to our bulges,
so the influence of younger stars must be considered.

5) For the very red photo-bulges, the size-luminosity relation reveals
a luminosity (i.e., surface brightness) increase at the level of $\sim
1$ mag by redshift $z \sim 1$. This luminosity brightening coupled
with a lack of color evolution is difficult to explain by simple
passive evolution.  One plausible alternate scenario consistent with
our data starts with a dominant ($\sim$95\%) {\it metal-rich},
early-formation ($z \ga 1.5-2.0$) population that is later polluted
with relatively mild and gradually decreasing star formation.

6) In support of this on-going star formation scenario, we find that
even among galaxies in which {\it both} components are very red (i.e.,
good E-S0 candidates), roughly 60\% (10/16) show O II emission lines.
An even larger fraction, 80\% (8/10), of the most luminous galaxies
show emission lines, indicating on-going star formation at the level
of $\sim 1 M\odot$yr$^{-1}$ per 10$^{10}M_\odot$ of stars.  Although
this rate is considerably greater than needed to explain the constancy
in bulge colors, the source of the star formation activity and the
division of new stars between disk and bulge remain too uncertain to
assess whether we have a true inconsistency.

7) Blue photo-bulges are only a small fraction of bulges (8\%). They
host star formation activity that ranges from mild to intense; possess
lower luminosities ($M_B$) on average than the redder photo-bulges;
often (3/4) reside in {\it redder} photo-disks; and are characterized
by restframe half-light $B$ surface brightnesses too low to enable
them to be the progenitors of the redder, more luminous photo-bulges.
Moreover, the narrow velocity widths ($<\sim 100$~\kms) measured from
some of their strong emission lines argue for low masses, and thus we
conclude that the very blue photo-bulges are in fact luminous,
centrally concentrated, star formation sites within disks that, on
average, have older stellar populations. These blue photo-bulges are
not likely to be the genuine progenitors of {\it luminous} bulges
today, but some are perhaps the
predecessors of small bulges in spirals.

8) Taking points 1 and 7 together, we find little evidence in our
deeper and more extensive data to support previous claims \citep[e.g.,
by ][]{abraham99, menanteau01, schade99} for a {\it significant} (30\%
to 50\%) population of either blue bulges or blue ellipticals at
redshifts 0.7 - 1.0.

9) The colors of photo-disks are almost always the same as or bluer
than that of the photo-bulges, but it is worth noting that we do find
a few very-red, luminous photo-disks at high redshift.  If these are
genuine disks, the implication is that at $z \sim 1$ not all massive
disks are young and that some old, massive S0's have already existed
in the field.

10) The integrated luminosity density ($B$) of {\it very red} photo-bulges
comprise $\sim$36\% of the total at high redshifts $z \sim 1$, a result
in need of better statistics before solid conclusions can be drawn.

11) Finally, we compare our data to improved heuristic formation
models of E-S0's and bulges by Bouwens \etal and find that neither the
{\it Late} nor {\it Simultaneous} bulge formation models match the
{\it B/T} and bulge color distributions. The {\it Early} monolithic
collapse model with old bulges, however, or the semi-analytic models
of, e.g., Kauffmann and collaborators, both provide predictions that
yield far superior and, overall, good matches to our data. The
bulge colors from the models being bluer than seen in the
observations will probably need more complicated physics, e.g.,
protracted but mild star formation from cooling flows from internal gas 
or infusion  of external gas.  

Despite the various conclusions afforded by the present sample, many
important issues remain to be resolved with improved data. On the
observational side, larger samples are clearly needed to improve the
statistics, especially of rarer subsamples, such as the fraction of
blue bulges or the frequency of AGN activity and their correlation
with bulge properties.  Even extending the sample to lower
redshifts may serve to strengthen or to challenge the somewhat
unexpected results we have thus far found at high redshifts. Direct
comparisons of the bulge colors of cluster galaxies versus those in
the field would be valuable. Color gradients and other photometric
structural information (e.g., light profiles which are not $r^{1/4}$,
the presence of nuclear point sources, measures of galaxy distortions)
need to be explored in detail as well as the level of biases in
photo-bulge measurements that may result from unusual morphologies.
Finally, diverse forms of correlations will help to improve our
understanding of the nature of bulges, especially between bulge
properties and other information, such as disk properties, morphology,
close neighbors and environment, or from other wavelengths (near-IR,
far-IR, submm, radio, X-ray), and especially the wealth of new local
data from the 2dF and SDSS.

\acknowledgements 

The authors thank the staffs of HST and Keck for their help in
acquiring the data, to the W. M. Keck Foundation for the telescopes,
to the Hawaiian people for use of their sacred mountain, and to Bev
Oke and Judy Cohen for LRIS that made the redshifts possible. We also
thank R. de Propris, P. Eisenhardt, and A. Graham for useful
discussions regarding the color-magnitude diagram and light profiles
of E-S0's and the referee for many constructive suggestions.  
Support for this work was provided by NASA through grants
AR-05801.01, AR-06402.01, AR-07532.01, and AR-08381.01 from the Space Telescope
Science Institute, which is operated by AURA, Inc., under NASA
contract NAS 5-26555; a research grant from the Committee on Research
from the University of California, Santa Cruz; reward funds for DEIMOS
from CARA; and by NSF grants AST 95-29098 and 0071198.  The project
was initiated and supported by the Science and Technology Center for
Particle Astrophysics during its 10 years of operation at the
University of California, Berkeley.


\appendix

\section {Selection Function for the Bulge Sample} 
\label{app_sel}

\subsection {Overview of Approach} 

Since selection effects can conceivably mimic real evolutionary
changes in the high-redshift galaxy population, it is important to
determine how they affect the DEEP/GSS sample in general and the bulge
sample in particular.  Our approach has two major components.  The
first is based on simulations to determine the incompleteness of our
photometric catalog from which the spectroscopic samples are derived.
The second is based on a purely empirical determination of any
incompleteness of the spectroscopically-confirmed sample by comparing
it to the full photometric catalog.  In both cases, simplifying
assumptions as detailed below are adopted.

In the most general case, the selection function can be quantified by
a weight for each object that is proportional to the inverse of the
effective areal coverage of the entire GSS sample and which combines
the selection functions that depend on multiple parameters.  For this
work on bulges, we have restricted the dependencies to a
small subset of possible parameters that relate most closely to our
analysis, namely, apparent flux, size (or surface brightness), {\it
pB/T}, and color.  A more detailed discussion of selection functions,
but for disks rather than bulges, is provided in \cite{simard99}. The
following summarizes the main components related to this study of high
redshift bulges.

\subsection{Distribution Functions} 

The {\it observed} distribution of  bulges within a
multi-dimensional space of {\it intrinsic} properties, $MP$, versus
redshift, $\Psi_{O}(MP, z)$, is the result of any inherent (i.e.,
within the Universe) distribution $\Psi_{U}(MP,z)$ modified by
observational selection effects, whose functions we designate as $S$.
The possible parameters included within $MP$ are many, but for this
work, the most relevant are the absolute luminosity of the bulge in
restframe $B$, $M_B$; the bulge effective or half-light radius in kpc,
$ R_e$ (or averaged surface brightness within the effective radius,
$\Sigma_{e}$); the bulge to total ratio, $B/T$; and the bulge
restframe color, $U-B$.  All of our radii, scalelengths, and surface
brightnesses refer to the non-circularized semimajor axis values.
Given that the completeness in the detection of faint disks has
already been found to be dependent on at least surface brightness as
well as apparent magnitude (see \citealt{simard99}), we might expect
these to be significant selection effects for bulges as well. Because
of their high surface brightness, the bias for bulges may be less than
that for disks, but bulges still show some dispersion in their sizes
and surface brightnesses in local samples \citep*{bender92,
burstein97}.

After adopting a cosmology and a set of spectral energy distributions
(SEDs) that span the range possessed by real galaxies, we can apply
the appropriate corrections to translate any values of $MP$ and
redshift $z$ to a set of observed parameters, $mp$, or vice versa. We
may thus, henceforth, speak of any function $f(MP,z)$ or $f(mp)$
interchangeably.

The path from $\Psi_{U}(MP,z)$ to $\Psi_{O}(MP,z)$ is given by:

\begin{eqnarray}
\Psi_{O}(MP,z) = S_{PS}(MP,z) S_{UP}(MP,z) \Psi_{U}(MP,z), 
\label{seleq}
\end{eqnarray}

\noindent The subscript $UP$ stands for ``Universe sample to
Photometric sample,'' and the subscript $PS$ stands for ``Photometric
sample to Spectroscopic sample'', where the spectroscopic sample
refers specifically to our photo-bulge sample.  The distribution of
intrinsic galaxy properties, $\Psi_{U}(MP,z)$, is not known {\it a
priori}.  Once the two selection functions in Eq.~\ref{seleq} have
been characterized, however, their product (denoted $S_{US}$
hereafter) yields the region of the $MP$ volume where real galaxies
would have been observed if they existed.  $S_{US}$ is particularly
valuable in making reliable comparisons of theoretical models to data.

\subsection{Spectroscopic Sample Selection}

The current DEEP/GSS sample has a total of 587 objects with both
reliable Keck redshifts and $HST$ structural parameters \citep[][:
GSS1]{vogt04}.  The purpose of the present paper is to study the
luminosities, colors, and volume densities of luminous
bulges at redshifts up to $z \sim 1$, so the sample was
further reduced to 86 galaxies by selecting only galaxies with
photo-bulges brighter than $I_{814}$ = 23.566 and redshifts $0.73 < z
< 1.04$.  Two AGN's (GSS ID: 142\_4838 and 273\_4925) which would have
met our constraints were excluded by requiring the effective radius of
the photo-bulge be greater than 0.03 arcsec (0.3 pixels).

The high redshift limits that define the sample were chosen 1) to
correspond roughly to the $0.75 < z < 1.0$ range adopted by CFRS in
their study of high redshift ellipticals \citep{schade99}; 2) with an
adjustment to a limit just below 0.75 and higher than 1.0 to include
two significant spikes of galaxies at these two limits, as seen in
Fig.~\ref{z_hist}; 3) to avoid redshifts higher than $z \sim 1.04$
where our incompleteness is likely to be serious because [O II]3727
\AA, often our only redshift indicator, falls into the 7600 \AA \
atmospheric ``A'' band absorption feature and then enters a dense and
very bright forest of atmospheric night sky OH lines.  Moreover, in
our chosen redshift range, the observed $I_{814}$ and $V_{606}$
filters correspond roughly to restframe $U-B$, while at higher
redshifts, we are observing further into the ultraviolet where local
galaxies have not been well observed.

The flux limit $I_{814} = 23.566$ for photo-bulges was chosen to
achieve a relatively high spectroscopic success rate and to ensure
moderate precision of structure decompositions and color
measurements. With no consideration of dependencies on redshift, the
DEEP/GSS redshift sample is statistically 85\% complete to bulge
$I_{814}$ = 23.566, meaning that reliable redshifts (quality greater
than 2.9) were obtained for 85\% of the targets observed
spectroscopically.  We emphasize that the final redshift sample of
bulges, however, is neither spatially complete nor uniformly sampled
throughout the GSS, since not all objects with bulges
brighter than $I_{814}$ = 23.566 have thus far been targeted.
Moreover, although the spectroscopic sample was largely chosen as a
magnitude limited sample, i.e. on the basis of $(V+I)/2 \sim R$
magnitudes, the number of targets at each magnitude interval was
purposely chosen to be relatively flat rather than rising towards
fainter fluxes like the counts.  Another reason the observed targets
do not represent a random sampling of the full photometric catalog is
that some candidates were chosen on other criteria, such as having a
clearly visible disk or very red or very blue colors.
The next two
sections address the determination of the actual selection functions.

\subsection{Determination of $S_{PS}$, the Selection Function from
the Full Photometric Catalog to the  Spectroscopic Sample }

Since we do not yet have redshifts for the entire sample of galaxies
in the 28 WFPC2 fields of the GSS, we make the simplifying assumption
that the redshift distribution of the GSS is spatially invariant
across the entire strip. This assumption implies that our existing
spectroscopic sample, regardless of its spatial distribution, has a
redshift distribution that is representative of that from the entire
GSS field.  With this simplifying assumption, we then define the
weight, $W$, for each object in our {\it high redshift} sample of
photo-bulges to be the inverse of the fractional coverage of the
entire GSS field size of 134 square arcmin.  Thus a weight of 5 for an
object implies that it occupies a portion of the observed $pB$ flux,
bulge fraction, color, and surface brightness volume where the
spectroscopic sample totals to 0.2 of the true {\it averaged} number
of galaxies in the Universe in the full 134 square arcmin field of
view. The full selection function is a product of two terms: one from
the true distribution to the photometric catalog, $S_{UP}(mp)$, and
one from the photometric catalog to the actual spectroscopic sample,
$S_{PS}(mp)$.

Here we discuss the determination of $S_{PS}$, while $S_{UP}$ is
described in the next section. As already mentioned, we restrict the
space of observed variables to {\it photo-bulge} flux, color, bulge
fraction ($ {\it pB/T}$), and surface brightness (or size). Some
dependence of the spectroscopic completeness on flux and surface
brightness is to be expected when considering the entire galaxy, but
is not as obvious when considering the flux or surface brightness of a
galaxy subcomponent, such as the photo-bulge.  For bulge color and
bulge fraction, completeness may be further complicated by possible
correlations of these with the relative ease of detecting reliable
spectroscopic features.  Galaxies with strong emission lines, for
example, are expected to be found in very blue galaxies,i.e.,
preferentially among those with small bulge fractions.  Given the
difficulty of accurately assessing all the factors that may affect the
degree to which our actual photo-bulge sample is representative when
compared to that averaged over the entire GSS, we take the following
rough empirical approach.  We ignore more subtle selection effects due
to small number fluctuations, systematic biases due to variable
densities of objects in different parts of the multiparameter space,
and covariances among the parameters.  We have also ignored the fact
that the six objects in the deeper pointing (i.e., those with ID's of
073\_XXXX) have their own selection functions - we have merely adopted
the single one derived for the other 27 GSS pointings.

To derive the selection function, we simply study the relative numbers
of various subsamples of galaxies in our full spectroscopic sample to
that found with the same observed properties in the full photometric
catalog of the GSS(i.e., the one with 587 objects).  No division by
redshifts was made, since we do not have redshifts for all objects in
the full photometric catalog. As previously noted, we have two
photometric catalogs, one that is the full catalog in which the images
in $V$ and $I$ for all objects in the entire GSS were processed
separately by GIM2D. The other catalog only has information for
galaxies in the {\it spectroscopic} sample and was processed with
GIM2D operating on both images simultaneously.  Since our subsample
selection and structural parameters are based on the simultaneous mode
of GIM2D, {\it ideally} we would have the entire GSS processed in this
mode to yield a full photometric catalog for the determination of the
selection function. In practice, we adopt the relatively simple
parameterization of the selection function that is derived instead
from the catalog processed in the separate image mode. We have not
found significant {\it systematic} differences in the measurements,
only improved precision (smaller errors) when the simultaneous mode of
GIM2D is used.

We first divided the spectroscopic sample into several ranges of
photo-bulge $I_{814}$ ($I_{pB}$): e.g., 18-20, 20-21, 21-23, and
23-23.566. For each, we obtained the ratio of the number of
spectroscopic objects relative to the entire GSS photometric catalog
and plotted these against $I_{pB}$.  A smooth fit as a function of
bulge flux then yields the simple selection function:

	{\rm W} = 2.7 for $I_{pB}$ between 20 and 21;

	{\rm W} = 2.2 for $I_{pB}$ between 21 and 21.5; and 

	{\rm W} = $2.0*I_{pB}$ - 41.5 for $I_{pB} > 21.5,$  
\noindent
where $I_{pB}$ is the apparent $I_{814}$ magnitude  of the photo-bulge
as measured in the catalog using separate fits
to the $HST$ \  $I$ and $V$ images (see Table 1). 

We then searched for deviations from these average weights due to each
of apparent galaxy size, color, and {\it pB/T} ratios separately
(i.e., no additional simultaneous subdivision by two or more
parameters) in each of several ranges.  We examined only the
subsamples with high quality redshifts. We deemed the above to be the
most relevant for this study. We thus ignored other possible
parameters for study, including those related to the original
selection of targets for spectroscopy (e.g., presence of close
neighbors, inclination of galaxies, etc.)  or possibly related to the
lack of success in obtaining a redshift (slit length, position within
slit, PA and ellipticity, airmass, etc.).

The findings are relatively simple. Using 0.1 arcsec (1 pixel)  intervals, we found
no variations due to size to within the 68\% confidence limit (68\% CL
corresponding to 1 $\sigma$ for a normal distribution) when
small-number statistics were explicitly taken into account. For color
($V_{606}-I_{814}$ of the photo-bulge) in 0.2 to 0.5 mag bins, we
again found no variations at a significance level greater than 68\%
CL. The biggest discrepancy was 4 objects observed, whereas 10.7 were
expected for the faintest bin with $I_{pB} = 23$ to 23.566 and $V-I$
color of 1.5 to 2.0.  This is a plausible bias because such red, faint
galaxies may have greater difficulty yielding reliable redshifts. For
{\it pB/T}, however, while we found no variation for the three
brightest flux ranges to within 95\% CL,
we did find significant variation in the weights in the faintest
photo-bulge flux range. For $ {\it  pB/T}$ larger than 0.8 (perhaps 
corresponding to
pure ellipticals), our sample had only 1 object while 7 were
expected.  
Small number statistics indicate
that this is still within our chosen threshold of 95\% CL and so the simple
estimate above was retained.  
For  {\it  pB/T} below 0.2, however, we expected
5.8 out of 44 total and found 13, a result that is significantly low  at
greater than the 99\% CL. We thus recommend adopting a weight $W$ =
2.1 instead of the average weight formula above {\it
when selection effects that are dependent on {\it  B/T} are  needed}. 
Only four objects within our {\it high redshift} photo-bulge
sample are affected by this deviation from the global average: GSS ID:
084\_5452, 094\_2210, 094\_7063, and 144\_1141. The cause of this
excess is attributed to our  bias in favor of good candidates for
rotation curve measurements, i.e. well formed,  bright spirals, which
have faint bulges and thus small  {\it  pB/T} ratios.

\subsection{Determination of $S_{UP}$, the Universe to Photometric Catalog 
Selection Function}

The selection function $S_{UP}(mp)$ contains the information needed to
convert any sample of galaxies on the sky to the photometric catalog
produced with SExtractor and reflects the adopted SExtractor detection
parameters (detection threshold in $\sigma$'s, minimum detection area,
etc.).  $S_{UP}(MP,z)$ of the intrinsic properties, MP, is then merely
a conversion of the selection function $S_{UP}(mp)$ using
$K$-corrections that depend on color and redshift and corrections for
size and luminosity distances that depend on the choice of cosmology.

Without a much deeper photometric catalog for direct empirical
measures of the selection function, we have chosen to determine
$S_{UP}(mp)$ from simulations created by \citet{simard02}.  We
generate 30,000 galaxy models with structural parameter values that
uniformly cover the following ranges: total galaxy brightness ---
$16.0$ $\leq$ $I \leq 25.0$; half-light major-axis radius of an
$r^{1/4}$ bulge --- $0\farcs0 \leq r_{e} \leq 4\farcs 0$; bulge to
total ratio --- $0 \leq {\it B/T} \leq 1.0$; bulge ellipticity: $0
\leq e \leq 0.7$; disk scale length --- $ 0\farcs01 \leq r_D \leq
10\farcs0$; and disk inclination angles --- $0 \leq i \leq 85$.  Note
that the simulations were performed only in the $I_{814}$ image, so
that the observed $V-I$ colors were used only to apply the appropriate
K corrections to link to the intrinsic properties. Each model galaxy
was added, one at a time, to an empty 20\arcsec $\times$ 20\arcsec
section of a typical $HST$ GSS image.  ``Empty'' here means that no
objects were detected by SExtractor in that sky section using the same
detection parameters used to construct the object catalog.  Using an
empty section of the GSS ensured that $S_{UP}(mp)$ was constructed
with the real background noise that was seen by the detection
algorithm.  The background noise included read-out, sky, and the
brightness fluctuations of very faint galaxies below the detection
threshold.  This last contribution to the background noise is
particularly hard to model theoretically without {\it a priori}
knowledge of the counts, light profiles, and clustering properties of
galaxies undetected by the SExtractor software; the current approach
bypassed this problem.  SExtractor was run on each simulation with the
same parameters that were used to build the SExtractor photometric
catalog.  The function $S_{UP}(mp)$ was taken to be the fraction of
galaxies successfully detected and measured by SExtractor at each
value of $(I, r_e, {\it B/T})$.

The results are simple: we should have detected 100\% of all galaxies
to our limit of bulge $I_{814} \leq 23.566$ for $r_e \leq 2\farcs5$
regardless of any of the other parameters that were varied. Even up to
$r_e = 4\farcs0$ (the largest in our sample is less than $1\farcs0$),
the completeness is expected to be 99\%.  Thus the final weight for
$S_{US}$ is the same as for $S_{PS}$ determined in the previous
section.
\clearpage

\section {Additional  Model Predictions vs. Data}
\label{app_models}

Besides the {\it B/T} vs. bulge-color plot of the models in the main
section (Fig.~\ref{model_bt_ub}), we present here the model plots
corresponding to the other data figures (Figs.~\ref{bt_hist},
\ref{ub_mag}, \ref{bt_ubbd}). As in Fig.~\ref{model_bt_ub}, the data
refer to the quality sample of 52 rather than the starting bulge
candidate sample of 86. A few comments are made for each figure.

\begin{figure}
\plotone{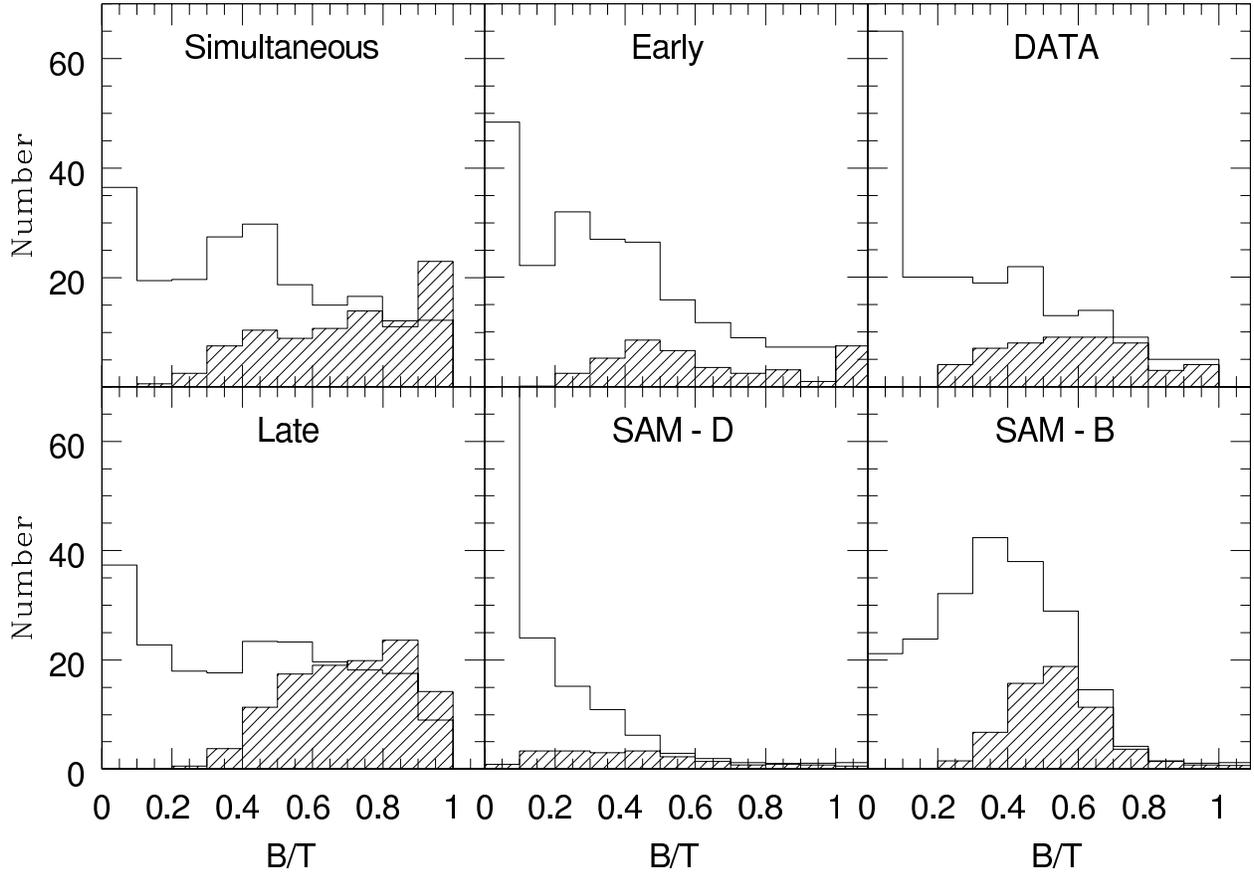}
\caption[model_bthist.eps] 
{B/T distributions for various models as
labeled vs the data.  For the data, the open histogram is the same as
in Fig.~\ref{bt_hist} for the high redshift spectroscopic sample.  The hatched
histogram differs by being the quality bulge sample of 52 rather
than the 86 shown in Fig.~\ref{bt_hist}.  
The model curves show the distributions
with the same selection criteria as adopted for the data.  The open
histogram shows a good match only between the {\it Early} model and the
data, while the hatched histograms show a fair match to the data by
the SAM-B and the {\it Early} models.
\label{model_bt_hist}}
\end{figure}

Fig. \ref{model_bt_hist},{\it {\it B/T} Histogram:} Both the open and
hatched histograms are discriminating.  The open histogram of the
actual data shows a strong peak at the bulgeless ($B/T \sim 0$) end, a
flat distribution to $B/T \sim 0.5$, and a linear drop to the high
bulge limit of $B/T = 1$.  The {\it Simultaneous} and {\it Late}
models of \citet{bouwens99} both show much flatter distributions,
while the SAM-B model shows a humped distribution peaking at $B/T \sim
0.35$.  The SAM-D model shows a sharp drop from the bulgeless end and
is thus also a poor match to the data. Only the {\it Early} of
\citet{bouwens99} model shows a reasonable match, though the bulgeless
end of the model shows too few galaxies.

The hatched portions of Fig.~\ref{model_bt_hist} show that both {\it
Late} and {\it Simultaneous} models predict a large population of
luminous bulges and that most of these would have $B/T > 0.7$. The
quality-sample data are a poor match showing roughly half the number
of luminous bulges and a fairly broad B/T distribution with a peak
near $B/T \sim0.6$. In contrast, the {\it Early} model predicts far
fewer luminous bulges, roughly half that observed, with a peak at
lower $B/T \sim 0.45$, and a small bunch at $B/T = 1$.  The SAM-B
model predicts numbers and a peak in B/T that are a good match to the
data, but the predicted spread is somewhat narrower than
observed. While the SAM-D models appeared to give the best matching
distribution in Fig. \ref{model_bt_ub} of B/T vs. color, here its
histogram shows the numbers are fewer and a shape skewed to lower B/T
than observed.

Qualitatively, the hatched histograms suggest the best fitting model
would be SAM-B with the {\it Early} almost as good while the open
histogram clearly favors the {\it Early} over the SAM-B. The SAM-D
gives a better match than SAM-B for the open distribution, but is
clearly inferior for the hatched bulge sample. Both the {\it
Simultaneous} and {\it Late} models are clearly ruled out in both
distributions.

\clearpage


\clearpage

\begin{figure}
\plotone{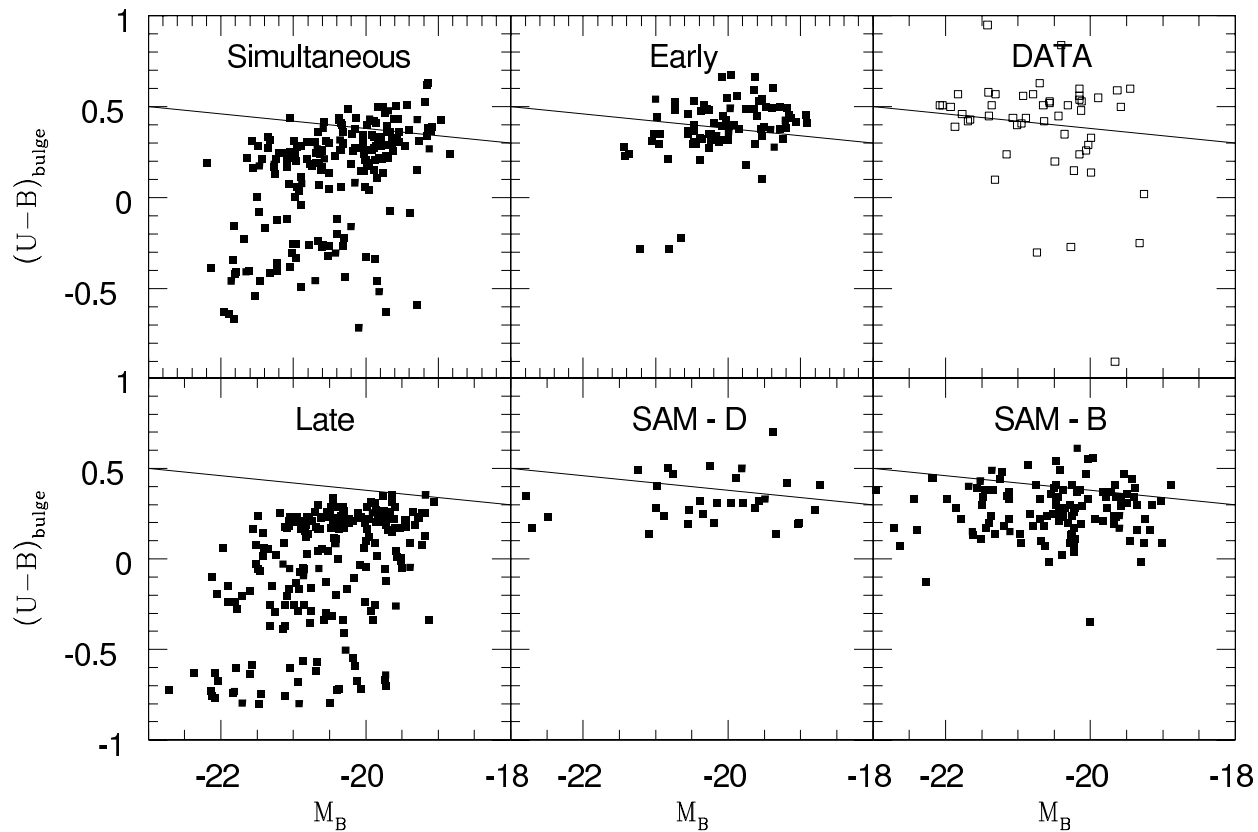}
\caption[model_colormag.eps] 
{Restframe $U-B$ color vs. luminosity
($M_B$) for models and the quality-sample data. The solid line serves
as a reference and is the observed locus for early-type {\it cluster}
galaxies at redshift $z \sim 0.83$ \citep{dokkum00}; see
Fig. \ref{ub_mag} for further details.
A large number of luminous, very-blue bulges is predicted by both the
{\it Simultaneous} and {\it Late} models, while the other three models
yield only a few such bulges, as does the observed sample.
\label{model_ub_mag}}
\end{figure}

Fig. \ref{model_ub_mag}, {\it Bulge Color vs Bulge Luminosity:} The
observational data exhibit a relatively tight band spanning about 3
magnitudes in luminosity and very red colors near $U-B \sim 0.5$ and a
much sparser spread of galaxies towards bluer colors. As seen in a
single Monte-Carlo realization of each of the three improved models by
\citet{bouwens99} and SAM-B and SAM-D (Fig.~\ref{model_ub_mag}), the
{\it Simultaneous} and {\it Late} models are poor matches to data,
while the remaining three models have overall distributions that are
qualitatively similar.  As might be expected, the {\it Late} models
have more luminous blue bulges than the {\it Simultaneous} models; but
both have many more luminous blue ($U-B < 0$) and very blue ($U-B <
0.25$) bulges than do the {\it Early} or SAM models. When the {\it
Early} model and SAMs are examined more closely, there are subtle but
significant deviations from the data. First, the SAM $U-B$ colors for
the bulges are bluer on average by 0.1 to 0.2 mag than that of the
observations.  This color difference is an independent confirmation of
our claim that photo-bulges appear to be too red for an easy
explanation with only passive evolution.  Second, except for one
bulge  in the SAM-B distribution with $M_B \sim -20$ and $U-B < 0$,
the other one in SAM-B and the three in the {\it Early} model are all
at the luminous end of the distribution. The observations show three
or four such blue bulges and all are in the lower half of the
luminosity range.

Overall, the {\it Early} and SAM models match the color-mag data well,
although both {\it Early} and SAM models predict bluer bulges than
observed. The {\it Simultaneous} and {\it Late} models produce far too
many very-luminous, blue bulges to be compatible with our data.  No
model seems to predict a {\it sloped} CM diagram as seen in the
integrated colors of distant red cluster galaxies \citep{dokkum00}.

\clearpage


\clearpage

\begin{figure}
\plotone{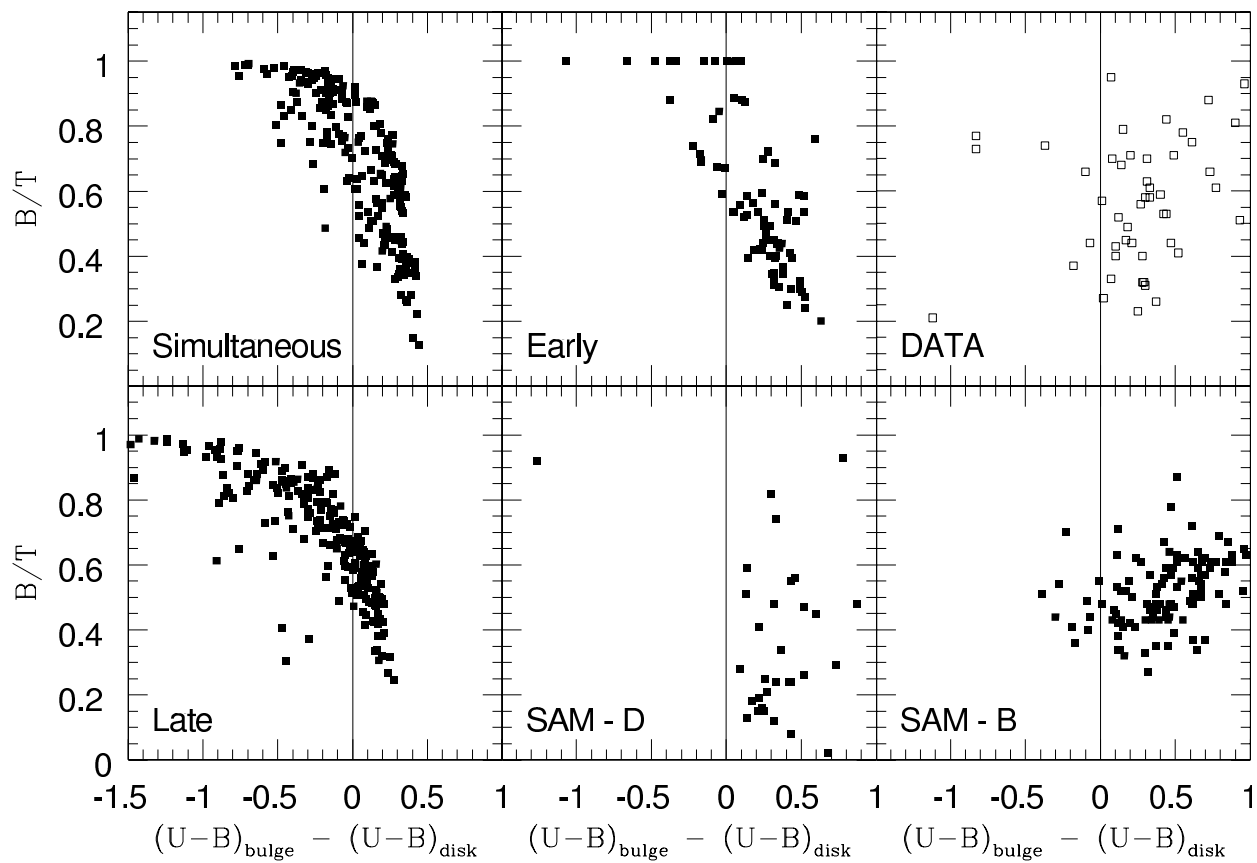}
\caption[model_btbdcolor.eps] 
{{\it B/T} in restframe $B$ vs the
{\it difference} in the restframe $U-B$ colors of the bulges and disks
for models (as labeled) and data for the quality sample.  Each set of
modeled objects is based on one Monte-Carlo realization. The vertical
line divides those bulges that are redder than the disk (on the
right) from those which are bluer (on the left).  Note that among
large {\it B/T} systems, the bulges are systematically bluer than
disks (left-hand side) for the {\it Simultaneous} and {\it Late}
models, and even in the {\it Early} models.  The other two models
match the data qualitatively, though differences in the distributions
are visible.
\label {model_bt_ubbd}}
\end{figure}

Fig. \ref{model_bt_ubbd}, {\it {\it  B/T} vs. Color Difference between Bulge and Disk:}
Fig.~\ref{model_bt_ubbd} shows that the majority of galaxies in
either the {\it Simultaneous} or {\it Late} models have bulges that
are bluer than disks.
In contrast, the {\it Early} and SAM models and the observations all
have bulges that are almost always as red or redder than any disk.
The {\it Early} model shows a systematic trend of bluer bulges  
galaxies with larger {\it B/T} that is not observed. Thus
overall, again the {\it Simultaneous} and {\it Late} models are
strongly excluded by the data, while the other three show
distributions that are fair to good matches, with the {\it Early}
model being somewhat less accurate than the other two.

\clearpage


\clearpage
\section {Comments on Individual Objects}
\label{app_notes}

Here we consolidate additional comments on individual objects.  Sizes
refer to major-axis half-light sizes, i.e., $r_e$ for photo-bulges and
1.67$\times r_d$ for photo-disks. Kinematic measurements have been
made by fitting Gaussians to the emission line profiles, stellar
templates to the absorption lines, and structure-based estimates of
the terminal velocity from 2-D measurement of rotation curves. CID
refers to the CFRS ID's of ellipticals (B/T = 1) as measured by
\cite{schade99}; a mention is made in each case of whether any
emission lines of O II were seen. By ``very red'' for the
subcomponents, we mean $U-B > 0.25$; by ``less red'', we mean $U-B$
between 0 and 0.25; by ``blue'', we mean $U-B$ between -0.25 and 0,
the latter corresponding to the average color of local Sbc galaxies;
and by ``very blue'' to mean $U-B < -0.25$. The * accompanying the
ID indicates candidates in the quality sample (see
Section~\ref{quality})

\vskip 0.3cm

062\_2060* -- CID: 14.1028 (specified twice in their published table)
with O II detected with EW of 31+/-10\AA. This is the 4th most
luminous system with both a very red photo-bulge and photo-disk
component and {\it pB/T} = 0.57.  Yet strong O II emission lines
(restframe $EW \sim$ 10\AA ) are seen with $\sigma \sim 150$~\kms
linewidth. $H_{\delta}$ is strong in absorption, indicating presence
of a young ($<$ 1 Gyr old) stellar population.

062\_6465 -- This object is in the {\it possible} blue E-S0 sample of
\citet{im01}. It has a redder, but larger photo-bulge  than
the very blue, tiny photo-disk.

062\_6859* -- CID: 14.1178 with no O II detected.  This object is in
the E/S0 sample of \citet{im02}. Both the photo-disk and photo-bulge
are very red.  O II emission may be present (restframe $EW \sim$
11.5\AA) and is unresolved.

064\_3021* -- CID: 14.0854 (specified twice in their published table)
with no detection of emission lines. Third most luminous galaxy with
the photo-bulge and photo-disk being both very red.  We find {\it
pB/T} = 0.68 and also find no evidence of emission lines.


073\_1809* -- Part of a complex, interacting pair or set of galaxies,
detected with ISOCAM and observed with an infrared spectrograph on
Keck to measure its $H_{\alpha}$ \citep[see][for details]{cardiel03}.

073\_4569 -- This less red galaxy has a very red, low luminosity ($M_B
> -20$) photo-bulge that is larger than the tinier and blue
photo-disk.

073\_7749 -- This object has a very red photo-bulge  that is
much larger than the less red (i.e., slightly bluer) photo-disk.

074\_6044* -- This is the second most luminous galaxy with both
components being very red. The photo-disk is much smaller than the
photo-bulge. Strong O II emission lines (restframe $EW \sim $8\AA )
are detected and their velocity widths are unresolved (i.e., $\sigma
\lesssim 60$~\kms)

074\_6844* -- This object has the limiting eccentricity of 0.70 and an
extremely tiny photo-bulge. Though the photo-bulge size is
suspect, its color is nevertheless measured to be intrinsically very
red.

084\_1138* -- CID: 14.1277 with no emission, but their visual
classification is Sab or later.  This is one of 3 very red pB in the quality sample
with unusual morphology. HST images show a blue tidal feature
or single wide spiral arm. We derive a {\it pB/T} $\sim 0.43$ and a
blue photo-disk. We find no strong emission lines.

084\_4515 -- This object has an extremely red ($U-B = 1.24$), low
luminosity ($M_B > -20$) photo-bulge that is larger than the tinier
and very blue photo-disk. This galaxy is an ISOCAM source and has been
observed spectroscopically in the near-infrared by \citet{cardiel03}.

092\_1339* -- CID: 14.1496 with strong O II emission. This galaxy is
the only {\it non-very-red} photo-bulge more luminous than $M_B = -21$
and is one of the two blue E-S0's in the quality sample (other is 294\_2078) 
studied by \citet{im01}.  With a $U-B =
0.10$, it is still red by our definition.  The photo-disk is
minor ({\it pB/T} = 0.88), both in luminosity and size. Multiple Keck
spectra yield a well-measured low velocity width of O II of only $\sim
85$~\kms.  Except for the low mass inferred from the kinematics, this
object would otherwise be the best candidate for a genuine blue
bulge. Such low mass systems, however, might be closer to the
luminous compact blue galaxies (c.f., \citealt{guzman98}), some of
which appear to be possible progenitors of dwarf ellipticals such as
NGC-205.

092\_2023 -- This object, with eccentricity of 0.69, has close to the
limiting value (0.70) imposed by the software modeling. More
interesting, it has an extremely tiny photo-bulge and is the
most luminous galaxy with $r_e < 0.1$ arcsec (1 pixel). Even when the whole galaxy is
considered, the half-light size remains so tiny ($< 1$ kpc), that it
lies in the extreme tail of the distribution of sizes for E-S0
\citep[see GSS9 or][] {bernardi03a}.  Yet, when its large velocity
dispersion of $\sim$200~\kms is taken into account, it is offset from
the local fundamental plane by roughly 2.5 mag; this amount matches
well the overall evolution seen at redshift $z \sim 1$
\citep{gebhardt03}.  This amount is far smaller than the inferred
offset of $\sim$4 mag from the size-luminosity relation and is a
caution that any inferred evolution from the size-luminosity relation
should be independently checked. Whether the photo-bulge size and thus
surface brightness are reliable, its color is, nevertheless, measured
to be very red, an expected result since the whole galaxy is very red.

092\_3358 -- One of 7 photo-bulges more luminous than $M_B = -20$ that
are {\it not very red}. It has uniformly blue colors (restframe $U-B
\sim 0$) for the entire galaxy, including a photo-disk that is
nominally smaller in half-light size than the photo-bulge. Strong O II
emission is detected, but with low velocity width $\sigma \sim
50$~\kms.

092\_6027 -- Its {\t pB/T} may be overestimated by 0.07 -- see discussion 
in Sec. \ref{GIM2D}.

092\_7241 -- The photo-disk is measured to be smaller and redder than
the photo-bulge, but the galaxy is blue overall, with the spectra
showing very strong emission lines of O II.  Two fainter neighbors lie
within 2 arcsec.

093\_1325 -- This very red galaxy has an intrinsically very red, low
luminosity ($M_B \sim -19.1$) photo-bulge that is much
larger in size than the equally red photo-disk.

093\_2268 -- Both components are of low-luminosity and blue, with the
photo-disk slightly smaller than the photo-bulge.

093\_2327* -- One of three out of 41 very red quality pB with unusual morphology,
in this case apparently having 4 very close interacting or merging satellite galaxies.

093\_2470* -- CID 14.1311 with no emission lines at the galaxy
redshift.  This system is, however, part of a quad-lens system
\citep{ratnatunga99, crampton96} where the background source at
redshift $z \sim 3.4$ is easily discerned via strong, broad emission
lines. Our {\it pB/T} $\sim 0.5$ suggests an S0 rather than pure
$r^{1/4}$ elliptical, though the photo-disk has colors close to that
of Sbc galaxies ($U-B \sim 0$).  This galaxy is the second most
luminous in the spectroscopic sample of 205 with high redshifts $z$
between 0.73 and 1.04.

093\_3251* -- CID: 14.1356 with strong O II emission and a visual
classification of Sab or later.  \cite{schade99} claimed this is a blue
pure elliptical. We measure {\it pB/T} $\sim 0.6$, a
very red photo-bulge, and a very blue disk. The galaxy is blue
overall, shows features resembling spiral-arms or tidal extensions,
and yields strong O II emission lines (restframe $EW \sim$ 32\AA) that
are unresolved in velocity width.

094\_1313* -- This object has an eccentricity of 0.69 (close to the
limiting value of 0.70) and a small, very red, pure photo-bulge. 
Its {\it pB/T} may be overestimated by 0.07 -- see discussion
in section \ref{GIM2D}. 

094\_2210 -- This system is distinguished in having among the lowest
 {\it pB/T} $\sim 0.11$ in our sample; a morphology with multiple
 blobs that might suggest a disk in early formation \citep{koo96}; and yet a very
 red photo-bulge. The rotation curve estimate of a terminal velocity
 $\sim 290$\kms by \citet{vogt96} suggests that this galaxy is
 massive.

094\_2660* -- This is the 5th most luminous galaxy with a very red
photo-bulge and photo-disk. Emission lines of O II are seen
(restframe $EW \sim $13\AA ) with a high velocity width $\sigma \sim
200$~\kms.

094\_4009 -- A very-blue, low-luminosity photo-bulge with a very-red,
equally bright photo-disk (restframe $U-B \sim 0.42$). The morphology
is peculiar; strong emission lines of O II are found; and the velocity
width is unresolved at $\sigma \sim 26$~\kms.

094\_4767* -- Among the bluer photo-bulges in the quality sample.  The
image, however, shows two compact concentrations of light of roughly
equal brightness and color and imbedded {\it at the edge} of a round
disk-like component. One of the compact subcomponents has been
identified as the photo-bulge, while the other has been regarded in
our detection system to be {\it a separate galaxy} (see GSS2 for an
image of the residuals to the GIM2D fit).  The photo-bulge in this
case is {\it unlikely} to be a genuine blue bulge and is instead
probably one of two blue, very actively star-forming regions of a
late-type galaxy.

094\_6234* -- This is one of three out of 41 very red {\it pB} with unusual
morphology -- in this case, several very close apparently interacting 
neighbors. Its {\it pB/T} may be overestimated by 0.09 as discussed 
in section \ref{GIM2D}. 
 
103\_2074* -- This is the 9th most luminous galaxy with both components
are very red.  The photo-bulge is much tinier and and of lower luminosity
than the photo-disk.  Emission lines of O II are detected
(restframe $EW \sim $6-12\AA ) with a large velocity width of $\sigma
\sim 195$~\kms. Its {\it pB/T} may be overestimated by 0.09 as discussed
in section \ref{GIM2D}.

103\_2974 -- The very red galaxy has a very red, low luminosity ($M_B
> -20$) photo-bulge that is larger than the less red photo-disk.

103\_4766* -- This very red galaxy has a photo-bulge with eccentricity
of 0.69, close to the forced limit of 0.70.  Though the size of its
extremely tiny (0.04 arcsec) photo-bulge may be suspect, it
appears very red. The photo-bulge is accompanied by a nominally
larger, but still very red photo-disk.

103\_7221* -- The 6th most luminous galaxy with both components being
very red. Emission lines of O II are seen at $EW \sim$ 6\AA with a
velocity width of $\sigma \sim 40$ \kms.

104\_6432* -- One of 7 non-very-red photo-bulges more luminous than
$M_B = -20$. It has the 3rd bluest photo-bulge among the quality
sample, but it has a very red, {\it smaller photo-disk} and a $ {\it
pB/T} = 0.73$. The small photo-disk better represents the center of
this galaxy and thus probably its true bulge. Thus the true
bulge is then actually very red ($U-B = 0.57$). The presence of
emission lines is uncertain. Its {\it pB/T} may be overestimated by 
0.07 as discussed in section \ref{GIM2D}.

112\_5966* -- The very red photo-bulge is very small (0.06 arcsec).

113\_3311* -- This is the 8th most luminous galaxy with both
components being very red.  The photo-disk is, however, much smaller
than the photo-bulge.  A companion galaxy 113\_2808 (lower right of
image panel) has the same redshift ($z = 0.8117$) and is seen as a
very peculiar arc-like or string-like galaxy with one end pointing
towards 113\_3311.  Strong emission lines of O II are detected ($EW
\sim $5\AA) with a moderate velocity width of $\sigma \sim
100$~\kms. $H_{\delta}$ is strong in absorption and higher order
Balmer lines are visible, both clues suggesting presence of a young
post-starburst phase.

113\_3646* -- Its {\it pB/T} may be overestimated by 0.07 as
discussed in section \ref{GIM2D}.

124\_2009 -- This low-luminosity, very-blue photo-bulge  has
an eccentricity of 0.69 (close to imposed limit of 0.70) and is
accompanied by a slightly smaller but more luminous and redder
photo-disk.

134\_4363* -- This object has an unusually high restframe $B$? {\it
B/T} = 0.98 with a very red (0.57) photo-bulge.

144\_1141 -- The very red photo-bulge is extremely small in this
photo-disk dominated system (${\it pB/T} = 0.15$).

152\_3226 -- This uniformly very-blue galaxy has a peculiar morphology
and is near a very bright galaxy about 1 arcsec away. The very blue
photo-bulge is larger than the very blue photo-disk.

152\_5051* -- The extremely red photo-bulge color may be affected by
a dusty edge-on disk. The overall photometry may suffer
significant contamination by a very bright projected neighbor.

153\_0432 -- The photo-bulge is larger than the photo-disk, but both
are very red. O II is detected at EW $\sim$ 10\AA\ and is unresolved.

153\_2422 -- Peculiar morphology, perhaps part of an interacting
system with 153\_2622.  Photo-bulge is much larger than the photo-disk
but both are very blue.  The O II emission is very strong (EW $\sim
47$\AA) and unresolved in width.

153\_2622 -- Peculiar morphology and other part of 153\_2422 system.

153\_5853 -- The photo-bulge is small, faint, and blue. This galaxy is
a good candidate to belong to the compact narrow emission line galaxy
(CNELG) class.  Its photo-B/T may be overestimated by a large systematic
error of 0.14 as discussed in section \ref{GIM2D}.

163\_4865* -- Its photo-bulge color is unphysically red ($U-B = 1.53$),
but has large random errors and may be affected by a dust.

164\_6109* -- One of two luminous ($M_B < -21$) photo-bulges that are
{\it not very red}, but only barely, with $U-B \sim 0.24$. The
photo-disk has blue colors and emission lines are detected with the
width unresolved, i.e. $\sigma < 50$~\kms.

174\_4356* -- Has the very bluest photo-bulge in the quality sample, a
low {\it pB/T} ratio, and a less red photo-disk.  The galaxy has
asymmetrical subcomponents.

183\_2970* -- Very blue, peculiar galaxy in a complex system. The
photo-bulge, the 2nd bluest among the quality sample, is however much
larger than the very blue photo-disk. Strong emission lines of EW
$\sim 60$\AA\ are unresolved though seen with a tilt in the 2-D
sky-substracted image of the spectrum.

184\_6971 -- This very blue galaxy has been assumed to be one part of
a very blue, close triple system. The photo-bulge is measuerd to be
much larger than the photo-disk.  Emission lines are very strong (EW
($H_{\beta} \sim 80$\AA) and resolved at about $\sigma \sim $80 \kms.

193\_1838* -- This galaxy has an especially prominent spiral
structure.  The photo-bulge color of $U-B = 0.24$ is just below the
very red threshold. The photo-disk is blue and the system has a low $
{\it pB/T} \sim 0.23$.

203\_4339* -- This is the 7th most luminous galaxy with both components
very red. Emission lines are observed ($EW \sim 2.5$\AA) with velocity
width barely resolved ($\sigma \sim 60$\kms).

212\_1030* -- One of 7 photo-bulges more luminous than $M_B = -20$
that are {\it not very red}. It has $U-B = 0.20$ and is in the blue
E-S0 sample of \citet{im01}. The object is one in a string of several
blobs (see Fig.~\ref{vi_panel}). Presence of emission lines is
uncertain.

222\_2555* -- This very red galaxy has a very red photo-bulge that is
measured to be larger than the less red photo-disk. The {\it pB/T} is
greater than 0.7 and thus still included in the quality sample.


273\_4427 -- This object is in the {\it possible} blue E-S0 sample of
\citet{im01}. The photo-disk is very blue, more luminous, and much
smaller than the photo-bulge. This galaxy is a good candidate to
belong to the compact narrow emission line galaxy (CNELG) class. The
strong emission lines have a velocity width of $\sigma \sim 86$~\kms.
 
273\_7619 -- The photo-bulge is {\it not very red}, but the galaxy has
a very tiny (0.1 arcsec disk scale length), very-red photo-disk. Thus
the true bulge is actually very red (see Fig.~\ref{vi_panel}).

274\_5920* -- This is the most luminous galaxy and has the most
luminous photo-disk in our spectroscopic
sample of 205 at redshifts between 0.73 and 1.04. The two
subcomponents are roughly equal in luminosity and both are very
red. No emission lines are detected.

282\_5737* -- This galaxy has a much redder, fainter, and smaller photo-disk
than photo-bulge.  

283\_5331* -- This very red galaxy has a very red photo-bulge. Though
the photo-bulge is larger than the blue photo-disk, the {\it pB/T} is
0.75 and thus the bulge was included in the quality sample.

283\_6152* -- This the the 10th most luminous galaxy in which the
photo-bulge and photo-disk are both very red. The
photo-bulge is tiny and has an eccentricity at the limit of 0.70.
Emission lines of $EW \sim 3$\AA\ are unresolved.

292\_0936 -- The {\it pB/T} is 0.12, the lowest in the total 86
bulge sample.  The photo-bulge is still very red ($U-B \sim 0.65$,
but has large errors of 0.25 mag in $U-B$) and the photo-disk is blue
($U-B = -0.07$).

292\_6262* -- Both components are very red with a {\it pB/T} of
0.32. No emission lines are seen.

294\_2078* -- This blue galaxy, like 092\_1339,  is in the blue E-S0 sample of
\citet{im01}, but it has a rotation curve and appears to be a
spiral. The photo-bulge is nonetheless very red while the photo-disk
is very blue.  This is an excellent example of the confusion regarding
bulge colors when subcomponents are not separated.

303\_1249* -- Both components are very red, with {\it pB/T} of
0.44. Emission lines are observed ($EW \sim 5$\AA) to be unresolved.

303\_4538 -- The blue photo-bulge is much larger than the very red
photo-disk, which is the more likely counterpart to the true bulge.

313\_4845 -- Its photo-B/T may be overestimated by 0.11 as discussed
in section \ref{GIM2D}.

313\_7453* -- This photo-bulge with $U-B = -0.25$ is the third bluest
in the quality sample (after 183\_2970 and 174\_4356), but it has
a redder photo-disk.  The morphology
shows a central kidney-bean shaped component.  The galaxy is
relatively compact for its blue color and may qualify as a compact
narrow emission line galaxy (CNELG). The O II emission line is strong
($EW \sim 20$\AA)  and is unresolved.


\clearpage

\begin{figure}
\caption
{\footnotesize $V_{606}$ and $I_{814}$ images of the full 86 bulge candidate sample.
Each image is $8.0\arcsec\times 8.0\arcsec$ with the candidate in the center
and with North and East in usual noon and 9 o'clock position.
Top-row labels give the GSS-ID, $I_{814}$ mag, 
and $V_{606}-I_{814}$ color of the photo-bulge. The
second row shows the redshift, $M_B$ of the photo-bulge ($h=$0.7,
$\Omega_m$=0.3, $\Omega_{\Lambda} = 0.7$), $pB/T$ ratio in restframe
$B$, and restframe $U-B$ color. Asterisks (*) indicate members of the
quality sample.
\label{vi_panel}}
\end{figure}

\clearpage

\begin{deluxetable}{l c c c l l l l l r r l l l l}
\tabletypesize{\scriptsize}
\tablecolumns{14}
\tablewidth{0pt}
\tablecaption{Galaxy Data for High Redshift Photo-Bulge 
Sample from Separate Color Fits\label{table1}}
\tablehead{
\colhead{No.} &
\colhead{Source ID} &
\colhead{R.A.} &
\colhead{Dec.} &
\colhead{$I_{814}$} &
\colhead{$I_{814}$} &
\colhead{$V-I$} &
\colhead{$V-I$} &
\colhead{$V-I$} &
\colhead{$r_e$} &
\colhead{$r_{1/2}$} &
\colhead{$z$} &
\colhead{$Qz$} &
\colhead{$W$} &
\colhead{Notes}\\
\colhead{\nodata} & \colhead{\nodata} & \colhead{J2000} &
\colhead{J2000}   & \colhead{Bulge}   & \colhead{Gal.}  &
\colhead{Bulge} & \colhead{Gal.}  & \colhead{Aper.}  & \colhead{px}    & 
\colhead{px}  & \colhead{\nodata} & \colhead{\nodata} & 
\colhead{\nodata} & 
\colhead{\nodata}\\
\colhead{(1)} & \colhead{(2)} & \colhead {(3)} &
\colhead{(4)} & \colhead{(5)} & \colhead {(6)} &
\colhead{(7)} & \colhead{(8)} & \colhead {(9)} &
\colhead{(10)} & \colhead{(11)} & \colhead {(12)} 
& \colhead {(13)} & \colhead {(14)} & \colhead {(15)} }
\startdata
1* &     052\_6543 &  14:17:48.28 &   52:31:17.2 &   21.20 &   20.31 &    1.80 &    1.84 & 1.95  &   1.93 &    6.34 &    0.755 &  4.0 &  2.2 & ef \\*      
2* &     062\_2060 &  14:17:45.99 &   52:30:32.1 &   22.04 &   21.16 &    2.04 &    2.12 & 2.21  &   2.30 &    3.53 &    0.985 &  4.0 &  2.6 & cde \\*      
3  &     062\_6465 &  14:17:41.22 &   52:30:26.9 &   22.62 &   21.89 &    3.99 &    1.69 & 1.61  &  4.25 &    2.67 &    1.020 &  2.9 &  3.7 & dg \\*      
4* &     062\_6859 &  14:17:40.93 &   52:30:21.1 &   22.52 &   21.91 &    2.41 &    2.39 & 2.21  &  1.61 &    3.12 &    0.987 &  2.9 &  3.5 & cf \\*      
5* &     064\_3021 &  14:17:52.03 &   52:29:29.3 &   21.60 &   21.13 &    1.61 &    1.97 & 2.10  &  3.36 &    5.04 &    0.997 &  3.0 &  1.7 & bcdf \\*      
6  &     064\_4412 &  14:17:53.41 &   52:29:41.1 &   23.73 &   22.33 &    1.47 &    1.36 & 1.89  &  7.14 &    9.38 &    0.988 &  3.0 &  6.0 & \\*      
7  &     064\_4442 &  14:17:54.04 &   52:29:11.3 &   23.86 &   21.97 &    3.87 &    1.04 & 1.54  &  6.29 &    6.31 &    0.878 &  3.0 &  6.2 & h \\*      
8  &     064\_4813 &  14:17:53.75 &   52:29:41.0 &   23.65 &   21.85 &    2.89 &    1.20 & 1.44  &  5.90 &    6.49 &    0.986 &  3.0 &  5.8 & h \\
\enddata
\end{deluxetable}

\begin{deluxetable}{l c r r r r r r r r r r r  l}
\tabletypesize{\scriptsize}
\tablecolumns{14}
\tablewidth{0pt}
\tablecaption{Data for Galaxy Components From Simultaneous 
Color Fits\label{table2}}
\tablehead{
\colhead{No.} &
\colhead{Source ID} &
\colhead{$I_{814}$} &
\colhead{$I_{814}$} &
\colhead{$I_{814}$} &
\colhead{$pB/T$} &
\colhead{$V-I$ } &
\colhead{$V-I$ } &
\colhead{$V-I$ } &
\colhead{$r_e$ } &
\colhead{$e$}   &
\colhead{$r_d$} &
\colhead{$i$} &
\colhead{$\chi^2_{814}$} \\ 
\colhead{\nodata} & \colhead{\nodata} & 
\colhead{Gal.}    & \colhead{Bulge} &
\colhead{Disk}   & \colhead{$I_{814}$}   
& \colhead{Gal.} & \colhead{Bulge}  & \colhead{Disk} &
\colhead{Bulge} & \colhead{Bulge} 
& \colhead{Disk} & \colhead{Disk} & \colhead{$\chi^2_{606}$}\\
\colhead{(1)} & \colhead{(2)} & \colhead {(3)} &
\colhead{(4)} & \colhead{(5)} 
& \colhead {(6)} & \colhead{(7)} 
& \colhead{(8)} & \colhead {(9)} & \colhead{(10)} 
& \colhead{(11)} & \colhead {(12)} 
& \colhead{(13)} & \colhead{(14)}
}
\startdata
1* &     052\_6543 &  20.31 &  21.18 &  20.95 &  0.45 &   1.87 &   2.01 &   1.77 &   0.20 &   0.47 &   0.65 &  13.67 &    1.034  \\ \nodata & \nodata & $^{+0.01 }_{-0.02 }$ & $^{+0.03 }_{-0.03 }$ & $^{+0.03 }_{-0.03 }$ & $^{+0.01 }_{-0.02 }$ & $^{+0.04 }_{-0.03 }$ & $^{+0.06 }_{-0.08 }$ & $^{+0.06 }_{-0.04 }$ & $^{+0.01 }_{-0.01 }$ & $^{+0.02 }_{-0.01 }$ & $^{+0.02 }_{-0.02 }$ & $^{+3.51 }_{-3.80 }$  &    1.009 \\      
2* &     062\_2060 &  21.11 &  21.72 &  22.01 &  0.57 &   2.23 &   2.25 &   2.21 &   0.30 &   0.46 &   0.26 &  65.60 &    1.087  \\ \nodata & \nodata & $^{+0.02 }_{-0.02 }$ & $^{+0.08 }_{-0.08 }$ & $^{+0.12 }_{-0.09 }$ & $^{+0.04 }_{-0.04 }$ & $^{+0.03 }_{-0.04 }$ & $^{+0.11 }_{-0.12 }$ & $^{+0.16 }_{-0.14 }$ & $^{+0.02 }_{-0.02 }$ & $^{+0.09 }_{-0.06 }$ & $^{+0.01 }_{-0.01 }$ & $^{+2.56 }_{-3.47 }$  &    1.003 \\      
3  &     062\_6465 &  21.91 &  22.62 &  22.69 &  0.51 &   1.69 &   3.58 &   1.00 &   0.30 &   0.10 &   0.15 &  42.91 &    1.008  \\ \nodata & \nodata & $^{+0.03 }_{-0.06 }$ & $^{+0.14 }_{-0.12 }$ & $^{+0.16 }_{-0.13 }$ & $^{+0.07 }_{-0.05 }$ & $^{+0.06 }_{-0.05 }$ & $^{+1.41 }_{-0.46 }$ & $^{+0.14 }_{-0.17 }$ & $^{+0.05 }_{-0.03 }$ & $^{+0.04 }_{-0.05 }$ & $^{+0.01 }_{-0.01 }$ & $^{+2.50 }_{-3.04 }$  &    1.035 \\      
4* &     062\_6859 &  21.89 &  22.33 &  23.07 &  0.66 &   2.34 &   2.26 &   2.49 &   0.20 &   0.27 &   0.32 &  62.92 &    0.982  \\ \nodata & \nodata & $^{+0.03 }_{-0.02 }$ & $^{+0.11 }_{-0.09 }$ & $^{+0.20 }_{-0.16 }$ & $^{+0.05 }_{-0.06 }$ & $^{+0.03 }_{-0.03 }$ & $^{+0.16 }_{-0.15 }$ & $^{+0.37 }_{-0.33 }$ & $^{+0.02 }_{-0.03 }$ & $^{+0.04 }_{-0.04 }$ & $^{+0.03 }_{-0.03 }$ & $^{+2.23 }_{-2.86 }$  &    1.011 \\
\enddata
\tablecomments{
Section \ref{simul_fit_section} discusses the use of simultaneous fitting to extract the measurements. Errors correspond to 68\% confidence limits from the
GIM2D fits and Monte Carlo samplings and do {\it not} include any systematic
or random errors that can be derived from simulations (see GSS2 and Section
~\ref{GIM2D}).\\
Col. (1): Sequence number ordered by source ID; galaxies with 
* belong to the quality sample (see section \ref{quality}). \\
Col. (2): Source ID is given by FFC-XXYY, where FF is the 
subfield,
C is the WFPC2 chip number, and XX and YY are the  
chip~coordinates in units of 10 pixels. \\
Col. (3): $I_{814}$ of entire galaxy based on GIM2D fits. \\
Col. (4): $I_{814}$ of photo-bulge component.  \\
Col. (5): $I_{814}$ of photo-disk component.  \\
Col. (6): pB/T ratio measured on the $I_{814}$ image; true random errors are $\gtrsim 0.1$. Section \ref{GIM2D} discusses systematic errors.\\
Col. (7): $V_{606}-I_{814}$ color of entire galaxy. \\
Col. (8): $V_{606}-I_{814}$ color of photo-bulge component. \\
Col. (9): $V_{606}-I_{814}$ color of photo-disk component. \\
Col. (10): Half-light major-axis radius of photo-bulge in arcsecs
on the  $I_{814}$ image; note that $r_e (1-e)^{1/2}$
gives the circularized effective radius,$r_{e,c}$ 
as measured with a circular aperture. \\
Col. (11): Bulge eccentricity (limited to a maximum of 0.7) 
is equal to $1-b/a$ (a and b  are the semi-major and semi-minor
axis, respectively). \\
Col. (12): Exponential scale length of photo-disk in 
arcsecs
from $I_{814}$ image.  \\
Col. (13): Photo-disk inclination angle (0 for face-on). \\
Col. (14): Reduced $\chi^2$ of simultaneous fits for the I814 and 
V606 images, with expected values near 1.0.}
\end{deluxetable}

\begin{deluxetable}{l c r r r r r r r r r r }
\tabletypesize{\scriptsize}
\tablecolumns{12}
\tablewidth{0pt}
\tablecaption{Derived Properties of Galaxy Components ($h$ = 0.7, 
$\Omega_m = 0.3$, $\Omega_{\Lambda} = 0.7$)\label{table3}}
\tablehead{
\colhead{No.} &
\colhead{Source ID} &
\colhead{$M_B$ } &
\colhead{$M_B$} &
\colhead{$M_B$} &
\colhead{$pB/T$} &
\colhead{$U-B$} &
\colhead{$U-B$} &
\colhead{$U-B$} &
\colhead{$R_e$} &
\colhead{$\langle \Sigma_e\rangle$} &
\colhead{$R_d$} \\
\colhead{\nodata} & \colhead{\nodata} 
& \colhead{Gal.} & \colhead{Bulge} & \colhead{Disk}   
& \colhead{$B$mag}   & \colhead{Gal.} & \colhead{Bulge}  &
\colhead{Disk}   & \colhead{Bulge}    & \colhead{Bulge} &
\colhead{Disk} \\
\colhead{(1)} & \colhead{(2)} & \colhead {(3)} &
\colhead{(4)} & \colhead{(5)} & \colhead {(6)} &
\colhead{(7)} & \colhead{(8)} & \colhead {(9)} &
\colhead{(10)} & \colhead{(11)} & \colhead {(12)} }
\startdata
1* &     052\_6543 & -21.83 & -20.95 & -21.19 &  0.45 &   0.31 &   0.41 &   0.24 &   1.47 &  18.52 &   4.78  \\* \nodata & \nodata & $^{+0.01 }_{-0.01 }$ & $^{+0.03 }_{-0.03 }$ & $^{+0.03 }_{-0.03 }$ & $^{+0.01 }_{-0.01 }$ & $^{+0.03 }_{-0.02 }$ & $^{+0.04 }_{-0.06 }$ & $^{+0.04 }_{-0.03 }$ & $^{+0.06 }_{-0.07 }$ & $^{+0.11 }_{-0.10 }$ & $^{+0.12 }_{-0.12 }$   \\*      
2* &     062\_2060 & -22.27 & -21.66 & -21.37 &  0.57 &   0.42 &   0.43 &   0.42 &   2.41 &  18.80 &   2.09  \\* \nodata & \nodata & $^{+0.02 }_{-0.02 }$ & $^{+0.09 }_{-0.10 }$ & $^{+0.14 }_{-0.11 }$ & $^{+0.06 }_{-0.04 }$ & $^{+0.01 }_{-0.02 }$ & $^{+0.05 }_{-0.06 }$ & $^{+0.07 }_{-0.07 }$ & $^{+0.15 }_{-0.14 }$ & $^{+0.18 }_{-0.16 }$ & $^{+0.10 }_{-0.07 }$   \\*      
3  &     062\_6465 & -21.53 & -21.08 & -20.53 &  0.66 &   0.18 &   0.88 &  -0.21 &   2.43 &  19.47 &   1.17  \\* \nodata & \nodata & $^{+0.05 }_{-0.07 }$ & $^{+0.18 }_{-0.13 }$ & $^{+0.21 }_{-0.18 }$ & $^{+0.09 }_{-0.11 }$ & $^{+0.03 }_{-0.03 }$ & $^{+0.18 }_{-0.13 }$ & $^{+0.08 }_{-0.11 }$ & $^{+0.38 }_{-0.23 }$ & $^{+0.27 }_{-0.27 }$ & $^{+0.05 }_{-0.05 }$   \\*      
4* &     062\_6859 & -21.53 & -21.07 & -20.38 &  0.66 &   0.48 &   0.44 &   0.54 &   1.57 &  18.45 &   2.53  \\* \nodata & \nodata & $^{+0.03 }_{-0.03 }$ & $^{+0.12 }_{-0.11 }$ & $^{+0.25 }_{-0.19 }$ & $^{+0.07 }_{-0.08 }$ & $^{+0.01 }_{-0.02 }$ & $^{+0.07 }_{-0.07 }$ & $^{+0.15 }_{-0.15 }$ & $^{+0.18 }_{-0.25 }$ & $^{+0.29 }_{-0.37 }$ & $^{+0.24 }_{-0.26 }$   \\
\enddata
\tablecomments{Errors are formal 68\% confidence limits  derived by Monte Carlo sampling
within the GIM2D fitting program. These errors do {\it not} include any systematic or random
errors derived by use of simulations (see Section ~\ref{GIM2D} and GSS2).\\
Col. (1): Sequence number ordered by source ID; galaxies with * belong to 
the quality sample (see Section \ref{quality}).\\
Col. (2): Source ID is given by FFC-XXYY, where FF is the 
subfield,
C is the WFPC2 chip number, and XX and YY are the  
chip coordinates in units of 10 pixels.\\
Col. (3): Absolute magnitude in $B$  of galaxy. \\
Col. (4): Absolute magnitude in $B$  of photo-bulge. \\
Col. (5): Absolute magnitude in $B$  of photo-disk. \\
Col. (6): Photo-bulge to total ratio in restframe $B$; random errors are
generally $\gtrsim 0.1$. Section \ref{GIM2D} discusses systematic errors.\\
Col. (7): Restframe $U-B$ color of galaxy. \\
Col. (8): Restframe $U-B$ color of photo-bulge. \\
Col. (9): Restframe $U-B$ color of photo-disk. \\
Col. (10): Effective (half-light) major-axis radius of photo-bulge
in kpc; note that $R_e (1-e)^{1/2}$
gives the circularized effective radius $R_{e,c}$ as measured with a circular aperture, 
where $e$ is from column 11 of Table 2. \\
Col. (11): Restframe $B$ surface brightness of photo-bulge
in units of mag per square arcsec averaged  within the major-axis effective 
radius. \\
Col. (12): Exponential scale length of photo-disk 
in kpc.  \\
}
\end{deluxetable}

\begin{deluxetable}{l l c c c c }
\tabletypesize{\scriptsize}
\tablecolumns{6}
\tablewidth{0pt}
\tablecaption{Summary of Galaxy Classifications\label{table4}}
\tablehead{
\colhead{Note} &
\colhead{Class or Group} &
\colhead{No. in } &
\colhead{\%  } &
\colhead{No. in} &
\colhead{\% } \\
\colhead{\nodata} &
\colhead{\nodata} & \colhead{Total Sample} & \nodata &
\colhead{Quality Sample}   & \nodata
}
\startdata
\nodata  & Sample size & 86  & 100 & 52 & 100 \\*
\nodata  & Very red pB & 58  & 67 & 41 & 79 \\*
\nodata  & Less red  pB & 11 & 13 & 7 & 13 \\*
1 & Both pB and pD very red  &  19  & 22 & 16 & 31 \\*      
2 & Both pB and pD less red or very-red & 44 & 51 & 34 & 65 \\*
3 & Very red pB and blue or very blue pD & 19   & 22 & 10 & 19 \\*
4 & Less red pB and blue or very blue pD & 6 & 7 & 4 & 8 \\*   
5 & Blue or very blue pB and red or very red pD &  5  & 6 & 2 & 4 \\
\enddata
\tablecomments{Color definitions for photo-bulge (pB) or photo-disk (pD):\\
very red ($U-B \geq 0.25$), less red ($0 \leq U-B < 0.25$), \\
blue ($-0.25 \leq U-B < 0.$), very blue ($U-B < -0.25$).\\
1 --  Best candidates for passively evolving  E-S0. \\
2 --  Probable candidates for largely passively evolving E-S0. \\
3 --  Very red bulge imbedded within a blue disk. \\
4 --  Red bulge imbedded within a blue disk. \\
5 --  Blue bulge within a red disk. \\
6 --  Both bulge and disk are blue. \\
7 --  $r_e \geq 1.7 \times r_d$ (see Table 2) so a likely reversal of actual bulge and disk. \\
8 --  Use Im \etal criterion of  pB/T$ > 0.4$  (RT2+RA2 $\leq$ 0.08). \\
9 -- pB/T $\sim$ 1 within 68\% confidence limits. \\
10 -- pB/T $\geq$ 0.8 within 68\% confidence limits. \\
11 --  Lower number is for bulge $U-B < 0$; larger number is for $U-B < 0.14$. \\
}
\end{deluxetable}

\begin{deluxetable}{r l r c c c l}
\tablecolumns{7}
\tablewidth{0pt}
\tablecaption{Summary of Median Restframe $U-B$ Colors\label{table5}}
\tablehead{
\colhead{Seq.} &
\colhead{Sample} &
\colhead{Sample } &
\colhead{Median} &
\colhead{$U-B$} &
\colhead{$U-B$} &
\colhead{Comments} \\
\colhead{No.} &
\colhead{\nodata} & \colhead{Size} & \colhead{z}
& \colhead{Bulge} & \colhead{Galaxy} & \colhead{\nodata}}
\startdata
1 & All High z Galaxies & 211 & 0.83 & 0.34 & -0.09 & 1 \\
2 & High z Red Galaxies &  60 & 0.83 & 0.51 &  0.31 & 2 \\
3 & Full Bulge Sample       &  86 & 0.85 & 0.46 & 0.23 & 3 \\
4 & Quality Bulges   &  52 & 0.82 & 0.49 & 0.31 & 4 \\
5 & Very Red Quality Bulges & 41 & 0.82 & 0.51 & 0.34 & 5 \\
6 & As previous with large B/T & 26 & 0.90 & 0.52 & 0.36 & 6 \\
7 & Early-type Bulges & 18 & 0.81 & 0.51 & 0.39 & 7 \\
8 & Cluster Early-type Galaxies & 30 & 0.83 & \nodata & 0.45 & 8 \\
9 & Local Early Spiral Bulges & 29 & 0.0 & 0.40 & \nodata & 9 \\
10 & Local E-S0 Galaxies & 403 & 0.0 & \nodata & 0.52 & 10 \\
\enddata
\tablecomments{\\
1 High z means 0.73 $< z <$ 1.04 of high quality redshift sample. \\
2 Red means  $U-B > 0.15$ for integrated galaxy colors.\\
3 Median redshift lies between 0.84 and 0.87. \\
4 See text for definition of quality bulge sample. \\
5 Very red means $U-B > 0.25$ for photo-bulge component. \\
6 As in sample 5 with restframe $pB/T > 0.5$. \\
7 Early-type are those within sample of \cite{im02}. \\
8 Early-type (E-S0) members of cluster MS1054-03 from \cite{dokkum00}.\\
9 Early-type spiral bulge colors from  \cite{peletier96}.\\
10 E-S0 from RC3 for $M_B = -20.5$ of color-mag relation from  \cite{schweizer92}.
}
\end{deluxetable}

\end{document}